\documentclass[10pt,journal]{IEEEtran}
\usepackage{cite}
\usepackage{graphicx}
\usepackage{amsmath}
\usepackage{times}
\usepackage{latexsym}
\usepackage{graphicx}
\usepackage{bm}
\usepackage{amssymb}
\usepackage[center]{caption2}
\usepackage{stfloats}
\usepackage{cases}
\usepackage{url}
\usepackage{array}
\usepackage{setspace}
\usepackage{fancyhdr}
\usepackage{braket}
\usepackage{color}
\usepackage{subfigure}
\usepackage{algorithm}
\usepackage{algorithmicx}
\usepackage{algpseudocode}
\usepackage{amsmath}
\DeclareMathAlphabet{\mathscr}{OT1}{pzc}{m}{it}

\newtheorem{theorem}{Theorem}

\newtheorem{corollary}{Corollary}

\newtheorem{remark}{Remark}
\def\proof{\noindent\hspace{2em}{\itshape Proof: }}
\def\endproof{\hspace*{\fill}~$\square$\par\endtrivlist\unskip}

\allowdisplaybreaks[4]
\newcommand{\RNum}[1]{\uppercase\expandafter{\romannumeral #1\relax}}

\makeatletter
\renewcommand{\maketag@@@}[1]{\hbox{\m@th\normalsize\normalfont#1}}%
\makeatother
\begin{document}
\title{Exploiting Movable Elements of Intelligent Reflecting Surface for Enhancement of Integrated Sensing and Communication}
\author{Xingyu Peng, Qin Tao, Yong Liang Guan, and Xiaoming Chen \vspace{-8mm}
\thanks{X. Peng and X. Chen are with the College of Information Science and Electronic Engineering, Zhejiang University, Hangzhou 310027, China (Email: $\{$peng$\_$xingyu, chen$\_$xiaoming$\}$@zju.edu.cn).}
\thanks{Qin Tao is with the School of Information Science and Technology, Hangzhou Normal University, Hangzhou 311121, China. (Email: taoqin@hznu.edu.cn).}
\thanks{Yong Liang Guan is with the School of Electrical and Electronic Engineering,
Nanyang Technological University, Singapore. (Email: eylguan@ntu.edu.sg).}
}
\maketitle
\begin{abstract}
In this paper,  we propose to exploit movable elements of intelligent reflecting surface (IRS) to enhance the overall performance of integrated sensing and communication (ISAC) systems. Firstly, focusing on a single-user scenario, we reveal the function of movable elements by performance analysis, and then design a joint beamforming and element position optimization scheme. Further, we extend it to a general multi-user scenario, and also propose an element position optimization scheme according to the derived performance expressions. Finally, simulation results confirm that the movement of IRS elements can improve the communication rate and the sensing accuracy, and especially broaden the coverage of ISAC. 
\end{abstract}
\vspace{-0mm}\section{Introduction}
In recent years, the growing demand for high-frequency signals in communication systems has intensified the competition between communication and sensing systems for spectrum resources. To address this, integrated sensing and communication (ISAC) has been identified as one of typical scenarios of sixth-generation (6G) wireless networks. ISAC facilitates simultaneous communication and sensing by leveraging shared hardware and spectrum,  delivering high-speed data transmission and precise sensing capabilities
\cite{ISAC1,ISAC2,ISAC3}. However, due to the sharing of resources, performance trade-offs between communication and sensing need to be carefully considered.

Movable antennas (MAs) \cite{MA-CHANNEL}, also known as the fluid antennas (FAs) \cite{FA2}, where the positions of each antenna can be adjusted by connecting the MA to radio frequency (RF) chains via flexible cables, open new possibilities for both communication and sensing. 
Therefore, MAs or FAs have gained prominence for their ability to fully exploit spatial variations in wireless channels within the movable regions through antenna position adjustments. 
Specifically, the work \cite{MA-CHANNEL}  characterized the capacity of MA-enabled MIMO systems, demonstrating that MIMO channel power can be significantly enhanced while reducing the channel condition numbers. Then, the work \cite{MA1} examined multi-user scenarios, showing that optimizing MA positions can leverage spatial diversity to boost desired signal power and suppress inter-user interference. In sensing systems, MA techniques have also shown notable potential for interference suppression and array gains \cite{MA-SEN,MA-SEN2,MA_Full}. The work \cite{MA-SEN2} applied MA arrays to the low-earth orbit satellite networks, enhancing satellite beam coverage and interference mitigation by the proposed low-complexity and low movement overhead position scheme.

To enhance the performance of ISAC with limited wireless resources, intelligent reflecting surfaces (IRSs) have been widely regarded as a promising assistance technology \cite{IRS_ISAC_MOV}.
By adjusting the phase shifts of the reflecting elements, the IRS provides substantial passive gain, enhancing the overall performance of ISAC systems.
Moreover, the IRS extends the coverage of both sensing and communication by creating virtual line-of-sight (VLoS) paths between transceivers \cite{WQQ}. This capability is particularly crucial for ISAC signals operating in the millimeter-wave (mmWave) or terahertz (THz) bands in dense urban or indoor environments. In such scenarios, users or sensing targets are likely to be located in signal dead zones, where direct links are practically infeasible despite the general LoS reliance of high-frequency systems \cite{mmWave}. Current researches leverage IRS to assist ISAC system in two main ways:
Firstly, IRSs can improve both communication and sensing performance via joint beamforming design  \cite{IRS-ISAC2,IRS-ISAC148,IRS-ISAC169,Twc}.
Secondly, IRSs can provide additional spatial degrees of freedom, enabling more accurate estimation of sensing-related parameters \cite{TLS-ESPRIT, mmWave, Tcom, estimation4}. These capabilities allow for one base-station (BS) setup to perform localization while maintaining communication with users, thus reducing system costs compared to the traditional multi-BS localization configurations.

However, traditional fixed-spacing element-based IRS systems require a large number of reflecting elements to effectively improve system performance, which complicates precise phase shift control. Besides, it was demonstrated in \cite{MA-IRS1} that the phase distribution offset problem of the conventional FPA-IRS can be addressed if the position of each IRS element can be flexibly adjusted exploiting the MA technology.
Moreover, the work \cite{MA-IRS2} and \cite{MA-IRS0} demonstrated that the IRS with movable elements outperforms the traditional fixed-spacing element-based IRS in communication systems, particularly in terms of signal-to-noise ratio (SNR) and outage probability compared to integrating MA with the BS.
Intuitively, by leveraging the extra spatial flexibility, an IRS with movable elements can reposition its elements to locations with more favorable channel conditions, thereby enhancing both sensing and communication performance. 
Moreover, 
an IRS with movable elements introduces an additional degree of freedom by allowing direct modification of the channel itself through element repositioning. This mobility enables adaptive optimization of the correlation between communication and sensing subspaces, providing a novel means to balance these two functionalities \cite{R1,R2}.
However, the phase shift design algorithms proposed in previous studies are no longer applicable due to the mobility of IRS elements. Furthermore, the presence of numerous local optima within the movable region can lead to performance degradation, rendering the expected gains negligible. 
Therefore, developing an effective strategy to balance communication and sensing performance remains a significant challenge.  

\vspace{-0mm}
\subsection{Contributions}
In this paper, we focus on IRS-assisted ISAC systems with movable elements.
We analyze the communication and sensing performance across different scenarios and propose two joint beamforming and element position design schemes to further leverage the potential spatial degrees of freedom provided by movable elements, effectively balancing communication and sensing performance.
The main contributions are summarized as follows:
\begin{itemize}
\item{For the single-user scenario, we first analyze the upper bounds on ISAC performance by exploiting movable elements of IRS. With the derived closed-form expressions, we propose a low-complexity algorithm for element position and beamforming design, which can achieve optimal communication and sensing performance by leveraging angular information, reducing the pilot overhead.}
\item{For the multi-user scenario, we derive the performance lower bounds on both communication and sensing, showing that the ISAC performance achieved by an IRS with movable elements is better than that of a traditional IRS with half-wavelength spacing. A memory-penalized projected gradient descent (MPPGD) algorithm is introduced to reduce the risk of suboptimal local maxima, thereby enhancing the overall performance of both sensing and communication.}
\end{itemize}

The remainder of the paper is organized as follows. Section \ref{se:se2} introduces the system model.  Section \ref{se:se3} and  Section \ref{se:se4} analyze the overall performance of IRS with movable elements, and propose effective beamforming and element position design algorithms for different scenarios. Numerical results and discussions are provided in Section \ref{se:se5}, and finally, Section \ref{se:se6} concludes the paper.
\vspace{-0mm}
\section{System Model}\label{se:se2}
\begin{figure}[htbp]
\centering
\includegraphics[width=3.3in]{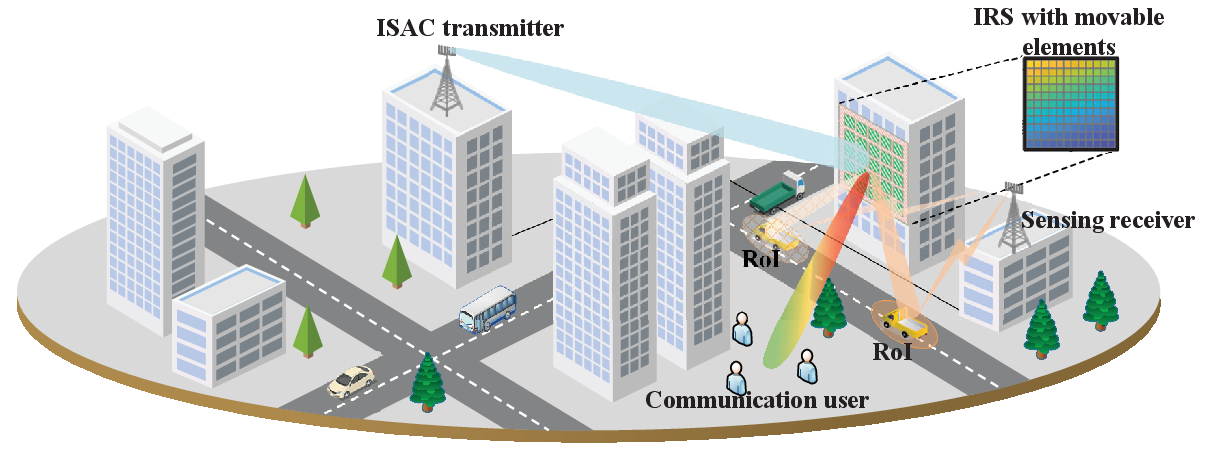}
\caption{System model: An IRS-assisted ISAC system with movable elements.}\label{fig:fig1}
\end{figure}

Consider an IRS-assisted ISAC system, as illustrated in Fig. \ref{fig:fig1}, consisting of one ISAC transmitter with $N_{\mathrm{B}}$ uniform linear array (ULA) antennas, an IRS with $N_{\mathrm{I}}$ movable reflecting elements, a dedicated sensing receiver with $N_{\mathrm{S}}$ ULA antennas, and $K$ communication user each with a single antenna, operating in the millimeter-wave (mmWave) spectrum band. 
Due to limited penetrability of mmWave signals, the direct paths between the ISAC transmitter and the users/target are obstructed \cite{IRS1}. To overcome this, an IRS with movable reflecting elements is deployed to create VLoS paths. Moreover, the IRS with movable elements\footnote{To avoid the coupling effect, the spacing between adjacent movable elements must exceed half the carrier wavelength \cite{MA-IRS1}.} is designed to assist the dedicated sensing receiver in detecting targets of interest within a region of interest (RoI) along a prescribed sensing direction, i.e., $\varphi _{{\rm{T}}}^{\rm{S}}$, in a static cluttered environment, while simultaneously enhancing user communication.
For convenience of analysis, we define a three-dimensional (3D) Cartesian coordinate, where the ISAC transmitter, sensing receiver, IRS, target, and user are located at positions $\mathbf{p}_{\mathrm{i}}=\left(x_{\mathrm{i}},y_{\mathrm{i}},z_{\mathrm{i}}\right)^T$ with the superscript ``$\mathrm{T}$'' denoting the transpose and subscript $i \in \left\{ {\mathrm{B},\mathrm{S},\mathrm{I},\mathrm{t},\mathrm{k}}\right \}$. 
The position of the $m$-th movable reflecting element of the IRS are represented by $\mathbf{r}^{\mathrm{I}}_m=\left(\mathrm{x}_m^{\mathrm{I}},\mathrm{y}_m^{\mathrm{I}}\right)^T \in \mathcal{C}$, respectively, where $\mathcal{C}$ is a square region of size $A \times A$.

Let ${{\bf{s}}_{\mathrm{c},k}\left(t\right)}$ and ${{\bf{s}}_{\mathrm{s}}\left(t\right)}$ denote the communication signal for the $k$-th user and the sensing signal with independent and identically distributed (i.i.d.) elements of zero mean and unit norm, respectively. Therefore, the signal $\mathbf{x}\left(t \right)$ transmitted by the ISAC transmitter can be expressed  as\footnote{The ISAC transmitter controls the phase shifts and positions of the IRS elements via the backhaul link. Therefore, perfect beam alignment is assumed and the movement errors are neglected.}\vspace{-0mm}
\begin{equation}\vspace{-0mm}
\mathbf{x}\left(t \right)= \sum\limits_{k=1}^{K}\mathbf{w}_{\mathrm{c},k}{{\bf{s}}_{\mathrm{c},k}\left(t\right)}+\mathbf{w}_\mathrm{s}{{\bf{s}}_{\mathrm{s}}\left(t\right)},
\end{equation}
satisfying $\mathbb{E} \left \{\left\| {\mathbf{x}\left(t \right)} \right\|^2\right\}=\mathrm{Tr}\left( {{\mathbf{W}}{{\bf{W}}^H}} \right) \le P$, where $P$ is the transmit power constraint,  $\mathbb{E}\left \{ \mathbf{a} \right \}$ denotes the expectation of $\mathbf{a}$, and $\mathrm{Tr}\left(\mathbf{A}\right)$ denotes the trace of matrix $\mathbf{A}$. The beamforming matrix at the ISAC transmitter is denoted as $\mathbf{W}=\left[\mathbf{w}_{\mathrm{c},1},\cdots,\mathbf{w}_{\mathrm{c}, K},\mathbf{w}_\mathrm{s}\right]$, where $\mathbf{w}_{\mathrm{c},k}$ and $\mathbf{w}_\mathrm{s}$ are communication beamforming for the $k$-th user and sensing beamforming vectors for target detection,\footnote{The radar waveforms aim to design the sensing beamforming through an optimal matrix $\mathbf{R}={\bf{W}}{{\bf{W}}^H}$\cite{ISAC1,R1}. Therefore, we can use $\mathbf{w}_{\mathrm{s}}$ to detect the target.} respectively.\footnote{Given that the clutter interference is substantial, a dedicated sensing signal and beamformer are employed for achieving stable performance.} \vspace{-0mm}\vspace{-0mm}
\subsection{Communication Model}
The channel matrix is determined by the signal propagation environment and the positions of the IRS elements. Therefore, the communication channel from the ISAC transmitter to the IRS and from the IRS to the $k$-th users can be expressed as the following field-response channel model \cite{MA-CHANNEL} \vspace{-0mm}
\begin{equation}\vspace{-0mm}\begin{array}{l}
{\bf{H}}_{{\rm{BI}}}=\mathbf{F}_{\mathrm{BI}}^H{\mathbf{\Sigma}_{\mathrm{BI}} \mathbf{G}_{\mathrm{BI}}},
\end{array}\end{equation}
and\vspace{-0mm}
\begin{equation}\vspace{-0mm}\begin{array}{l}
{\bf{h}}^H_{{\mathrm{IU},k}}=\mathbf{1}_{\mathrm{IU}}^H{\mathbf{\Sigma}_{\mathrm{IU},k} \mathbf{G}_{\mathrm{IU},k}},
\end{array}\end{equation}
where the superscript ``$\mathrm{H}$'' denotes the conjugate transpose, and the field response matrices (FRM) of the transmitter antennas, i.e., from the transmitter to the IRS $\mathbf{G}_{\mathrm{BI}}$, and movable elements, i.e., from the IRS to the $k$-th users $\mathbf{G}_{\mathrm{IU},k}$ and received at the IRS $\mathbf{F}_{\mathrm{BI}}$, are respectively defined as $\mathbf{G}_{\mathrm{BI}}\triangleq[\mathbf{g}_{\mathrm{BI}}\left({y}^{\rm{B}}_{1}\right),\mathbf{g}_{\mathrm{BI}}\left(y^{\rm{B}}_{2}\right),\ldots,\mathbf{g}_{\mathrm{BI}}\left(y^{\rm{B}}_{N_{\mathrm{B}}}\right)]$, 
$\mathbf{G}_{\mathrm{IU},k}\triangleq[\mathbf{g}_{{\mathrm{IU},k}}\left(\mathbf{r}_{1}^{\mathrm{I}}\right),\mathbf{g}_{\mathrm{IU},k}\left(\mathbf{r}_{2}^{\mathrm{I}}\right),\ldots,\mathbf{g}_{\mathrm{IU},k}\left(\mathbf{r}_{N_{\mathrm{I}}}^{\mathrm{I}}\right)]$,
 and $\mathbf{F}_{\mathrm{BI}}\triangleq[\mathbf{f}_{\mathrm{BI}}\left(\mathbf{r}_{1}^{\mathrm{I}}\right),\mathbf{f}_{\mathrm{BI}}\left(\mathbf{r}_{2}^{\mathrm{I}}\right),\ldots,\mathbf{f}_{\mathrm{BI}}\left(\mathbf{r}_{N_{\mathrm{I}}}^{\mathrm{I}}\right)]$,
where $y^{\rm{B}}_{{m}}$ is the position of the $m$-th antenna at ISAC transmitter. $\mathbf{\Sigma}_{\mathrm{BI}}= \mathrm{diag}\left(\alpha_{\mathrm{BI,}1},\cdots,\alpha_{\mathrm{BI},L_{\mathrm{BI}}} \right)$ and $\mathbf{\Sigma}_{\mathrm{IU},k}= \mathrm{diag}\left(\alpha_{\mathrm{IU},k,1},\cdots,\alpha_{\mathrm{IU},k,L_{\mathrm{IU}}} \right)$ denote the path-response matrices (PRM), with each diagonal element following a circularly symmetric complex Gaussian (CSCG) distribution, i.e., $\mathcal{CN} \left(0,\frac{\sigma_{0}^2}{L}\right)$, where $L$ and $\frac{\sigma_{0}^2}{L}$ denote the number of paths and channel power of each path, the transmit and receive field-response vectors (FRVs) for the communication channel are respectively given by \vspace{-0mm}
\begin{equation}\vspace{-0mm}\begin{array}{l}
\mathbf{g}(\mathbf{r}_{m})\triangleq\left[e^{j\frac{2\pi}{\lambda}\rho_{1}(\mathbf{r}_m)},e^{j\frac{2\pi}{\lambda}\rho_{2}(\mathbf{r}_m)},\ldots,e^{j\frac{2\pi}{\lambda}\rho_{L}(\mathbf{r}_m)}\right]^{T},
\end{array}\end{equation}
and\vspace{-0mm}
\begin{equation}\vspace{-0mm}
\mathbf{f}(\mathbf{r}_m)\triangleq\left[e^{j\frac{2\pi}{\lambda}\rho_{1}(\mathbf{r}_m)},e^{j\frac{2\pi}{\lambda}\rho_{2}(\mathbf{r}_m)},\ldots,e^{j\frac{2\pi}{\lambda}\rho_{L}(\mathbf{r}_m)}\right]^{T},
\end{equation}
where $\lambda $ is the carrier wavelength, $j$ denotes the imaginary part, $\rho_{l}^{\mathrm{B}}(\mathbf{r}_m^{\mathrm{B}})=y_m^{\mathrm{B}}\varphi_l^{{{\mathrm{B}}}}$, $\rho_{{l,k}}^{\mathrm{IU}}(\mathbf{r}_m^{\mathrm{I}})=x_m^{\mathrm{I}}\varphi_{l,k}^{{{\mathrm{IU}}}}+y_m^{\mathrm{I}}\omega_{l,k}^{{{\mathrm{IU}}}}$, and $\rho_{l}^{\mathrm{BI}}(\mathbf{r}_m^{\mathrm{I}})=x_m^{\mathrm{I}}\varphi_l^{{{\mathrm{BI}}}}+y_m^{\mathrm{I}}\omega_l^{{{\mathrm{BI}}}}$, where the virtual angles of departure (AoDs) and angles of arrival (AoAs) of the $l$-th path  \cite{MA-CHANNEL} are defined as\vspace{-0mm}
\begin{equation}\label{eq:eq6}\vspace{-0mm}
\varphi_l^{{{{i}}}}=\cos\phi _l^{{i}}\cos\psi _l^{{i}},\;
\omega_l^{{{{i}}}}=\cos\phi _l^{{i}}\sin\psi _l^{{i}}, \forall {i} \in \{\mathrm{B},\mathrm{IU},\mathrm{BI}\}
\end{equation}
where $\{\phi _l^{\mathrm{B}},\psi _l^{\mathrm{B}}\}$ and $\{\phi _{l,k}^{\mathrm{IU}},\psi _{l,k}^{\mathrm{IU}}\}$ are the elevation and azimuth effective AoDs of the $l$-th path that from the ISAC transmitter to IRS, and from IRS to the $k$-th user. $\{\phi _l^{\mathrm{BI}},\psi _l^{\mathrm{BI}}\}$ represents the elevation and azimuth effective AoAs at the IRS. 

The signal sent by the ISAC transmitter is first reflected by the IRS, and then arrived at the users. Thus, the received signal at the $k$-th user can be expressed as\vspace{-0mm}
\begin{equation}\label{eq:eqyc}\vspace{-0mm}\scalebox{1}{$
\hspace{-2mm}\begin{array}{l}
\mathbf{y}_{\mathrm{c},k}\left(t\right)= {\bf{h}}_{{\mathrm{IU},k}}^{{H}}{\bf{\Theta }}{{\bf{H}}_{{\rm{BI}}}}{{\bf{w}}_{\mathrm{c},k}}{\mathbf{s}_{\mathrm{c},k}}\left(t\right)
+  \sum\limits_{i \neq k}^{K}{\bf{h}}_{{\mathrm{IU},k}}^{{H}}{\bf{\Theta }}{{\bf{H}}_{{\rm{BI}}}}{{\bf{w}}_{\mathrm{c},i}}{\mathbf{s}_{\mathrm{c},i}\left(t\right)}\\
+  {\bf{h}}_{{\mathrm{IU},k}}^{{H}}{\bf{\Theta }}{{\bf{H}}_{{\rm{BI}}}}{{\bf{w}}_{s}}{\mathbf{s}_{\mathrm{s}}\left(t\right)}+ \mathbf{n}_{\mathrm{c}}\left(t\right),
\end{array}$}
\end{equation}
where the third term corresponds to the interference caused by the sensing signals and $\mathbf{n}_{\mathrm{c}}\left(t\right)$ denotes the additive white Gaussian noise (AWGN) with power $\sigma^2$.  $\boldsymbol{\Theta}=\mathrm{\mathrm{diag}}  \left(\boldsymbol{\xi}\right)$ is the phase shift matrix of the IRS, where $\boldsymbol{\xi}=\left [ e^{\mathrm{j}\theta _{1}},\cdots , e^{\mathrm{j}\theta _{N_{\mathrm{I}}}} \right ] ^T$ is the phase shift vector and $\theta _{m}\in [0,2\pi], \forall m=[1,\cdots,N_{\mathrm{I}}]$ is the phase shift of the $m$-th element.  
\vspace{-0mm}
\subsection{Sensing Model}
Meanwhile, the signal reflected by the IRS also arrives at the sensing receiver. If a target exists, the signal is reflected by the target.\footnote{For scenarios with multiple targets, sequential detection can be achieved through a narrow beam with time-division approach to reduce the interference from multiple targets.} Otherwise, the signal is reflected by the clutters and arrives at the sensing receiver.\footnote{The transceivers are connected via wired or wireless backhaul links, enabling the sensing receiver to acquire prior knowledge of environmental clutter characteristics \cite{bi-static}.} Due to the presence of clutters, the binary hypothesis testing problem for target detection can be modeled as \vspace{-0mm}
\begin{equation}\vspace{-0mm}
\begin{aligned}
&\mathbf{y}_{\mathrm{s}}\left(t\right)=\begin{cases}
 \mathbf{y}_{\mathrm{s,C}}\left(t\right)+ \mathbf{n}_{\mathrm{s}}\left(t\right),\mathcal{H}_0\\
\mathbf{y}_{\mathrm{s,T}}\left(t\right)+\mathbf{y}_{\mathrm{s,C}}\left(t\right)+ \mathbf{n}_{\mathrm{s}}\left(t\right),\mathcal{H}_1,\end{cases}
\end{aligned}
\end{equation}
where $\mathbf{n}_{\mathrm{s}}\left(t\right)$ is the AWGN with power $\sigma^2$, hypotheses $\mathcal{H}_0$ and $\mathcal{H}_1$ correspond to the absence and presence of the target, respectively. 
The target-related signal, i.e., $\mathbf{y}_{\mathrm{s,T}}\left(t\right)$, and the clutter-related interference signal, i.e., $\mathbf{y}_{\mathrm{s,C}}\left(t\right)$, can be expressed as\vspace{-0mm}
\begin{equation}\vspace{-0mm}
\mathbf{y}_{\mathrm{s,T}}\left(t\right)=\mathbf{r}^H_{\mathrm{co}}{{\alpha _{{\rm{IS,T}}}}{\bf{a}}_{{\rm{S}}}^*(\varphi _{{\rm{T}}}^{\rm{S}}){\bf{a}}_{\rm{I}}^T(\varphi _{{\rm{T}}}^{\rm{IS}},\omega _{{\rm{T}}}^{\rm{IS}}){\bf{\Theta }}{{\bf{H}}_{{\rm{BI}}}}\mathbf{x}\left(t\right)},
\end{equation}
\vspace{-0mm}and\vspace{-0mm}
\begin{equation}\vspace{-0mm}
\hspace{-2mm}\mathbf{y}_{\mathrm{s,C}}\left(t\right)={\mathbf{r}^H_{\mathrm{co}}\sum\limits_{c=0}^{C}{\alpha _{{\rm{IS,c}}}}{\bf{a}}_{{\rm{S}}}^*(\varphi _{{\rm{c}}}^{\rm{S}}){\bf{a}}_{\rm{I}}^T(\varphi _{{\rm{c}}}^{\rm{IS}},\omega _{{\rm{c}}}^{\rm{IS}})
{\bf{\Theta }}{{\bf{H}}_{{\rm{BI}}}}\mathbf{x}\left(t\right)},
\end{equation}
where  ${\left\| {\bf{r}}_{\mathrm{co}} \right\|^2} = 1$ is the sensing beamforming vector, and $C$ is the number of clutters. Here, $c=0$ represent the direct link between the IRS and sensing receiver. Moreover, $\alpha_{\mathrm{IS,c}}$ and $\alpha_{\mathrm{IS,T}}$ denote the complex gain for the $c$-th clutter and target, respectively.
The array response for the $l$-th path can be expressed as \vspace{-0mm}
\begin{equation}\vspace{-0mm}\begin{array}{l}
\mathbf{a}_{{{\mathrm{S}}}}(\varphi^{{\mathrm{S}}}_l)\hspace{-1mm} =\hspace{-1mm}{{\left[ {{e}^{j \frac{2\pi}{\lambda}\rho_l^{{\mathrm{S}}}(\mathbf{r}_1^{\mathrm{\mathrm{S}}})}},{{e}^{j \frac{2\pi}{\lambda}\rho_l^{{\mathrm{S}}}(\mathbf{r}_2^{{\mathrm{S}}})}},\hspace{-1mm}\cdots \hspace{-1mm},{{e}^{j \frac{2\pi}{\lambda}\rho_l^{{\mathrm{S}}}(\mathbf{r}_{N_{{\mathrm{S}}}}^{{\mathrm{S}}})}}\right]^{T}}}\hspace{-1mm},\end{array}
\end{equation}
and
\vspace{-0mm}
\begin{equation}\vspace{-0mm}\begin{array}{l}
{\bf{a}}_{\rm{I}}(\varphi_l^{{\mathrm{IS}}},\omega_l^{{\mathrm{IS}}}) \hspace{-1mm}=\hspace{-1mm}{{\left[ {{e}^{j \frac{2\pi}{\lambda}\rho_l^{{\mathrm{IS}}}(\mathbf{r}_1^{\mathrm{I}})}},{{e}^{j \frac{2\pi}{\lambda}\rho_l^{{\mathrm{IS}}}(\mathbf{r}_2^{\mathrm{I}})}},\hspace{-1mm}\cdots \hspace{-1mm},{{e}^{j \frac{2\pi}{\lambda}\rho_l^{{\mathrm{IS}}}(\mathbf{r}_{N_{\mathrm{I}}}^{\mathrm{I}})}}\right]^{T}}}\hspace{-2mm}, \end{array}
\end{equation}
where $\rho_{l}^{\mathrm{IS}}(\mathbf{r}_m^{\mathrm{I}})=x_m^{\mathrm{I}}\varphi_l^{{{\mathrm{IS}}}}+y_m^{\mathrm{I}}\omega_l^{{{\mathrm{IS}}}}$,
$\rho_{l}^{\mathrm{S}}(\mathbf{r}_m^{\mathrm{S}})=y_m^{\mathrm{S}}\omega_l^{{{\mathrm{S}}}}$, and $y^{\rm{S}}_{{m}}$ is the position of the $m$-th antenna at the sensing receiver. 
$\varphi^{{\mathrm{S}}}_l$ and $\{\varphi_l^{{\mathrm{IS}}},\omega_l^{{\mathrm{IS}}}\}$ are the virtual AoAs of the $l$-th path from the IRS to the sensing receiver, and AoDs of the $l$-th path from the IRS to the clutters, respectively.

\section{Single User Scenario}\label{se:se3}
In this section, we first explore the single-user scenario to highlight the advantages of IRS with movable elements. To gain some useful insights, we analyze the achievable performance upper bounds on the ISAC systems and propose an angular information-based scheme for beamforming and movable element position design to achieve these performance limits.
For the ease of analysis, we consider the LoS path between the ISAC transmitter, IRS, and user that is widely adopted in millimeter wave systems \cite{IRS1}, i.e., $L_{\mathrm{BI}}=L_{\mathrm{IU}}=1$. Moreover, due to the large power disparity among the direct link and NLoS links between the IRS and the sensing receiver, the direct link is considered as a significant source of clutter interference for target detection \cite{SIC_radar}.
\vspace{-0mm}\subsection{Performance Bound Analysis}
Based on Eq. $\left(\ref{eq:eqyc}\right)$, the received signal-to-interference-plus-noise ratio (SINR) at the $k$-th user is given by
\begin{equation}\vspace{-0mm}
\begin{array}{l}
{\gamma_{\mathrm{c}}}\left( {{{\bf{W}}},{\bf{r}^{\mathrm{I}}},{\bf{\Theta }}} \right) 
= \frac{{\left | {\alpha _{{\rm{IU}}}}{\alpha _{{\rm{BI}}}}{\bf{a}}_{\rm{I}}^T(\varphi^ {\rm{IU}},\omega^{\rm{IU}})
{\bf{\Theta }}{\bf{a}}_{\rm{I}}^*(\varphi ^{\rm{BI}},\omega ^{\rm{BI}})
{\bf{a}}_{{\rm{B}}}^T(\varphi^{\rm{B}}){{\bf{w}}_c} \right | ^2}}
{{  \left | {{\alpha _{{\rm{IU}}}}{\alpha _{{\rm{BI}}}}{\bf{a}}_{\rm{I}}^T(\varphi^ {\rm{IU}},\omega^{\rm{IU}}){\bf{\Theta }}{\bf{a}}_{\rm{I}}^*(\varphi ^{\rm{BI}},\omega ^{\rm{BI}})
{\bf{a}}_{{\rm{B}}}^T(\varphi^{\rm{B}}){{\bf{w}}_{s}}} \right | ^2 + {\sigma ^2}}}.
\end{array}
\end{equation}

It has been proved that the detection probability is directly related to the signal-to-clutter-and-noise-ratio (SCNR) \cite{3,SCNR-P}. Therefore, the SCNR is adopted as the sensing performance metric, which is given by
\begin{equation}
\hspace{-3mm}\begin{array}{l}
\gamma_{\mathrm{s}}\left( {{{\bf{W}}},{\bf{r}^{\mathrm{I}}},{\bf{\Theta }}},{\bf{r}_{\mathrm{co}}} \right)= \\
\frac{{{{\left\| {\alpha _{{\rm{IS,T}}}}{\alpha _{{\rm{BI}}}}{\mathbf{r}_{\mathrm{co}}^H}{\bf{a}}_{{\rm{S}}}^*(\varphi _{{\rm{T}}}^{\rm{S}}){\bf{a}}_{\rm{I}}^T(\varphi _{{\rm{T}}}^{\rm{IS}}\hspace{-0.5mm},\hspace{-0.5mm}\omega _{{\rm{T}}}^{\rm{IS}}){\bf{\Theta }}
{\bf{a}}_{\rm{I}}^*(\varphi ^{\rm{BI}}\hspace{-0.5mm},\hspace{-0.5mm}\omega ^{\rm{BI}}){\bf{a}}_{{\rm{B}}}^T(\varphi^{\rm{B}}){\mathbf{W}} \right\|}^2}}}{{{{\left\| {{{\alpha _{{\rm{IS,}}0}}{\alpha _{{\rm{BI}}}}{{\mathbf{r}_{\mathrm{co}}^H}}{\bf{a}}_{{\rm{S}}}^*(\varphi _{0}^{\rm{S}}){\bf{a}}_{\rm{I}}^T(\varphi _{0}^{\rm{IS}}\hspace{-0.5mm},\hspace{-0.5mm}\omega _{0}^{\rm{IS}}){\bf{\Theta }}
{\bf{a}}_{\rm{I}}^*(\varphi ^{\rm{BI}}\hspace{-0.5mm},\hspace{-0.5mm}\omega ^{\rm{BI}}){\bf{a}}_{{\rm{B}}}^T(\varphi^{\rm{B}})
{\mathbf{W}}}} \right\|}^2} + {\sigma ^2}}}.
\end{array}
\end{equation}
\begin{theorem}\label{theorem:0}
For the considered single-user scenarios, the upper bound for SCNR and SINR can be derived as\vspace{-0mm}
\begin{equation}\label{equp}\vspace{-0mm}\begin{array}{l}
\gamma_{\mathrm{s}}\left( {{{\bf{W}}},{\bf{r}^{\mathrm{I}}},{\bf{\Theta }}},{\bf{r}_{\mathrm{co}}} \right)\leq\gamma_{\mathrm{s}}^{\mathrm{upp}}=\frac{P{N_{\mathrm{B}}}{N_{\mathrm{S}}}}{\sigma^2}{\left| {\alpha _{{\rm{IS,T}}}}{\alpha _{{\rm{BI}}}}\right|^2}N_{\mathrm{I}}^2,\end{array}
\end{equation}
and\vspace{-0mm}
\begin{equation}\label{equp2}\vspace{-0mm}\begin{array}{l}
{\gamma_{\mathrm{c}}}\left( {{{\bf{W}}},{\bf{r}^{\mathrm{I}}},{\bf{\Theta }}} \right)\leq \gamma_{\mathrm{c}}^{\mathrm{upp}}=\frac{P{N_B}}{\sigma^2}{{\left| {{\alpha _{{\rm{BI}}}}{\alpha _{{\rm{IU}}}}} \right|}^2}N_{\mathrm{I}}^2,\end{array}
\end{equation}
\proof
 See Appendix \ref{app:theorem:0}.
\endproof
\end{theorem}

Theorem \ref{theorem:0} derives the upper bounds for the ISAC system of both sensing and communication performance. Notably, the upper bound for communication performance is similar to that of the traditional fixed-spacing IRS-assisted communication-only system \cite{WQQ}, while the bound for sensing performance matches that of the IRS-assisted sensing-only system using perfect SIC method to eliminate the interference from the direct path \cite{SIC_radar}.
\begin{remark}\label{remark:0}
As observed in Appendix \ref{app:theorem:0}, the traditional fixed-spacing IRS-assisted ISAC system struggles to simultaneously achieve performance upper bounds due to conflicting phase alignment strategies at the IRS, i.e., the PPR conditions cannot be satisfied. In contrast, an IRS with movable elements offers positional flexibility, enabling better reconfiguration of the communication and sensing channels. This enhanced configurability facilitates a more effective performance trade-off between communication and sensing.
\end{remark}
\vspace{-0mm}\subsection{Design of Element Position and Beamforming}
In this subsection, we propose an angular information-based joint element position and beamforming optimization scheme to simultaneously achieve the upper bounds for both communication and sensing. 
Specifically, the algorithm can be divided into three steps:

{\textbf{Step 1: Optimal transmit beamforming design.}} 
As proven in Appendix \ref{app:theorem:0}, the communication and sensing performance can be maximized simultaneously by letting \vspace{-0mm}
\begin{equation}\label{w}\vspace{-0mm}\begin{array}{l}
{{\bf{w}}_{\mathrm{c}}} = \sqrt P \frac{{\bf{a}}_{{\rm{B}}}^*(\varphi^{\rm{B}})}{{\left\| {\bf{a}}_{{\rm{B}}}^*(\varphi^{\rm{B}}) \right\|}},{{\bf{w}}_{\mathrm{s}}}=\bf{0}.\end{array}
\end{equation}
Thus, the SCNR and SINR can be rewritten into \vspace{-0mm}
\begin{equation}\label{eq:eqw1}\vspace{-0mm}
\begin{array}{l}
\hspace{-5mm}{\gamma_{\mathrm{s}}}\left( {{\bf{r}_m^{\mathrm{I}}},{\bf{\Theta }}},{\bf{r}_{\mathrm{co}}} \right) =\\
\hspace{-5mm}{\frac{P{N_{\mathrm{B}}}{\left| {\alpha _{{\rm{IS,T}}}}{\alpha _{{\rm{BI}}}}{\mathbf{r}_{\mathrm{co}}^H}{\bf{a}}_{{\rm{S}}}^*(\varphi _{{\rm{T}}}^{\rm{S}})\right|^2}
  {\left| {\sum\limits_{m = 1}^{{N_{\rm{I}}}} {{e^{j\left[ {\frac{{2\pi }}{\lambda }\left(\rho_{\rm{T}}^{\mathrm{IS}}(\mathbf{r}_m^{\mathrm{I}})-\rho^{\mathrm{BI}}(\mathbf{r}_m^{\mathrm{I}})\right) + {\theta _m}} \right]}}} } \right|^2}
}{{P{N_{\mathrm{B}}}{\left| {\alpha _{{\rm{IS,}}0}}{\alpha _{{\rm{BI}}}}{{\mathbf{r}_{\mathrm{co}}^H}}{\bf{a}}_{{\rm{S}}}^*(\varphi _{0}^{\rm{S}}) \right|^2}
{\left| {\sum\limits_{m = 1}^{{N_{\rm{I}}}} {{e^{j\left[ {\frac{{2\pi }}{\lambda }\left(\rho_{\rm{I}}^{\mathrm{IS}}(\mathbf{r}_m^{\mathrm{I}})-\rho^{\mathrm{BI}}(\mathbf{r}_m^{\mathrm{I}})\right) + {\theta _m}}\right]}}} } \right|^2} + {\sigma ^2}}}},
\end{array}
\end{equation}
and
\begin{equation}\label{eq:eqw2}\vspace{-0mm}
\begin{array}{l}
\hspace{-2mm}{\gamma_{\mathrm{c}}}\left( {{\bf{r}_m^{\mathrm{I}}},{\bf{\Theta }}} \right)  ={\frac{{P{N_B}{{\left| {{\alpha _{{\rm{BI}}}}{\alpha _{{\rm{IU}}}}} \right|}^2}   }}{{{\sigma ^2}}}}{\left| {\sum\limits_{m = 1}^{{N_{\rm{I}}}} {{e^{j\left[ {\frac{{2\pi }}{\lambda }\left(\rho^{\mathrm{IU}}(\mathbf{r}_m^{\mathrm{I}})-\rho^{\mathrm{BI}}(\mathbf{r}_m^{\mathrm{I}})\right) + {\theta _m}} \right]}}} } \right|^2}.
\end{array}
\end{equation}\vspace{-0mm}
\begin{algorithm}[htbp]
		\caption{The proposed beamforming and position design scheme for single user scenario}\label{algo:1}\scalebox{0.9}{ 
        \parbox{0.54\textwidth}{ 
\begin{algorithmic}[1]
		\State Initialize the integer matrix $\mathbf{K}^0$, the position matrix $\mathbf{X}^0$,  and set the outer and inner iteration index $t_1=t_2=1$.		
		\State  Update the transmit beamforming vector $\mathbf{w}_{\mathrm{c}}$ and $\mathbf{w}_{\mathrm{s}}$ by (\ref{w}).
		\State  Update the phase shift vector $\boldsymbol{\xi}$ by (\ref{theta}).
				\Repeat
					\State Generate the circulant shift matrix $\mathbf{A}^{t_1}$ and $\theta_0^{t_1}$.
					\State  Calculate the position matrix $\mathbf{X}^{t_1}_{t_2}$ by Eq. (\ref{X}).
					\Repeat
						\For {$m=1$ to $N_{\mathrm{I}}$}
							\If {$\max\left({x^{\mathrm{I}}_{m}},{y^{\mathrm{I}}_{m}}\right) > u_{\mathrm{max}}$}
								\State $\mathbf{k}_{m}=\mathbf{k}_{m}-\mathbf{1}$.
						   \ElsIf {$\min\left({x^{\mathrm{I}}_{m}},{y^{\mathrm{I}}_{m}}\right) < u_{\mathrm{min}}$}
								\State $\mathbf{k}_{m}=\mathbf{k}_{m}+\mathbf{1}$.
							\EndIf
						\EndFor	
						\State  Calculate the position matrix $\mathbf{X}^{t_1}_{t_2}$ by Eq. (\ref{X}).
						\State  Set $t_2=t_2+1$.
					\Until  {$t_2$ reaches the predefined maximum number $T_2$ or 
$\mathbf{r}^{\mathrm{I}}_m \in \mathcal{C}, \forall m\in[1, N_{\mathrm{I}}]$ are satisfied.}
					\State  Set $t_1=t_1+1$.
				\Until 
the $\mathbf{r}^{\mathrm{I}}_m \in \mathcal{C}, \forall m\in[1, N_{\mathrm{I}}]$ and $\left \| \mathbf{r}^{\mathrm{I}}_m - \mathbf{r}^{\mathrm{I}}_n \right\| \ge D, \forall n \neq m\in[1, N_{\mathrm{I}}]$ are satisfied.
	\end{algorithmic}}}
	\end{algorithm} \vspace{-0mm}

{\textbf{Step 2: PPR conditions.}}
According to Eq. (\ref{eq:eqw1}) and Eq. (\ref{eq:eqw2}), by adjusting the element position and phase shift, the performance upper bound can be achieved when the phase-position-related (PPR) conditions are satisfied, i.e.,\vspace{-0mm}
\begin{equation}\vspace{-0mm}\label{eq:eq20}
\left\{ \begin{array}{l}
{\frac{{2\pi }}{\lambda }\left(\rho_{\rm{T}}^{\mathrm{IS}}(\mathbf{r}_m^{\mathrm{I}})-\rho^{\mathrm{BI}}(\mathbf{r}_m^{\mathrm{I}})\right) + {\theta _m}}=\theta_0+2n_m\pi,\\
{\frac{{2\pi }}{\lambda }\left(\rho^{\mathrm{IU}}(\mathbf{r}_m^{\mathrm{I}})-\rho^{\mathrm{BI}}(\mathbf{r}_m^{\mathrm{I}})\right) + {\theta _m}}=\theta_0+2l_m\pi\\
{\frac{{2\pi }}{\lambda }\left(\rho_{\rm{I}}^{\mathrm{IS}}(\mathbf{r}_m^{\mathrm{I}})-\rho^{\mathrm{BI}}(\mathbf{r}_m^{\mathrm{I}})\right) + {\theta _m}}=\frac{2\pi\left(m-1\right)}{N_{\mathrm{I}}},\\
\qquad n_m,l_m \in \mathcal{Q}, \forall m \in \{1,\cdots,N_{\mathrm{I}}\},
\end{array} \right.
\end{equation}
where the first and third equations ensure optimal sensing performance, while the second equation guarantees optimal communication performance. $Q$ denotes the set of integers, and $\theta_0$ represents the an arbitrary phase shift within the range $\left[0,2\pi\right]$. According to Eq. (\ref{eq:eq20}), we first design a phase alignment strategy to achieve the communication performance upper bound, i.e., the second equation of PPR, where\vspace{-0mm}
\begin{equation}\label{theta}\vspace{-0mm}\begin{array}{l}
\boldsymbol{\xi}=\mathrm{diag}\left({\bf{a}}_{\rm{I}}(\varphi^ {\rm{IU}},\omega^{\rm{IU}})
{\bf{a}}_{\rm{I}}^H(\varphi ^{\rm{BI}},\omega ^{\rm{BI}})\right),\end{array}
\end{equation}
and the linear combination of the PPR condition can be reformulated as the position-related (PR) equations\vspace{-0mm}
\begin{equation}\vspace{-0mm}
\left\{ \begin{array}{l}
\frac{{2\pi }}{\lambda }\left(\rho_{\rm{T}}^{\mathrm{IS}}(\mathbf{r}_m^{\mathrm{I}})-\rho_{\rm{I}}^{\mathrm{IS}}(\mathbf{r}_m^{\mathrm{I}}\right)
+\frac{2\pi\left(m-1\right)}{N_{\mathrm{I}}}=\theta_0+2n_m\pi,\\
{\frac{{2\pi }}{\lambda }\left(\rho^{\mathrm{IU}}(\mathbf{r}_m^{\mathrm{I}})-\rho_{\rm{I}}^{\mathrm{IS}}(\mathbf{r}_m^{\mathrm{I}})\right)}+\frac{2\pi\left(m-1\right)}{N_{\mathrm{I}}}=\theta_0+2l_m\pi,
\end{array} \right.
\end{equation}
which can be rewritten into a more compact form\vspace{-0mm}
\begin{equation}\vspace{-0mm}\begin{array}{l}
\mathbf{BX}=\theta_0\mathbf{I}+2\pi(\mathbf{K}-\mathbf{A}),\end{array}
\end{equation}
where all elements of matrix $\mathbf{K}=\left[\mathbf{k}_1,\cdots,\mathbf{k}_{N_{\mathrm{I}}}\right]\in \mathbb{C} ^{2\times N_{\mathrm{I}}}$ are the minimum integers that make $\mathbf{r}_{m}^{\mathrm{I}}, \forall m \in \{1,\cdots,N_{\mathrm{I}}\}$, $\mathbf{A}$ is a circulant shift matrix formed by the vector $\mathbf{c}^T=\left[  0,\cdots,  \frac{{N_{\mathrm{I}}} - 1}{{{N_{\mathrm{I}}}}}\right]$, \vspace{-0mm}
\begin{equation}\vspace{-0mm}
\mathbf{B}=\frac{2\pi}{\lambda}\left[ \begin{array}{l}
{\varphi_{\mathrm{T}}^{{{\mathrm{IS}}}}-\varphi_{\mathrm{I}}^{{{\mathrm{IS}}}},\omega_{\mathrm{T}}^{{{\mathrm{IS}}}}-\omega_{\mathrm{I}}^{{{\mathrm{IS}}}}}\\
{\varphi^{{{\mathrm{IU}}}}-\varphi_{\mathrm{I}}^{{{\mathrm{IS}}}},\omega^{{{\mathrm{IU}}}}-\omega_{\mathrm{I}}^{{{\mathrm{IS}}}}}
\end{array} \right],
\end{equation}
and\vspace{-0mm}
\begin{equation}\vspace{-0mm}\begin{array}{l}
\mathbf{X}=\left[ 
{\mathbf{r}_{1}^{\mathrm{I}}},\cdots,{\mathbf{r}_{N_{\mathrm{I}}}^{\mathrm{I}}}
\right].\end{array}
\end{equation}

Thus, the position of each movable element can be determined by the inverse of matrix $\mathbf{B}$, namely\vspace{-0mm}
\begin{equation}\label{X}\vspace{-0mm}
\mathbf{X}=\mathbf{B}^\dag \left[\theta_0\mathbf{I}+2\pi(\mathbf{K}-\mathbf{A})\right].
\end{equation}

{\textbf{Step 3: Optimal sensing beamforming design.}
}Under the PPR condition, the sensing performance can be rewritten into\vspace{-0mm}
\begin{equation}\vspace{-0mm}
\begin{array}{l}
{\gamma_{\mathrm{s}}}\left( {\bf{r}_{\mathrm{co}}} \right) =\frac{P{N_{\mathrm{B}}N_{\mathrm{I}}^2}}{{{ {\sigma ^2}}}}{\left| {\alpha _{{\rm{IS,T}}}}{\alpha _{{\rm{BI}}}}{\mathbf{r}_{\mathrm{co}}^H}{\bf{a}}_{{\rm{S}}}^*(\varphi _{{\rm{T}}}^{\rm{S}})\right|^2},
\end{array}
\end{equation}
where the SCNR can be maximized to achieve the full array gain over the desired direction, i.e., 
\begin{equation}\label{r}\vspace{-0mm}\begin{array}{l}
\mathbf{r}_{\mathrm{co}}={{\bf{a}}_{{\rm{S}}}^*(\varphi _{{\rm{T}}}^{\rm{S}})}/{\left\|{\bf{a}}_{{\rm{S}}}^*(\varphi _{{\rm{T}}}^{\rm{S}})\right\|}.\end{array}
\end{equation}

After the above three steps without alternating optimization, the potential spatial resources can be exploited, thereby achieving performance upper bounds for both communication and sensing. The overall algorithm is summarized in \textbf{Algorithm} \ref{algo:1}, where $u_{\mathrm{min}}=-A$, and $u_{\mathrm{max}}=A$ denote the lower and upper bounds of the movable region, respectively. 
\vspace{-0mm}
\section{Multi-user Scenario}\label{se:se4}
Recognizing the effectiveness of IRS with movable elements for enhancing ISAC performance in a single-user scenario, in this section, we extend it to a multi-user scenario. In such a scenario, multi-user and clutter interference introduce new analytical challenges, rendering the previous analysis inapplicable. To address this, we derive the performance lower bounds on the ergodic rate for both sensing and communication to reveal the impact of element position. Furthermore, we propose a joint ISAC beamforming and IRS element position design scheme to effectively exploit the spatial degrees of freedom within the movable region, thereby enhancing the overall system performance.
\vspace{-0mm}
\subsection{Performance Analysis}
Due to interference from multi-user signals and clutters, analyzing the overall system performance using instantaneous channel state information (CSI) is challenging.\footnote{CSI can be acquired by exploiting the sparse representation of the channel responses \cite{R4,R5}, thus perfect CSI is considered.} To address this, the ergodic rate \cite{MIMO-CAPACITY} is adopted as the performance metric to characterize the impact of movable element positions.

The ergodic rate of the $k$-th user and the SCNR 
at sensing receiver can be respectively expressed as \vspace{-0mm}
\begin{equation}\vspace{-0mm}\begin{array}{l}
R_{\mathrm{c},k}=\mathbb{E}\left\{ \hat{R}_{\mathrm{c},k}  \right\},\hat{R}_{\mathrm{c},k}=\log_{2}\left(1+{\gamma_{\mathrm{c},k}}\left( {{{\bf{W}}},{\bf{r}^{\mathrm{I}}},{\bf{\Theta }}} \right)\right),\end{array}
\end{equation}
and\vspace{-0mm}
\begin{equation}\vspace{-0mm}\begin{array}{l}
M_{\mathrm{s}}=\mathbb{E}\left\{ \hat{M}_{\mathrm{s}} \right\},\hat{M}_{\mathrm{s}}= \log_{10}\left(\gamma_{\mathrm{s}}\left( {{{\bf{W}}},{\bf{r}^{\mathrm{I}}},{\bf{\Theta }}},{\bf{r}_{\mathrm{co}}} \right)\right),\end{array}
\end{equation}
where \vspace{-0mm}
\begin{equation}\label{eq:eq14}\vspace{-0mm}
\begin{array}{l}
{\gamma_{\mathrm{c},k}}\left( {{{\bf{W}}},{\bf{r}^{\mathrm{I}}},{\bf{\Theta }}} \right) 
= \frac{{\left | {\bf{h}}_{{\mathrm{IU},k}}^{{H}}{\bf{\Theta }}{{\bf{H}}_{{\rm{BI}}}}{{\bf{w}}_{\mathrm{c},k}}\right | ^2}}
{{ \sum\limits_{i \neq k}^{K} \left | {\bf{h}}_{{\mathrm{IU},k}}^{{H}}{\bf{\Theta }}{{\bf{H}}_{{\rm{BI}}}}{{\bf{w}}_{\mathrm{c},i}} \right | ^2 + \left|  {\bf{h}}_{{\mathrm{IU},k}}^{{H}}{\bf{\Theta }}{{\bf{H}}_{{\rm{BI}}}}{{\bf{w}}_{s}}   \right|^2+ {\sigma ^2}}},
\end{array}
\end{equation}
and \vspace{-0mm}
\begin{equation}\vspace{-0mm}
\begin{array}{l}
\gamma_{\mathrm{s}}\left( {{{\bf{W}}},{\bf{r}^{\mathrm{I}}},{\bf{\Theta }}},{\bf{r}_{\mathrm{co}}} \right)= 
\frac{{{{\left\| {{\alpha _{{\rm{IS,T}}}}\mathbf{r}^H_{\mathrm{co}}{\bf{a}}_{{\rm{S}}}^*(\varphi _{{\rm{T}}}^{\rm{S}}){\bf{a}}_{\rm{I}}^T(\varphi _{{\rm{T}}}^{\rm{IS}},\omega _{{\rm{T}}}^{\rm{IS}}){\bf{\Theta }}{{\bf{H}}_{{\rm{BI}}}}}
{\mathbf{W}} \right\|}^2}}}
{\left\| \sum\limits_{c=0}^{C}{\alpha _{{\rm{IS,c}}}}\mathbf{r}^H_{\mathrm{co}}{\bf{a}}_{{\rm{S}}}^*(\varphi _{{\rm{c}}}^{\rm{S}}){\bf{a}}_{\rm{I}}^T(\varphi _{{\rm{c}}}^{\rm{IS}},\omega _{{\rm{c}}}^{\rm{IS}})
\bf{\Theta }\bf{H}_{\rm{BI}}\mathbf{W} \right\|^2 + \sigma ^2}.
\end{array}
\end{equation}
\begin{theorem}\label{theorem:1}
The ergodic rate of the $k$-th user is lower bounded by\vspace{-0mm}
\begin{equation}\label{eq:eqclb}\vspace{-0mm}\begin{array}{l}
R_{\mathrm{c,k}}^{\mathrm{lb}}= \log_{2}\left(1+ {\gamma_{\mathrm{c},k}^{\mathrm{lb}}}\left( {{{\bf{W}}},{\bf{r}^{\mathrm{I}}},{\bf{\Theta }}} \right) \right) ,\end{array}
\end{equation}
where \vspace{-0mm}
\begin{equation}\vspace{-0mm}
\begin{array}{l}
\gamma_{\mathrm{c},k}^{\mathrm{lb}}\left( {{{{\bf{W}}},{\bf{r}^{\mathrm{I}}},{\bf{\Theta }}}} \right)
=\frac{B^{\mathrm{c}}_k}{{A^{\mathrm{c}}_k  -B^{\mathrm{c}}_k + C^{\mathrm{c}}_k  + {\sigma ^2}}},
\end{array}
\end{equation}
  $A^{\mathrm{c}}_k=\sum\limits_{i = 1}^K {\mathbb{E}\left\{ {{{\left| {{\bf{h}}_{{\rm{IU,}}k}^{{H}}{\bf{\Theta }}{{\bf{H}}_{{\rm{BI}}}}{{\bf{w}}_{{\rm{c}},i}}} \right|}^2}} \right\}}$, $B^{\mathrm{c}}_k=  {{\left| {\mathbb{E}\left\{ {{\bf{h}}_{{\rm{IU,}}k}^{{H}}{\bf{\Theta }}{{\bf{H}}_{{\rm{BI}}}}{{\bf{w}}_{{\rm{c}},k}}} \right\}} \right|}^2}$, and $C^{\mathrm{c}}_k=\mathbb{E}\left\{ {{{\left| {{\bf{h}}_{{\rm{IU,}}k}^{{H}}{\bf{\Theta }}{{\bf{H}}_{{\rm{BI}}}}{{\bf{w}}_{\rm{s}}}} \right|}^2}} \right\}$.

The performance lower bound for sensing in high SCNR region can be approximated as \vspace{-0mm}
\begin{equation}\vspace{-0mm}\begin{array}{l}
M_{\mathrm{s}}^{\mathrm{lb}}\approx \log_{10}\left(\gamma_{\mathrm{s}}^{\mathrm{lb}}\left( {{{\bf{W}}},{\bf{r}^{\mathrm{I}}},{\bf{\Theta }}},{\bf{r}_{\mathrm{co}}} \right) \right),\end{array}
\end{equation}
where\vspace{-0mm}
\begin{equation}\vspace{-0mm}
\begin{array}{l}
\gamma_{\mathrm{s}}^{\mathrm{lb}}\left( {{{\bf{W}}},{\bf{r}^{\mathrm{I}}},{\bf{\Theta }}},{\bf{r}_{\mathrm{co}}} \right)
=\frac{B^{\mathrm{s}}}{A^{\mathrm{s}}- B^{\mathrm{s}}  + {\sigma ^2}},
\end{array}
\end{equation}
$A^{\mathrm{s}}=\mathbb{E}\left\{ {{{\left\| {\sum\limits_{c = 0}^C {{\alpha _{\mathrm{IS},c}}{{\bf{r}}_{\mathrm{co}}^H}{\bf{a}}_{{\rm{S}}}^*(\varphi_{c}^{\rm{IS}}){\bf{a}}_{\rm{I}}^T(\varphi_{c}^{\rm{IS}},\omega _{c}^{\rm{IS}}){\bf{\Theta }}{{\bf{H}}_{{\rm{BI}}}}{\bf{W}}} } \right\|}^2}} \right\} $, and $B^{\mathrm{s}}={{{\left\| {\mathbb{E}\left\{ {\alpha _{{\rm{IS,T}}}}{{{\bf{r}}^H_{\mathrm{co}}}{\bf{a}}_{{\rm{S}}}^*(\varphi _{{\rm{T}}}^{\rm{S}}){\bf{a}}_{\rm{I}}^T(\varphi _{{\rm{T}}}^{\rm{IS}},\omega _{{\rm{T}}}^{\rm{IS}}){\bf{\Theta }}{{\bf{H}}_{{\rm{BI}}}}{\bf{W}}} \right\} } \right\|}^2}}$.
\proof
Due to page limitations, the proof is omitted. It can be found in \cite{MIMO-CAPACITY} and derived through some mathematical transformations, where the quality holds when the desired signal over unknown channel part is zero.
\endproof
\end{theorem}

Theorem \ref{theorem:1} derives the deterministic lower bound of communication and  sensing performance, where $\mathbb{E}$ represents the expectations over the random channel realizations. The lower bounds depend on the precoding, sensing beamforming, IRS phase shift vectors, and position of the movable elements. To further elucidate the impact of the movable elements on ISAC system performance, the bound analysis is provided.
\begin{corollary}\label{coro:1}
To gain more useful insights, the maximum ratio transmission (MRT) and maximal ratio combining (MRC) method are adopted \cite{MIMO-CAPACITY}, i.e., $\mathbf{w}_{\mathrm{c},k}=\eta_k\left( {\bf{h}}_{{\rm{IU,}}k}^{{H}} {\bf{\Theta }} {\bf{H}}_{{\rm{BI}}} \right)^H$, $\mathbf{w}_{\mathrm{s}}=\eta_{\mathrm{s}}\left( {\alpha _{{\rm{IS,T}}}}{\bf{a}}_{\rm{I}}^T(\varphi _{{\rm{T}}}^{\rm{IS}},\omega _{{\rm{T}}}^{\rm{IS}}){\bf{\Theta }}{{\bf{H}}_{{\rm{BI}}}} \right)^H$, ${{\bf{r}}_{\mathrm{co}}}=\frac{{\bf{a}}_{{\rm{S}}}^*(\varphi _{{\rm{T}}}^{\rm{S}})}{\left\| {\bf{a}}_{{\rm{S}}}(\varphi _{{\rm{T}}}^{\rm{S}})  \right\|}$, where ${\eta _{\mathrm{c},k}} = \sqrt{\frac{{ {{p_{\mathrm{c},k}}} }}{{ {\mathbb{E}\left\{ {{{\left\| {{\bf{h}}_{{\rm{IU,}}k}^{{H}}{\bf{\Theta }}{{\bf{H}}_{{\rm{BI}}}}} \right\|}^2}} \right\}} }}}$, and ${\eta _{\mathrm{s}}} = \sqrt{\frac{{ {{p_{\mathrm{s}}}} }}{{ {\mathbb{E}\left\{ {{{\left\| {{\alpha _{{\rm{IS,T}}}}{\bf{a}}_{\rm{I}}^T\left( {\varphi _{\rm{T}}^{{\rm{IS}}},\omega _{\rm{T}}^{{\rm{IS}}}} \right){\bf{\Theta }}{{\bf{H}}_{{\rm{BI}}}}} \right\|}^2}} \right\}} }}}$ are the power coefficient for the communication user and sensing receiver.
The channel expectations for communication can be rewritten into\vspace{-0mm}
\begin{equation}\vspace{-0mm}\scalebox{1}{$
\begin{array}{l}
A^{\mathrm{c}}_k=\sum\limits_{i = 1}^K {{{\left| {{\eta _i}} \right|}^2}} \sum\limits_{n = 1}^{{N_{\mathrm{B}}}} \sum\limits_{n_{1} = 1}^{{N_{\mathrm{B}}}}\sum\limits_{m = 1}^{{N_{\mathrm{I}}}} \sum\limits_{m_{1} = 1}^{{N_{\mathrm{I}}}}  \sum\limits_{m_{2} = 1}^{{N_{\mathrm{I}}}} \sum\limits_{m_{3} = 1}^{{N_{\mathrm{I}}}} \\
\qquad \times{{{{\bf{ \bar{G}}}}_{k,i}}\left[ {{{\bf{r}}_m^{\mathrm{I}}},{{\bf{r}}_{m_{1}}^{\mathrm{I}}},{{\bf{r}}_{m_{2}}^{\mathrm{I}}},{{\bf{r}}_{m_{3}}^{\mathrm{I}}}} \right]{{{\bf{ \bar {F}}}}_{n,n_{1}}}\left[ {{{\bf{r}}_m^{\mathrm{I}}},{{\bf{r}}_{m_{1}}^{\mathrm{I}}},{{\bf{r}}_{m_{2}}^{\mathrm{I}}},{{\bf{r}}_{m_{3}}^{\mathrm{I}}}} \right]},
\end{array}$}
\end{equation}
\begin{equation}\vspace{-0mm}\scalebox{1}{$
\begin{array}{l}
B^{\mathrm{c}}_k = \eta _k^2N_B^2{\left( {\sum\limits_{m = 1}^{{N_{\mathrm{I}}}} {\sum\limits_{m_{1} = 1}^{{N_{\mathrm{I}}}} {{{\bf{G}}_k}\left[ {{{\bf{r}}_m},{{\bf{r}}_{m_{1}}}} \right]{\bf{F}}\left[ {{{\bf{r}}_m},{{\bf{r}}_{m_{1}}}} \right]} } } \right)^2},
\end{array}$}
\end{equation}
and\vspace{-0mm}
\begin{equation}\vspace{-0mm}\scalebox{1}{$
\begin{array}{l}
C^{\mathrm{c}}_k={\left| {{\eta _{\mathrm{s}}}} \right|^2}\frac{{\sigma _{{\rm{IS}}}^{\rm{2}}}}{C+1}\sum\limits_{n = 1}^{{N_{\mathrm{B}}}}\sum\limits_{n_{1} = 1}^{{N_{\mathrm{B}}}} \sum\limits_{m = 1}^{{N_{\mathrm{I}}}} \sum\limits_{m_{1} = 1}^{{N_{\mathrm{I}}}}  \sum\limits_{m_{2} = 1}^{{N_{\mathrm{I}}}} \sum\limits_{m_{3} = 1}^{{N_{\mathrm{I}}}} \\
\qquad \times{{{\bf{G}}_k}\left[ {{{\bf{r}}_m^{\mathrm{I}}},{{\bf{r}}_{m_{2}}^{\mathrm{I}}}} \right]{{{\bf{ \bar {F}}}}_{n,n_{1}}}\left[ {{{\bf{r}}_m^{\mathrm{I}}},{{\bf{r}}_{m_{1}}^{\mathrm{I}}},{{\bf{r}}_{m_{2}}^{\mathrm{I}}},{{\bf{r}}_{m_{3}}^{\mathrm{I}}}} \right]}      {{\bf{J}}_{\mathrm{T}}}\left[ {{{\bf{r}}_{m_{1}}^{\mathrm{I}}},{{\bf{r}}_{m_{3}}^{\mathrm{I}}}} \right].
\end{array}$}
\end{equation}

Similarly, the expectations for sensing can be expressed by\vspace{-0mm}
$$\begin{array}{l}A^{\mathrm{s}}= \sum\limits_{k = 1}^K \eta _k^2\sum\limits_{c = 0}^C {{\left| {{\bf{a}}_{\rm{S}}^T\left( {\theta _{\rm{T}}^{{\rm{IS}}}} \right){\bf{a}}_{\rm{S}}^*\left( {\theta _c^{{\rm{IS}}}} \right)} \right|}^2}\sum\limits_{n,n_{1} = 1}^{{N_{\mathrm{B}}}}  \sum\limits_{m,m_{1} = 1}^{{N_{\mathrm{I}}}} \sum\limits_{m_{2},m_{3} = 1}^{{N_{\mathrm{I}}}} \\
\times{\frac{{\sigma _{{\rm{IS}}}^{\rm{2}}}}{{C + 1}}} {{{\bf{ \bar {F}}}}_{n,n_{1}}}\left[ {{{\bf{r}}_m^{\mathrm{I}}},{{\bf{r}}_{m_{1}}^{\mathrm{I}}},{{\bf{r}}_{m_{2}}^{\mathrm{I}}},{{\bf{r}}_{m_{3}}^{\mathrm{I}}}} \right]{\bf{G}}_k^*\left[ {{{\bf{r}}_{m_{1}}^{\mathrm{I}}},{{\bf{r}}_{m_{3}}^{\mathrm{I}}}} \right]    {{\bf{J}}_i}\left[ {{{\bf{r}}_{m_{2}}^{\mathrm{I}}},{{\bf{r}}_m^{\mathrm{I}}}} \right] \\
 + \eta _{\rm{s}}^2\sum\limits_{i = 1}^C {{\left| {{\bf{a}}_{\rm{S}}^T\left( {\theta _{\rm{T}}^{{\rm{IS}}}} \right){\bf{a}}_{\rm{S}}^*\left( {\theta _c^{{\rm{IS}}}} \right)} \right|}^2}\sum\limits_{n = 1}^{{N_{\mathrm{B}}}} \sum\limits_{n_{1} = 1}^{{N_{\mathrm{B}}}}\sum\limits_{m = 1}^{{N_{\mathrm{I}}}} \sum\limits_{m_{1} = 1}^{{N_{\mathrm{I}}}}  \sum\limits_{m_{2} = 1}^{{N_{\mathrm{I}}}} \sum\limits_{m_{3} = 1}^{{N_{\mathrm{I}}}} \\
\end{array}$$
\begin{equation}\vspace{-0mm}\scalebox{1}{$
\begin{array}{l}
\times{\frac{{\sigma _{{\rm{IS}}}^{\rm{4}}}}{{{{\left( {C + 1} \right)}^2}}}{{{\bf{ \bar {F}}}}_{n,n_{1}}}\left[ {{{\bf{r}}_m^{\mathrm{I}}},{{\bf{r}}_{m_{1}}^{\mathrm{I}}},{{\bf{r}}_{m_{2}}^{\mathrm{I}}},{{\bf{r}}_{m_{3}}^{\mathrm{I}}}} \right]}       {{{\bf{ J}}}_{\mathrm{T},i}}\left[ {{{\bf{r}}_m^{\mathrm{I}}},{{\bf{r}}_{m_{1}}^{\mathrm{I}}},{{\bf{r}}_{m_{2}}^{\mathrm{I}}},{{\bf{r}}_{m_{3}}^{\mathrm{I}}}} \right]\\
 + 2\eta _{\rm{s}}^2N_{\rm{S}}^2\frac{{\sigma _{{\rm{IS}}}^{\rm{4}}}}{{{{\left( {C + 1} \right)}^2}}}\sum\limits_{n = 1}^{{N_{\mathrm{B}}}}\sum\limits_{n_{1} = 1}^{{N_{\mathrm{B}}}} \sum\limits_{m = 1}^{{N_{\mathrm{I}}}} \sum\limits_{m_{1} = 1}^{{N_{\mathrm{I}}}}  \sum\limits_{m_{2} = 1}^{{N_{\mathrm{I}}}} \sum\limits_{m_{3} = 1}^{{N_{\mathrm{I}}}} \\
\times{{{{\bf{ \bar {F}}}}_{n,n_{1}}}\left[ {{{\bf{r}}_m^{\mathrm{I}}},{{\bf{r}}_{m_{1}}^{\mathrm{I}}},{{\bf{r}}_{m_{2}}^{\mathrm{I}}},{{\bf{r}}_{m_{3}}^{\mathrm{I}}}} \right]}      {{{\bf{ J}}}_{T,T}}\left[ {{{\bf{r}}_m^{\mathrm{I}}},{{\bf{r}}_{m_{1}}^{\mathrm{I}}},{{\bf{r}}_{m_{2}}^{\mathrm{I}}},{{\bf{r}}_{m_{3}}^{\mathrm{I}}}} \right],
\end{array}$}
\end{equation}
and\vspace{-0mm}
\begin{equation}\vspace{-0mm}\scalebox{1}{$
\begin{array}{l}
B^{\mathrm{s}}
 = N_{\rm{S}}^2\eta _{\rm{s}}^2\frac{{\sigma _{{\rm{IS}}}^{\rm{4}}}}{{L_{{\rm{IS}}}^2}}{\left| {\sum\limits_{n = 1}^{{N_{\mathrm{B}}}} {\sum\limits_{m = 1}^{{N_{\mathrm{I}}}} {\sum\limits_{m_{1} = 1}^{{N_{\mathrm{I}}}} {{\bf{F}}\left[ {{{\bf{r}}_m^{\mathrm{I}}},{{\bf{r}}_{m_{1}}^{\mathrm{I}}}} \right]{{\bf{J}}_{\mathrm{T}}}\left[ {{{\bf{r}}_{m_{1}}^{\mathrm{I}}},{{\bf{r}}_m^{\mathrm{I}}}} \right]} } } } \right|^2},
\end{array}$}
\end{equation}
where ${\bf{G}}_k$, ${\bf{F}}$, ${{{\bf{ \bar{G}}}}_{k,k}}$, ${{{\bf{ \bar{G}}}}_{k,i}}$, ${{{\bf{ \bar {F}}}}_{n,n}}$, ${{{\bf{ \bar {F}}}}_{n,n_{1}}}$, ${{\bf{J}}_c}$, and ${{{\bf{\bar {J}}}}_{\mathrm{T},c}}$ are given by Eq. (\ref{eq:eqsybeg}) $\sim$ Eq. (\ref{eq:eqsyend}). 
\end{corollary}
\proof
 See Appendix \ref{app:coro:1}.
\endproof

\begin{figure}[t]\vspace{-0mm}
\centering
\subfigure[Comparison of sensing performance.]
{\includegraphics[width=2.3in]{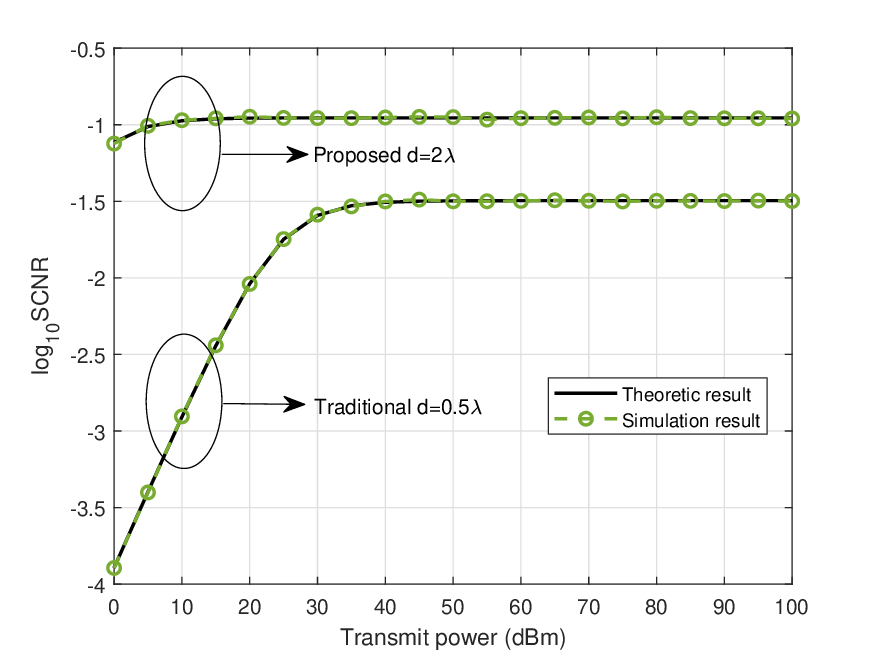}}\vspace{-0mm}
\subfigure[Comparison of communication performance.]
{\includegraphics[width=2.3in]{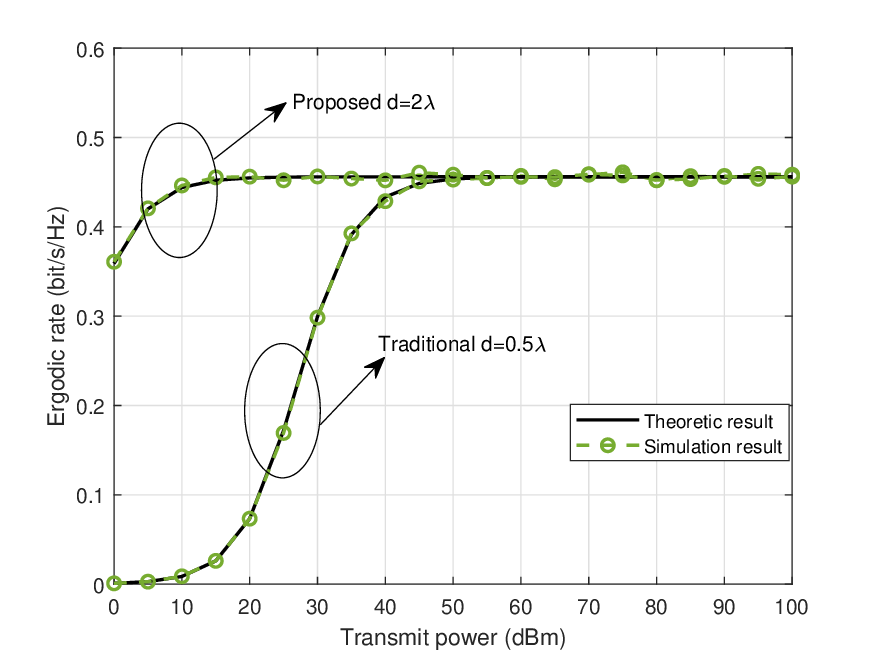}}\vspace{-0mm}
\caption{Performance comparison between movable elements and traditional half-wavelength spacing elements, where $N_{\mathrm{B}}=N_{\mathrm{S}}=16$, $N_{\mathrm{I}}=8$, $L^{c}_{k}=L^{s}_{1}=2$, $L^{s}_{2}=1$. 
}\label{fig:fig0}\vspace{-8mm}
\end{figure}

\begin{corollary}\label{coro:2}
For a 1D IRS-assisted multi-user with a LoS path ISAC system,  where the phase shifts of the IRS with movable elements are the same as those of a traditional IRS with fixed-spacing $d=\frac{\lambda}{2}$, the communication performance of the former surpasses that of the latter when $\varphi^{\mathrm{IU}}_{k}-\varphi^{\mathrm{BI}}=\frac{2}{N_{I}}L^{c}_{k}$. Meanwhile, the sensing performance improves when $\varphi^{\mathrm{IS}}_{\mathrm{T}}-\varphi^{\mathrm{BI}}=\frac{2}{N_{I}}L^{s}_{1}$ and $\varphi^{\mathrm{IS}}_{0}-\varphi^{\mathrm{BI}}=2L^{s}_{2}$, where $L^{c}_{k}$, $L^{s}_{1}$, and $L^{s}_{2}$ are the integers.
\begin{proof}
See Appendix \ref{app:coro:2}
\end{proof}
\end{corollary}
\begin{remark}
Corollary \ref{coro:2} demonstrates one example to show the prospect of an IRS with movable elements in the multi-user scenario. Under the same IRS phase shift and transceiver beamforming, the ISAC performance of an IRS with movable elements outperforms that of a traditional IRS with half-wavelength spacing. To illustrate this, Fig. \ref{fig:fig0} presents the lower bound of the communication and sensing across different spacings. It is seen that by adjusting the spacing of the movable-element IRS to $d=2\lambda$, both communication and sensing performance exceed those of the traditional fixed-spacing IRS with $d=\frac{\lambda}{2}$. It demonstrates that adjusting the positions of the movable elements, significant opportunities arise for enhancing the overall performance of the ISAC system. \vspace{-0mm}
\end{remark}
\vspace{-0mm}
\subsection{Joint Design of Element Position and Beamforming}
So far, we have derived the performance bound based on a heuristic beamforming scheme. While this approach offers valuable insights into the impact of movable element positions on system performance, beamforming designs relying on statistical CSI may lead to significant performance degradation and fail to fully exploit the potential of IRS with movable elements.
In this subsection, we attempt to maximize the sensing performance while guaranteeing the communication performance by jointly optimizing the transmit beamforming at the ISAC transmitter, IRS phase shifts, movable element positions, and the sensing beamforming vector.\footnote{By employing joint beamforming, the signal quality received at the users and the sensing receiver is significantly improved. As such, the transmitted data can be correctly decoded at the user, and higher-precision target state parameters such as distance, Doppler, and angle information can be estimated.} 
The optimization problem is formulated as follows
\begin{subequations}\vspace{-0mm}
\begin{flalign}\vspace{-0mm}
&{}&{\text{(P1)}}\quad {\mathop {\max }\limits_{  {{{\bf{W}}},{\mathbf{r}_{m}^{\mathrm{I}}},{\bf{\Theta }}},{\bf{r}_{\mathrm{co}}}   } }&\quad \hat{M}_{\mathrm{s}}
\label{eq:eq35a}& \\
&{}&{\mathrm{s.t.}}&\quad\hat{R}_{\mathrm{c},k} \ge \gamma_{0,k} \label{eq:eq35b}&\\
&{}&{}&\quad {\left\| {\bf{r}}_{\mathrm{co}} \right\|^2} = 1,\label{eq:eq35c}&
\\&{}&{}&\quad \left| {{{\left[ {\boldsymbol{\xi }} \right]}_m}} \right| = 1, \forall m\in[1, N_{\mathrm{I}}],\label{eq:eq35d}&
\\&{}&{}&\quad \mathbf{r}^{\mathrm{I}}_m \in \mathcal{C}, \forall m\in[1, N_{\mathrm{I}}],\label{eq:eq35e}&
\\&{}&{}&\quad \mathrm{Tr}\left( {{\mathbf{W}}{{\bf{W}}^H}} \right) \le P,\label{eq:eq35f}&
\\&{}&{}&\quad \left \| \mathbf{r}^{\mathrm{I}}_m - \mathbf{r}^{\mathrm{I}}_n \right\| \ge D, \forall n \neq m\in[1, N_{\mathrm{I}}] ,\label{eq:eq35g}&
\end{flalign}
\end{subequations}
where $\gamma_{0,k}$ denotes the rate threshold of the $k$-th user,  Eq. (\ref{eq:eq35e}), and Eq. (\ref{eq:eq35g}) impose constraints on the movable area and the minimum distance $D$ constraint between the two adjacent movable elements, respectively.
Due to the non-convexity of both the objective function and constraints, solving (P1) is challenging. To this end, we propose an effective algorithm.

Firstly, the objective and constraints functions in Eq. (\ref{eq:eq35a}) and Eq. (\ref{eq:eq35b}) can be rewritten into SCNR and SINR, which can be recast by exploiting the quadratic-transform technique to the fractional function \cite{FP}\vspace{-0mm}
\begin{equation}\vspace{-0mm}
\begin{array}{l}
\gamma_{\mathrm{s}}^1\left( {{{\bf{W}}},{\bf{r}^{\mathrm{I}}},{\bf{\Theta }}},{\bf{r}_{\mathrm{co}}},\mathbf{x}_{\mathrm{fp}} \right)\\
=  2\mathrm{Re}\left\{{{{ {{\alpha _{{\rm{IS,T}}}}\mathbf{r}^H_{\mathrm{co}}{\bf{a}}_{{\rm{S}}}^*(\varphi _{{\rm{T}}}^{\rm{S}}){\bf{a}}_{\rm{I}}^T(\varphi _{{\rm{T}}}^{\rm{IS}},\omega _{{\rm{T}}}^{\rm{IS}}){\bf{\Theta }}{{\bf{H}}_{{\rm{BI}}}}}
{\mathbf{Wx}} }}}\right\}\\
-\left\|\mathbf{x}_{\mathrm{fp}}\right\|^2\left({\left\| \sum\limits_{c=0}^{C}{\alpha _{{\rm{IS,c}}}}\mathbf{r}^H_{\mathrm{co}}{\bf{a}}_{{\rm{S}}}^*(\varphi _{{\rm{c}}}^{\rm{S}}){\bf{a}}_{\rm{I}}^T(\varphi _{{\rm{c}}}^{\rm{IS}},\omega _{{\rm{c}}}^{\rm{IS}})
\bf{\Theta }\bf{H}_{\rm{BI}}\mathbf{W} \right\|^2 \hspace{-1mm}+ \hspace{-1mm}\sigma ^2}\hspace{-1mm}\right),
\end{array}
\end{equation}\vspace{-0mm}
and\vspace{-0mm}
\begin{equation}\vspace{-0mm}
\begin{array}{l}
{\gamma_{\mathrm{c},k}^1}\left( {{{\bf{W}}},{\bf{r}^{\mathrm{I}}},{\bf{\Theta }}},{y}_k \right) = 2\mathrm{Re}\{{ {\bf{h}}_{{\mathrm{IU},k}}^{{H}}{\bf{\Theta }}{{\bf{H}}_{{\rm{BI}}}}{{\bf{w}}_{\mathrm{c},k}}y_k^*}\}\\
-\left|y_k^* \right|^2\left({{ \sum\limits_{i \neq k}^{K} \left | {\bf{h}}_{{\mathrm{IU},k}}^{{H}}{\bf{\Theta }}{{\bf{H}}_{{\rm{BI}}}}{{\bf{w}}_{\mathrm{c},i}} \right | ^2 + \left|  {\bf{h}}_{{\mathrm{IU},k}}^{{H}}{\bf{\Theta }}{{\bf{H}}_{{\rm{BI}}}}{{\bf{w}}_{s}}   \right|^2+ {\sigma ^2}}}\right),
\end{array}
\end{equation}
where $\mathbf{x}_{\mathrm{fp}}$ and $y_k$ are the auxiliary variables, and the optimal solutions for the two auxiliary variables can be obtained by solving $\partial \gamma_{\mathrm{s}}^1\left( {{{\bf{W}}},{\bf{r}^{\mathrm{I}}},{\bf{\Theta }}},{\bf{r}_{\mathrm{co}}},\mathbf{x}_{\mathrm{fp}} \right)/\partial \mathbf{x}_{\mathrm{fp}} =0$ and $\partial {\gamma_{\mathrm{c},k}^1}\left( {{{\bf{W}}},{\bf{r}^{\mathrm{I}}},{\bf{\Theta }}},{y}_k \right) /\partial {y}_k=0$, which yields\vspace{-0mm}
\begin{equation}\label{eq:eqx}\vspace{-0mm}
\begin{array}{l}
\mathbf{x}_{\mathrm{fp}}=
 \frac{\left({{ {{\alpha _{{\rm{IS,T}}}}\mathbf{r}^H_{\mathrm{co}}{\bf{a}}_{{\rm{S}}}^*(\varphi _{{\rm{T}}}^{\rm{S}}){\bf{a}}_{\rm{I}}^T(\varphi _{{\rm{T}}}^{\rm{IS}},\omega _{{\rm{T}}}^{\rm{IS}}){\bf{\Theta }}{{\bf{H}}_{{\rm{BI}}}}}
{\mathbf{W}} }}\right)^H}
{\left\| \sum\limits_{c=0}^{C}{\alpha _{{\rm{IS,c}}}}\mathbf{r}^H_{\mathrm{co}}{\bf{a}}_{{\rm{S}}}^*(\varphi _{{\rm{c}}}^{\rm{S}}){\bf{a}}_{\rm{I}}^T(\varphi _{{\rm{c}}}^{\rm{IS}},\omega _{{\rm{c}}}^{\rm{IS}})
\bf{\Theta }\bf{H}_{\rm{BI}}\mathbf{W} \right\|^2 + \sigma ^2},
\end{array}
\end{equation}
and\vspace{-0mm}
\begin{equation}\label{eq:eqy}\vspace{-0mm}
\begin{array}{l}
{y}_k=\frac{ {\bf{h}}_{{\mathrm{IU},k}}^{{H}}{\bf{\Theta }}{{\bf{H}}_{{\rm{BI}}}}{{\bf{w}}_{\mathrm{c},k}}}
{{ \sum\limits_{i \neq k}^{K} \left | {\bf{h}}_{{\mathrm{IU},k}}^{{H}}{\bf{\Theta }}{{\bf{H}}_{{\rm{BI}}}}{{\bf{w}}_{\mathrm{c},i}} \right | ^2 + \left|  {\bf{h}}_{{\mathrm{IU},k}}^{{H}}{\bf{\Theta }}{{\bf{H}}_{{\rm{BI}}}}{{\bf{w}}_{s}}   \right|^2+ {\sigma ^2}}}. 
\end{array}
\end{equation}

By leveraging the above transformations, the new objective and constraint function $\gamma_{\mathrm{s}}^1\left( {{{\bf{W}}},{\bf{r}^{\mathrm{I}}},{\bf{\Theta }}},{\bf{r}_{\mathrm{co}}},\mathbf{x}_{\mathrm{fp}} \right)$ and ${\gamma_{\mathrm{c},k}^1}\left( {{{\bf{W}}},{\bf{r}^{\mathrm{I}}},{\bf{\Theta }}},{y}_k \right)$ are convex. Hence, the original problem (P1) can be divided into four subproblems and iteratively optimized until convergence.  
\subsubsection{Optimize ${\bf{r}_{\mathrm{co}}}$ with given $\mathbf{r}_{m}^{\mathrm{I}}$, $\bf{\Theta}$, and $\mathbf{W}$, problem (P1) can be reformulated as}
\begin{subequations}\vspace{-0mm}
\begin{flalign}
&{}&{\text{(P2)}}\quad {\mathop {\min }\limits_{  {\bf{r}_{\mathrm{co}}}   } }&\quad {\mathbf{r}_{\mathrm{co}}^H}\mathbf{A}^{\mathrm{r}}{\bf{r}_{\mathrm{co}}} -2\mathrm{Re}\left\{{\mathbf{r}_{\mathrm{co}}^H}\mathbf{b}^{\mathrm{r}} \right\} , \label{eq:eq48a}& \\
&{}&\:\,{\mathrm{s.t.}}&\quad{\left\| {\bf{r}}_{\mathrm{co}} \right\|^2} = 1, \label{eq:eq48b}&
\end{flalign}
\end{subequations}
\vspace{-0mm}where\vspace{-0mm}\vspace{-0mm}
\begin{equation}\vspace{-0mm}\vspace{-0mm}
\begin{array}{l}
\mathbf{A}^{\mathrm{r}}=\left\|\mathbf{x}_{\mathrm{fp}}\right\|^2\sum\limits_{c=0}^{C}\sum\limits_{c_1=0}^{C}{\alpha _{{\rm{IS,c}}}}{\alpha _{{\rm{IS,c_1}}}^*}{\bf{a}}_{{\rm{S}}}^*(\varphi _{{\rm{c}}}^{\rm{S}}){\bf{a}}_{\rm{I}}^T(\varphi _{{\rm{c}}}^{\rm{IS}},\omega _{{\rm{c}}}^{\rm{IS}})
\bf{\Theta }\bf{H}_{\rm{BI}}\\
\mathbf{W}
\mathbf{W}^H\mathbf{H}_{\rm{BI}}^H\mathbf{\Theta }^H{\mathbf{a}}_{\rm{I}}^*(\varphi _{{\rm{c_1}}}^{\rm{IS}},\omega _{{\rm{c_1}}}^{\rm{IS}}){\mathbf{a}}_{{\rm{S}}}^T(\varphi _{{\rm{c_1}}}^{\rm{S}}),
\end{array}
\end{equation}
and\vspace{-0mm}
\begin{equation}\vspace{-0mm}
\begin{array}{l}
\mathbf{b}^{\mathrm{r}}={{{ {{\alpha _{{\rm{IS,T}}}}{\bf{a}}_{{\rm{S}}}^*(\varphi _{{\rm{T}}}^{\rm{S}}){\bf{a}}_{\rm{I}}^T(\varphi _{{\rm{T}}}^{\rm{IS}},\omega _{{\rm{T}}}^{\rm{IS}}){\bf{\Theta }}{{\bf{H}}_{{\rm{BI}}}}}
{\mathbf{Wx}} }}}.
\end{array}
\end{equation}
Since $\mathbf{A}^{\mathrm{r}}$ is positive definite, (P2) is a convex problem that can be directly solved by standard solvers such as CVX but with high complexity. We propose a low-complexity algorithm solving ${\bf{r}_{\mathrm{co}}}$ by deriving the closed-form expression based on the Lagrangian dual decomposition method \cite{Lagrangian}. The Lagrangian of the problem can be first derived as\vspace{-0mm}
\begin{equation}\vspace{-0mm}
\begin{array}{l}
\hspace{-2mm}\mathcal{L}\left( {\bf{r}_{\mathrm{co}}}, \lambda_{{r_\mathrm{co}}} \right)={\mathbf{r}_{\mathrm{co}}^H}\mathbf{A}^{\mathrm{r}}{\bf{r}_{\mathrm{co}}} -2\mathrm{Re}\left\{{\mathbf{r}_{\mathrm{co}}^H}\mathbf{b}^{\mathrm{r}} \right\}+ \lambda_{{r_\mathrm{co}}}\left( {\left\| {\bf{r}}_{\mathrm{co}} \right\|^2} - 1 \right),
\end{array}
\end{equation}
where $\lambda_{{r_\mathrm{co}}}$ is the Lagrangian multiplier corresponding to the unit constraint, which is determined by\vspace{-0mm}
\begin{equation}\vspace{-0mm}
\begin{array}{l}
\lambda_{{r_\mathrm{co}}} = \min \left\{ \lambda_{{r_\mathrm{co}}} \ge 0: {\left\| {\bf{r}}_{\mathrm{co}} \right\|^2} = 1 \right\},
\end{array}
\end{equation}
which can be determined efficiently through bisection search. Moreover, KKT conditions are used to solve the dual problem, and the optimal solutions can be obtained by $\partial \mathcal{L}\left( {\bf{r}_{\mathrm{co}}}, \lambda_{{r_\mathrm{co}}} \right)/\partial {\bf{r}_{\mathrm{co}}} =0$, which yields\vspace{-0mm}
\begin{equation}\label{eq:eqrco}\vspace{-0mm}\begin{array}{l}
 {\bf{r}_{\mathrm{co}}}=\left( \mathbf{A}^{\mathrm{r}}+\lambda_{{r_\mathrm{co}}}\mathbf{I}_{N_{\mathrm{S}}} \right)^{-1}\mathbf{b}^{\mathrm{r}}.\end{array}
\end{equation}
\subsubsection{Optimize ${\mathbf{r}_{m}^{\mathrm{I}}}$ with given $ {\bf{r}_{\mathrm{co}}}$, $\bf{\Theta}$, and $\mathbf{W}$, the problem can be recast as}\begin{subequations}\vspace{-0mm}
\begin{flalign}
&{}&{\text{(P3)}}\quad {\mathop {\max }\limits_{  {{\mathbf{r}_{m}^{\mathrm{I}}}}   } }&\quad \gamma_{\mathrm{s}}\left( {{{\bf{W}}},{\bf{r}^{\mathrm{I}}},{\bf{\Theta }}},{\bf{r}_{\mathrm{co}}} \right), \label{eq:eq54a}& \\
&{}&\:\,{\mathrm{s.t.}}&\quad{\gamma_{\mathrm{c},k}}\left( {{{\bf{W}}},{\bf{r}^{\mathrm{I}}},{\bf{\Theta }}} \right) \ge 2^{\gamma_{0,k}}-1 \label{eq:eq54b}&
\\&{}&{}&\quad \mathbf{r}^{\mathrm{I}}_m \in \mathcal{C}, \forall m\in[1, N_{\mathrm{I}}],\label{eq:eq54c}&
\\&{}&{}&\quad \left \| \mathbf{r}^{\mathrm{I}}_m - \mathbf{r}^{\mathrm{I}}_n \right\| \ge D, \forall n \neq m\in[1, N_{\mathrm{I}}] ,\label{eq:eq54d}&
\end{flalign}
\end{subequations}\vspace{-0mm}

The constraints on the movement range and minimum spacing in (P3) are non-convex, posing significant optimization challenges.  Although the authors in \cite{MA-CHANNEL} proposed an SCA-based approach to optimize the position of movable antennas one by one, this method is prone to being trapped in locally optimal solutions due to the abundance of local maxima solutions in the movable region. To overcome this,  we propose an efficient memory penalized projected gradient descent (MPPGD) algorithm that optimizes the positions of all movable elements simultaneously.
The penalty function supplements the objective function with a weighted penalty for violating the constraint, therefore, the optimization problem can be further written as\vspace{-0mm}
\begin{subequations}\vspace{-0mm}
\begin{flalign}
&{}&{\text{(P3-1)}}\quad {\mathop {\max }\limits_{  {\bf{r}_{\mathrm{co}}}   } }&\quad \mathcal{Q}\left(  {{{\bf{W}}},{\bf{r}^{\mathrm{I}}},{\bf{\Theta }}},{\bf{r}_{\mathrm{co}}},\mathbf{x}_{\mathrm{fp}}\right), \label{eq:eq55a}& \\
&{}&\:\,{\mathrm{s.t.}}&\quad\mathbf{r}^{\mathrm{I}}_m \in \mathcal{C}, \forall m\in[1, N_{\mathrm{I}}],\label{eq:eq55b}&
\end{flalign}
\end{subequations}\vspace{0mm}
where \vspace{-0mm}
\begin{equation}
\begin{array}{l}
\mathcal{Q}\left(  {{{\bf{W}}},{\bf{r}^{\mathrm{I}}},{\bf{\Theta }}},{\bf{r}_{\mathrm{co}}},\mathbf{x}_{\mathrm{fp}}\right)=\gamma_{\mathrm{s}}^1\left( {{{\bf{W}}},{\bf{r}^{\mathrm{I}}},{\bf{\Theta }}},{\bf{r}_{\mathrm{co}}},\mathbf{x}_{\mathrm{fp}} \right)\\
\qquad -\rho_p\mathcal{B}\left( \mathbf{r}^{\mathrm{I}} \right)+\sum_{k=1}^K\rho_{\mathrm{c},k}\min\left\{0,{\mathcal{F}_{\mathrm{c},k}}\left( {{\bf{r}^{\mathrm{I}}}} \right)\right\},
\end{array}
\end{equation}
with ${\mathcal{F}_{\mathrm{c},k}}\left( {{\bf{r}^{\mathrm{I}}}} \right)={\gamma_{\mathrm{c},k}^1}\left( {{{\bf{W}}},{\bf{r}^{\mathrm{I}}},{\bf{\Theta }}},{y}_k \right)-\gamma_{0,k}$, \vspace{-0mm}
\begin{equation}\vspace{-0mm}
\mathcal{B}\left( \mathbf{r}^{\mathrm{I}} \right)=\left\{
\begin{array}{l}
0,\quad\mathrm{if}\; \|\mathcal{P}\left( \mathbf{r}^{\mathrm{I}} \right)\|_1= 0\\
\left \| \mathbf{r}^{\mathrm{I}}_m - \mathbf{r}^{\mathrm{I}}_n \right\| - D,\mathrm{otherwise},
\end{array}\right.
\end{equation}
where $\left\| \cdot \right\|_1$ denotes the L-1 norm and $\mathcal{P}\left( \mathbf{r}^{\mathrm{I}} \right)$ is a set with each entry representing a pair of movable elements that violate the minimum inter-element distance constraint, i.e., \vspace{-0mm}
\begin{equation}\vspace{-0mm}
[\mathcal{P}\left( \mathbf{r}^{\mathrm{I}} \right)]_{m,n}=\left\{
\begin{array}{l}
{{1,\quad\mathrm{if}\quad \|\mathbf{r}_{m}^{\mathrm{I}}-\mathbf{r}_{n}^{\mathrm{I}}\|<D,1\leq n<m\leq N_{\mathrm{I}}}}\\
{{0,\quad\mathrm{otherwise}.}}
\end{array}\right.
\end{equation}
During the iteration process, the penalty factors, i.e.,$\rho_{\mathrm{c},j}$ and $\rho_p$, enforce the two constraints ${\mathcal{F}_{\mathrm{c},k}}\left( {{\bf{r}^{\mathrm{I}}}} \right)$ and $\mathcal{B}\left( \mathbf{r}^{\mathrm{I}} \right)$to zeros, thus, ensuring the constraint Eq. (\ref{eq:eq54b}) and Eq. (\ref{eq:eq54d}) hold.

Therefore, the projected gradient descent method can be adopted, and the solution of $\mathbf{r}^{\mathrm{I}}$ can be updated by \vspace{-0mm}
\begin{equation}\label{eq:eqri}\vspace{-0mm}\begin{array}{l}
\mathbf{r}^{\mathrm{I}}=\mathrm{Proj}_{\mathbf{r}^{\mathrm{I}}}\left(  \mathbf{r}^{\mathrm{I}}+{\lambda}_{r^{\mathrm{I}}}\nabla_{r^{\mathrm{I}}} \mathcal{Q}\left(  {{{\bf{W}}},{\bf{r}^{\mathrm{I}}},{\bf{\Theta }}},{\bf{r}_{\mathrm{co}}},\mathbf{x}_{\mathrm{fp}}\right)\right),\end{array}
\end{equation}
where $\mathrm{Proj}_{\mathbf{r}^{\mathrm{I}}}\left( \mathbf{a}\right)$ is the projection function that ensures all the movable elements only move in the respective feasible regions,\vspace{-0mm}
\begin{equation}\vspace{-0mm}\begin{array}{l}
\left[\mathrm{Proj}_{\mathbf{r}^{\mathrm{I}}}\left( \mathbf{a}\right)\right]=\left\{\begin{array}{rll}u_{\max},&\mathrm{if}&\left[\mathbf{a}\right]_i>u_{\max}\\\left[\mathbf{a}\right]_i,&\mathrm{if}&u_{\min}\leq\left[\mathbf{a}\right]_i\leq u_{\max}\\u_{\min},&\mathrm{if}&\left[\mathbf{a}\right]_i<u_{\min}\end{array}\right.\end{array}
\end{equation}
and the gradient of the objective function in Eq. (\ref{eq:eq55a}) can be derived as\vspace{-0mm}
\begin{equation}\vspace{-0mm}
\begin{array}{l}
\nabla_{r^{\mathrm{I}}} \mathcal{Q}\left(  {{{\bf{W}}},{\bf{r}^{\mathrm{I}}},{\bf{\Theta }}},{\bf{r}_{\mathrm{co}}},\mathbf{x}_{\mathrm{fp}}\right)\\
=\left\lceil\frac{\partial  \mathcal{Q}\left(  {{{\bf{W}}},{\bf{r}^{\mathrm{I}}},{\bf{\Theta }}},{\bf{r}_{\mathrm{co}}},\mathbf{x}_{\mathrm{fp}}\right)}{\partial \mathbf{r}^{\mathrm{I}}_1},\cdots,\frac{\partial  \mathcal{Q}\left(  {{{\bf{W}}},{\bf{r}^{\mathrm{I}}},{\bf{\Theta }}},{\bf{r}_{\mathrm{co}}},\mathbf{x}_{\mathrm{fp}}\right)}{\partial \mathbf{r}^{\mathrm{I}}_{N_{\mathrm{I}}}}\right\rceil^T\\
=\frac{\partial\gamma_{\mathrm{s}}^1\left( {{{\bf{W}}},{\bf{r}^{\mathrm{I}}},{\bf{\Theta }}},{\bf{r}_{\mathrm{co}}},\mathbf{x}_{\mathrm{fp}} \right)}{\partial \mathbf{r}^{\mathrm{I}}}-\rho_p \frac{\partial\mathcal{B}\left( \mathbf{r}^{\mathrm{I}} \right)}{\partial \mathbf{r}^{\mathrm{I}}}\\
+\sum_{k=1}^K \rho_{\mathrm{c},k}\begin{cases}0,\quad\text{if}\quad 0\leq{\mathcal{F}_{\mathrm{c},k}}\left( {{\bf{r}^{\mathrm{I}}}} \right).\\ 
\frac{\partial{\mathcal{F}_{\mathrm{c},k}}\left( {{\bf{r}^{\mathrm{I}}}} \right)}{\partial \mathbf{r}^{\mathrm{I}}},\text{otherwise}.\end{cases}
\end{array}
\end{equation}

Due to the highly nonlinear nature of $\mathcal{Q}\left(  {{{\bf{W}}},{\bf{r}^{\mathrm{I}}},{\bf{\Theta }}},{\bf{r}_{\mathrm{co}}},\mathbf{x}_{\mathrm{fp}}\right)$, multiple local maxima may arise with respect to the gradient descent step size ${\lambda}_{r^{\mathrm{I}}}$ \cite{ZLP}. To address this, we introduce a memory-based feasible solution method to mitigate the risk of suboptimal local maxima. Firstly, we gradually decrease ${\lambda}_{r^{\mathrm{I}}}$ from ${\lambda}_{r^{\mathrm{I}}}^{\max}$ by a small positive step size $0<\eta_1<1$, i.e., ${\lambda}_{r^{\mathrm{I}}}=\eta_1{\lambda}_{r^{\mathrm{I}}}$. Then, during the $t$-th iteration, if the solution is feasible and the objective function $\mathcal{Q}\left(  {{{\bf{W}}},{\bf{r}^{\mathrm{I}}},{\bf{\Theta }}},{\bf{r}_{\mathrm{co}}},\mathbf{x}_{\mathrm{fp}}\right)$ increases, we store the current solution as $\mathbf{r}^{\mathrm{I}}_{\mathrm{fes}}=\left(\mathbf{r}^{\mathrm{I}}\right)^{t}$ and ${\lambda}_{r^{\mathrm{I}}}^{\mathrm{fes}}={\lambda}_{r^{\mathrm{I}}}$, 
 and proceed with $S$ subsequent iterations. If the solution becomes infeasible or the objective function fails to improve, we revert to the last feasible solution $\mathbf{r}^{\mathrm{I}}_{\mathrm{fes}}$, reduce the gradient step size ${\lambda}_{r^{\mathrm{I}}}=\eta_2{\lambda}_{r^{\mathrm{I}}}^{\mathrm{fes}}$, where $0<\eta_2<1$,  and continue iterating until convergence. 
This memory-based feasible solution method ensures that $\gamma_{\mathrm{s}}\left( {{{\mathbf{W}^t}},\left({\bf{r}^{\mathrm{I}}_{\mathrm{fes}}}\right)^t,{\mathbf{\Theta }^t}},{\mathbf{r}_{\mathrm{co}}^t} \right)\leq\gamma_{\mathrm{s}}\left( {{{\mathbf{W}^t}},\left({\bf{r}^{\mathrm{I}}}\right)^t,{\mathbf{\Theta }^t}},{\mathbf{r}_{\mathrm{co}}^{(t+1)}} \right)$. As a result, it reduces the risk of being trapped in low-performance solutions compared to memoryless gradient descent methods.
\subsubsection{Optimize ${{\bf{W}}} $ with given  $ {\bf{r}_{\mathrm{co}}}$, $\mathbf{r}^{\mathrm{I}}$, and $\bf{\Theta}$, the problem can be reformulated as}
\begin{subequations}
\begin{flalign}
&{}&{\text{(P4)}}\quad {\mathop {\max }\limits_{  {{\bf{W}}}   } }&\quad \sum_{k=1}^{K}-\mathbf{w}_{\mathrm{c},k}^H\mathbf{A}^{\mathrm{w}}\mathbf{w}_{\mathrm{c},k}+2\mathrm{Re}\left\{\left(\mathbf{c}_{\mathrm{c},k}^{\mathrm{w}}\right)^H\mathbf{w}_{\mathrm{c},k}\right\}\notag\\
&{}&{}&-\mathbf{w}_{\mathrm{s}}^H\mathbf{A}^{\mathrm{w}}\mathbf{w}_{\mathrm{s}}+2\mathrm{Re}\left\{\left(\mathbf{c}_{\mathrm{s}}^{\mathrm{w}}\right)^H\mathbf{w}_{\mathrm{s}}\right\}, \label{eq:eq54a}& \\
&{}&\:\,{\mathrm{s.t.}}& 2\mathrm{Re}\left\{\left(\mathbf{b}_{k}^{\mathrm{w}}\right)^H\mathbf{w}_{\mathrm{c},k}\right\}-\sum_{i\neq k}^{K}\mathbf{w}_{\mathrm{c},i}^H\mathbf{B}_{k}^{\mathrm{w}}\mathbf{w}_{\mathrm{c},i} \notag \\
&{}&{}&-\mathbf{w}_{\mathrm{s}}^H\mathbf{B}_{k}^{\mathrm{w}}\mathbf{w}_{\mathrm{s}}-\left| y_{k} \right|^2\sigma^2 \ge 2^{\gamma_{0,k}}-1 \label{eq:eq60b}&
\\&{}&{}&\quad \mathrm{Tr}\left( {{\mathbf{W}}{{\bf{W}}^H}} \right) \le P,\label{eq:eq54c}&
\end{flalign}
\end{subequations}
\vspace{-0mm}where\vspace{-0mm}
\begin{equation}\vspace{-0mm}
\begin{array}{l}
\mathbf{A}^{\mathrm{w}}=\left\|\mathbf{x}_{\mathrm{fp}}\right\|^2  \sum\limits_{c=0}^{C}\sum\limits_{c_1=0}^{C}{\alpha _{{\rm{IS,c}}}^*}{\alpha _{{\rm{IS},c_1}}}\mathbf{H}_{\rm{BI}}^H\mathbf{\Theta }^H{\bf{a}}_{\rm{I}}^*(\varphi _{{\rm{c}}}^{\rm{IS}},\omega _{{\rm{c}}}^{\rm{IS}}){\bf{a}}_{{\rm{S}}}^T(\varphi _{{\rm{c}}}^{\rm{S}})\\
\mathbf{r}_{\mathrm{co}}
\mathbf{r}^H_{\mathrm{co}}{\bf{a}}_{{\rm{S}}}^*(\varphi _{{\rm{c_1}}}^{\rm{S}}){\bf{a}}_{\rm{I}}^T(\varphi _{{\rm{c_1}}}^{\rm{IS}},\omega _{{\rm{c_1}}}^{\rm{IS}})
\bf{\Theta }\bf{H}_{\rm{BI}},
\end{array}
\end{equation}
\begin{equation}\vspace{-0mm}
\begin{array}{l}
\mathbf{B}^{\mathrm{w}}_{k}={{\bf{H}}_{{\rm{BI}}}^H}{\bf{\Theta }}^H{\bf{h}}_{{\mathrm{IU},k}}
{\bf{h}}_{{\mathrm{IU},k}}^{{H}}{\bf{\Theta }}{{\bf{H}}_{{\rm{BI}}}},
\end{array}
\end{equation}
\begin{equation}\vspace{-0mm}
\begin{array}{l}
\left(\mathbf{b}_{k}^{\mathrm{w}}\right)^H={\bf{h}}_{{\mathrm{IU},k}}^{{H}}{\bf{\Theta }}{{\bf{H}}_{{\rm{BI}}}}y_k^*,
\end{array}
\end{equation}
\begin{equation}\vspace{-0mm}
\begin{array}{l}
\left(\mathbf{c}_{\mathrm{c},k}^{\mathrm{w}}\right)^H={{{ {{\alpha _{{\rm{IS,T}}}}\mathbf{r}^H_{\mathrm{co}}{\bf{a}}_{{\rm{S}}}^*(\varphi _{{\rm{T}}}^{\rm{S}}){\bf{a}}_{\rm{I}}^T(\varphi _{{\rm{T}}}^{\rm{IS}},\omega _{{\rm{T}}}^{\rm{IS}}){\bf{\Theta }}{{\bf{H}}_{{\rm{BI}}}}}
{\mathbf{x}_{\mathrm{fp}}}\left[k\right] }}},
\end{array}
\end{equation}
and\vspace{-0mm}
\begin{equation}\vspace{-0mm}
\begin{array}{l}
\left(\mathbf{c}_{\mathrm{s}}^{\mathrm{w}}\right)^H={{{ {{\alpha _{{\rm{IS,T}}}}\mathbf{r}^H_{\mathrm{co}}{\bf{a}}_{{\rm{S}}}^*(\varphi _{{\rm{T}}}^{\rm{S}}){\bf{a}}_{\rm{I}}^T(\varphi _{{\rm{T}}}^{\rm{IS}},\omega _{{\rm{T}}}^{\rm{IS}}){\bf{\Theta }}{{\bf{H}}_{{\rm{BI}}}}}
{\mathbf{x}_{\mathrm{fp}}}\left[K+1\right] }}},
\end{array}
\end{equation}
where ${\mathbf{x}_{\mathrm{fp}}}\left[k\right]$ denotes the $k$-th element of $\bf{x}$.
Since $\mathbf{A}^{\mathrm{w}}$ is positive definite, (P4) is a convex problem that can be efficiently solved by standard toolbox such as CVX.
\subsubsection{Optimize ${{\boldsymbol{\xi }}} $ with given $ {\bf{r}_{\mathrm{co}}}$, $\mathbf{r}^{\mathrm{I}}$, and $\mathbf{W}$, the problem can be reformulated as}\vspace{-0mm}
\begin{subequations}
\begin{flalign}
&{}&{\text{(P5)}}\quad {\mathop {\max }\limits_{  {{\boldsymbol{\xi }}}   } }&\quad -\boldsymbol{\xi}^H\left(\mathbf{A}^{\mathrm{\theta}}\right)^*\boldsymbol{\xi}+2\mathrm{Re}\left\{\boldsymbol{\xi}^H \left(\mathbf{c}^{\mathrm{\theta}}\right)^*\right\} \label{eq:eq60a}& \\
&{}&\:\,{\mathrm{s.t.}}& 2\mathrm{Re}\left\{\boldsymbol{\xi}^H \left(\mathbf{b}^{\mathrm{\theta}}_{\mathrm{c},k}\right)^* \right\}-\boldsymbol{\xi}^H\left(\mathbf{B}_{\mathrm{c},k}^{\mathrm{\theta}}+\mathbf{B}_{\mathrm{s}}^{\mathrm{\theta}}\right)^*\boldsymbol{\xi}\notag \\
&{}&{}&-\left| y_{k} \right|^2\sigma^2 \ge 2^{\gamma_{0,k}}-1 \label{eq:eq60b}&
\\&{}&{}&\quad \left| {{{\left[ {\boldsymbol{\xi }} \right]}_m}} \right| = 1, \forall m\in[1, N_{\mathrm{I}}],\label{eq:eq60c}&
\end{flalign}
\end{subequations}
\vspace{-0mm}where\vspace{-0mm}
\begin{equation}\vspace{-0mm}
\begin{array}{l}
\mathbf{A}^{\mathrm{\theta}}=\left\|\mathbf{x}_{\mathrm{fp}}\right\|^2  \sum\limits_{c=0}^{C}\sum\limits_{c_1=0}^{C}{\alpha _{{\rm{IS,c}}}^*}{\alpha _{{\rm{IS},c_1}}}
\mathbf{r}^H_{\mathrm{co}}{\bf{a}}_{{\rm{S}}}^*(\varphi _{{\rm{c_1}}}^{\rm{S}})\mathrm{diag}\left({\bf{a}}_{\rm{I}}^T(\varphi _{{\rm{c_1}}}^{\rm{IS}},\omega _{{\rm{c_1}}}^{\rm{IS}})\right)
\\
\mathbf{H}_{\rm{BI}}\mathbf{W}\mathbf{W}^H\mathbf{H}_{\rm{BI}}^H\mathrm{diag}\left({\bf{a}}_{\rm{I}}^*(\varphi _{{\rm{c}}}^{\rm{IS}},\omega _{{\rm{c}}}^{\rm{IS}})\right){\mathbf{a}}_{{\rm{S}}}^T(\varphi _{{\rm{c}}}^{\rm{S}})\mathbf{r}_{\mathrm{co}},
\end{array}
\end{equation}
\begin{equation}\vspace{-0mm}
\begin{array}{l}
\mathbf{B}^{\mathrm{\theta}}_{\mathrm{c},k}\hspace{-0.5mm}=\hspace{-0.5mm}\left| y_{k} \right|^2\sum\limits_{i\neq k}^K\mathrm{diag}\left({\bf{h}}^{{H}}_{{\mathrm{IU},k}}\right){{\bf{H}}_{{\rm{BI}}}}{{\bf{w}}_{\mathrm{c},i}} {{\bf{w}} ^H_{\mathrm{c},i}}{{\bf{H}}^H_{{\rm{BI}}}}\mathrm{diag}\left({\bf{h}}_{{\mathrm{IU},k}}\right),
\end{array}
\end{equation}
\begin{equation}\vspace{-0mm}
\begin{array}{l}
\mathbf{B}^{\mathrm{\theta}}_{\mathrm{s}}=\left| y_{k} \right|^2\mathrm{diag}\left({\bf{h}}_{{\mathrm{IU},k}}^{{H}}\right){{\bf{H}}_{{\rm{BI}}}}{{\bf{w}}_{\mathrm{s}}} {{\bf{w}}^H_{\mathrm{s}}} {{\bf{H}}^H_{{\rm{BI}}}}\mathrm{diag}\left({\bf{h}}_{{\mathrm{IU},k}}\right),
\end{array}
\end{equation}
\begin{equation}\vspace{-0mm}
\begin{array}{l}
\mathbf{b}_{\mathrm{c},k}^{\mathrm{\theta}}=\mathrm{diag}\left({\bf{h}}_{{\mathrm{IU},k}}^{{H}}\right){{\bf{H}}_{{\rm{BI}}}}{{\bf{w}}_{\mathrm{c},k}}y_k^*,
\end{array}
\end{equation}
and\vspace{-0mm}
\begin{equation}\vspace{-0mm}
\begin{array}{l}
\mathbf{c}^{\mathrm{\theta}}={{{ {{\alpha _{{\rm{IS,T}}}}\mathbf{r}^H_{\mathrm{co}}{\bf{a}}_{{\rm{S}}}^*(\varphi _{{\rm{T}}}^{\rm{S}})\mathrm{diag}\left({\bf{a}}_{\rm{I}}^T(\varphi _{{\rm{T}}}^{\rm{IS}},\omega _{{\rm{T}}}^{\rm{IS}})\right){{\bf{H}}_{{\rm{BI}}}}}
{\mathbf{Wx}} }}}.
\end{array}
\end{equation}

The constraint of (P5) is non-convex due to the unit-constraint in Eq. (\ref{eq:eq60c}). To this end, we employ the consensus alternative direction method of multipliers approach \cite{CADMM}, which allows for parallel updating of the optimization variables. To facilitate this process, we introduce auxiliary variables $\{\mathbf{z}_k\}_{k=1}^{K+1}=\boldsymbol{\xi}$, and the optimization problem can be rewritten as\vspace{-0mm}
\begin{subequations}\vspace{-0mm}
\begin{flalign}
&{}&{\text{(P5-1)}} {\mathop {\min }\limits_{{\boldsymbol{\xi}},\{\mathbf{z}_k\}_{k=1}^{K+1}} }& \boldsymbol{\xi}^H\left(\mathbf{A}^{\mathrm{\theta}}\right)^*\boldsymbol{\xi}-2\mathrm{Re}\left\{\boldsymbol{\xi}^H \left(\mathbf{c}^{\mathrm{\theta}}\right)^*\right\}\notag\\
&{}&{}&{}+{\rho \sum\limits_{i = 1}^{K+1} {{{\left\| {{{\bf{z}}_i} - {\boldsymbol{\xi }} + {{\boldsymbol{\mu}}_i}} \right\|}^2}} } \label{eq:eq74a},& \\
&{}&\hspace{-3mm}{\mathrm{s.t.}}&2\mathrm{Re}\left\{\mathbf{z}_k^H \left(\mathbf{b}^{\mathrm{\theta}}_{\mathrm{c},k}\right)^* \right\}-\mathbf{z}_k^H\left(\mathbf{B}_{\mathrm{c},k}^{\mathrm{\theta}}+\mathbf{B}_{\mathrm{s}}^{\mathrm{\theta}}\right)^*\mathbf{z}_k\notag \\
&{}&{}&-\left| y_{k} \right|^2\sigma^2 \ge 2^{\gamma_{0,k}}-1\label{eq:eq74b},&\\
&{}&{}& \left| {{{\left[ {\mathbf{z }_{K+1}} \right]}_m}} \right| = 1, \forall m\in[1,N_{\mathrm{I}}],\label{eq:eq74c}&
\end{flalign}
\end{subequations}
where $\rho$ and $\{\boldsymbol{\mu}_i\}_{i=1}^{K+1}$  are the penalty factor and the dual variable vector. Thus, the optimization problem can be solved by iteratively operating the following four steps until convergence.

\text{\bf{Step 1:}} Fixing $\{\mathbf{z}_k\}_{k=1}^{K+1}$, the problem can be reformulated as
\begin{equation}\vspace{-0mm}
\begin{aligned}
\text{(P5-2)}\quad\min_{\boldsymbol{\xi}}&\quad \boldsymbol{\xi}^H\left(\mathbf{A}^{\mathrm{\theta}}\right)^*\boldsymbol{\xi}-2\mathrm{Re}\left\{\boldsymbol{\xi}^H \left(\mathbf{c}^{\mathrm{\theta}}\right)^*\right\}\notag\\
&+{\rho \sum\limits_{i = 1}^{K+1} {{{\left\| {{{\bf{z}}_i} - {\boldsymbol{\xi }} + {{\boldsymbol{\mu}}_i}} \right\|}^2}} },
\end{aligned}
\end{equation}
which can be optimally solved by utilizing the first-order optimality condition\vspace{-0mm}
\begin{equation}\label{eq:eqxi}\vspace{-0mm}\begin{array}{l}
{\boldsymbol{\xi }} = {\left[ {\left(K+1\right)\rho {{\bf{I}}_{{N_I}}}+ \left(\mathbf{A}^{\mathrm{\theta}}\right)^*} \right]^{ - 1}}\left[ \left(\mathbf{c}^{\mathrm{\theta}}\right)^*+{\rho \sum\limits_{i = 1}^{K+1} {\left( {{{\bf{z}}_i} + {{\boldsymbol{\mu}}_i}} \right)} } \right].\end{array}
\end{equation}

\text{\bf{Step 2:}} Fixing $\boldsymbol{\xi}$ and $\mathbf{z}_{K+1}$, the problem can be reformulated as\vspace{-0mm}
\begin{equation}\vspace{-0mm}
\begin{array}{*{20}{l}}
\text{(P5-3)}&{\mathop {\min }\limits_{{{\bf{z}}_{K+1}}} }&{\rho {{\left\| {{{\bf{z}}_{K+1}} - {\boldsymbol{\xi }} + {{\boldsymbol{\mu}}_{K+1}}} \right\|}^2}},\\
&{s.t.}&{\left| {{{\left[ {{{\bf{z}}_{K+1}}} \right]}_m}} \right| = 1}, \forall m\in[1,N_{\mathrm{I}}],
\end{array}
\end{equation}
which can be solved by\vspace{-0mm}
\begin{equation}\label{eq:eqzK1}\vspace{-0mm}
{{\bf{z}}_{K+1}} = \exp \left( {j\angle \left( {{\boldsymbol{\xi }} - {{\boldsymbol{\mu}}_{K+1}}} \right)} \right).
\end{equation}

\text{\bf{Step 3:}} Fixing $\boldsymbol{\xi}$ and $\{\mathbf{z}_k\}_{k=1}^K$, the problem can be reformulated as\vspace{-0mm}
\begin{equation}\vspace{-0mm}
\begin{array}{*{20}{l}}
\text{(P5-4)}&\hspace{-3mm}{\mathop {\min }\limits_{{{\bf{z}}_2}} }&\hspace{-3mm}{\rho \sum\limits_{i = 1}^{K} {{{\left\| {{{\bf{z}}_i} - {\boldsymbol{\xi }} + {{\boldsymbol{\mu}}_i}} \right\|}^2}} },\\
&\hspace{-3mm}{s.t.}&\hspace{-3mm}\mathrm{Re}\left\{\mathbf{z}_k^H \left(\mathbf{b}^{\mathrm{\theta}}_{\mathrm{c},k}\right)^* \right\}-\mathbf{z}_k^H\left(\mathbf{B}_{\mathrm{c},k}^{\mathrm{\theta}}+\mathbf{B}_{\mathrm{s}}^{\mathrm{\theta}}\right)^*\mathbf{z}_k \\
&{}&-\left| y_{k} \right|^2\sigma^2 \ge 2^{\gamma_{0,k}}-1,
\end{array}
\end{equation}
which can be solved by Lagrangian dual decomposition method \cite{Lagrangian} \vspace{-0mm}
\begin{equation}\label{eq:eqzk}\vspace{-0mm}
{{\bf{z}}_k} = {\left[ { {{\bf{I}}_{{N_I}}}+ \lambda_{\mathrm{\theta},k} \left|y_k\right|^2\left(\mathbf{B}^{\mathrm{\theta}}_{\mathrm{c},k}+\mathbf{B}^{\mathrm{\theta}}_{\mathrm{s}}\right)^*} \right]^{ - 1}}\left( \lambda_{\mathrm{\theta},k}\left(\mathbf{b}^{\mathrm{\theta}}_{\mathrm{c},k}\right)^*+\boldsymbol{\xi} \right),
\end{equation}
with the Lagrangian multiplier $\lambda_{\mathrm{\theta},k}$ determined through bisection search \vspace{-0mm}
\begin{equation}\vspace{-0mm}
\begin{array}{l}
\lambda_{\mathrm{\theta},k} = \min \left\{\lambda_{\mathrm{\theta},k} \ge 0: \right.\\
\left. \mathrm{Re}\left\{\mathbf{z}_k^H \left(\mathbf{b}^{\mathrm{\theta}}_{\mathrm{c},k}\right)^* \right\}-\mathbf{z}_k^H\left(\mathbf{B}_{\mathrm{c},k}^{\mathrm{\theta}}+\mathbf{B}_{\mathrm{s}}^{\mathrm{\theta}}\right)^*\mathbf{z}_k -\left| y_{k} \right|^2\sigma^2 \ge \gamma_{0,k}\right\}.
\end{array}
\end{equation}

\text{\bf{Step 4:}} Update the dual variables by \vspace{-0mm}
\begin{equation}\label{eq:eqmu}\vspace{-0mm}
{{\boldsymbol{\mu}}_i} = {{\bf{z}}_i} - {\bf{\xi }} + {{\boldsymbol{\mu}}_i}.
\end{equation}

For the above four steps, it can be guaranteed to converge to a set of stationary solutions as proved in \cite{CADMM} and achieves a set of solutions that satisfy the constraint $\boldsymbol{\xi}=\{\mathbf{z}_k\}_{k=1}^{K+1}$.

With the solutions for problem (P2) $\sim$ (P5) obtained
above, we now finalize the proposed alternating optimization
algorithm to solve (P1). The overall algorithm is summarized in \textbf{Algorithm} \ref{algo:2}, where $\epsilon$, $\epsilon_1$, and $\epsilon_2$ are the predefined convergence threshold, $I_{\max}$ is the
maximum iteration index for optimizing the position of movable elements.  
\vspace{-0mm}\subsection{Algorithm Analysis}
In this subsection, we analyze the proposed beamforming and position design algorithm from the perspectives of convergence behavior and computational complexity.
\subsubsection {Convergence Behavior}
\ \newline
\indent{
The objective function $\gamma_{\mathrm{s}}\left( {{{\bf{W}}},{\bf{r}^{\mathrm{I}}},{\bf{\Theta }}},{\bf{r}_{\mathrm{co}}} \right)$ is  upper bounded due to the finite transmit power, and the optimization variables are decoupled properly by dividing the original problem (P1) into (P2) $\sim$ (P5). For optimizing the sensing beamforming vector $\mathbf{r}_\mathrm{co}$, 
the optimal solution is derived and the condition, i.e., $\gamma_{\mathrm{s}}\left( {{{\mathbf{W}^t}},\left({\bf{r}^{\mathrm{I}}}\right)^t,{\mathbf{\Theta }^t}},{\mathbf{r}_{\mathrm{co}}^t} \right)\leq\gamma_{\mathrm{s}}\left( {{{\mathbf{W}^t}},\left({\bf{r}^{\mathrm{I}}}\right)^t,{\mathbf{\Theta }^t}},{\mathbf{r}_{\mathrm{co}}^{(t+1)}} \right)$ can be satisfied. Besides, the MPPGD method for optimizing $\mathbf{r}_{\mathrm{I}}$ and $\mathbf{W}$ guarantees the non-decrease of the objective function, i.e., $\gamma_{\mathrm{s}}\left( {{{\mathbf{W}^t}},\left({\bf{r}^{\mathrm{I}}}\right)^t,{\mathbf{\Theta }^t}},{\mathbf{r}_{\mathrm{co}}^{(t+1)}} \right)\leq \gamma_{\mathrm{s}}\left( {{{\mathbf{W}^{(t+1)}}},\left({\bf{r}^{\mathrm{I}}}\right)^t,{\mathbf{\Theta }^t}},{\mathbf{r}_{\mathrm{co}}^{(t+1)}} \right)$ $\leq \gamma_{\mathrm{s}}\left( {{{\mathbf{W}^{(t+1)}}},\left({\bf{r}^{\mathrm{I}}}\right)^{(t+1)},{\mathbf{\Theta }^t}},{\mathbf{r}_{\mathrm{co}}^{(t+1)}} \right)$.
Similarly, for optimizing the phase shift vector $\boldsymbol{\xi}$, the condition can be held by the consensus ADMM approach \cite{CADMM}, i.e., $\gamma_{\mathrm{s}}\left( {{{\mathbf{W}^{(t+1)}}},\left({\bf{r}^{\mathrm{I}}}\right)^{(t+1)},{\mathbf{\Theta }^t}},{\mathbf{r}_{\mathrm{co}}^{(t+1)}} \right)\leq \gamma_{\mathrm{s}}\left( {{{\mathbf{W}^{(t+1)}}},\left({\bf{r}^{\mathrm{I}}}\right)^{(t+1)},{\mathbf{\Theta }^{(t+1)}}},{\mathbf{r}_{\mathrm{co}}^{(t+1)}} \right)$. Therefore, the proposed algorithm is non-decreasing and upper-bounded, thus the convergence of the proposed algorithm can be guaranteed.}
\subsubsection {Computational Complexity}
\ \newline
\indent{For Algorithm 1, the overall computational complexity is primarily determined by the calculation of matrices $\mathbf{K}$ and $\mathbf{X}$, which is $\mathcal{O}\left(T_2N_{\mathrm{I}} \right)$.
Therefore, the overall complexity of {Algorithm} \ref{algo:1} is $\mathcal{O}\left(T_1 T_2 N_{\mathrm{I}} \right)$, where $T_1$ and $T_2$ are the predefined maximum number of the outer and inner loop.}\vspace{-0mm}

\indent{The computational complexity of  Algorithm 2 depends on four parts. Firstly, optimize the sensing beamforming vector by calculating the matrix $\mathbf{A}^{\mathrm{r}}$ and its matirx inversion in Eq. (\ref{eq:eqrco}) and the computational complexity is $\mathcal{O}\left(N_{\mathrm{S}}N_{\mathrm{I}}^2+N_{\mathrm{S}}N_{\mathrm{I}}N_{\mathrm{B}}+N_{\mathrm{S}}N_{\mathrm{B}}K+C^2{ {{N_{\mathrm{S}}^2}} }+K^2{ {{N_{\mathrm{S}}^2}} } +{ {{N_{\mathrm{S}}^3}} }\right)$.
Then, the worst-case computational complexity of optimizing $\mathbf{r}^{\mathrm{I}}$ is no longer than $\mathcal{O}\left( \frac{I_{\max}MA^2}{\eta_1\eta_2}(N_{\mathrm{Bsen}}+N_{\mathrm{I}}^2+N_{\mathrm{I}}N_{\mathrm{B}}+N_{\mathrm{B}}K)\right)$ \cite{ZLP}, where $M$ denotes the number of feasible solutions. The computational complexity of solving (P4) is determined by calculating the matrix $\mathbf{A}^{\mathrm{w}}$, which is $\mathcal{O}\left( {N_{\mathrm{S}}}+{N_{\mathrm{I}}^2}+{N_{\mathrm{I}}}{N_{\mathrm{B}}}+C^2{N_{\mathrm{B}}^2}+{N_{\mathrm{B}}^3}K^3 \right)$. For optimizing $\boldsymbol{\xi}$, the computational complexity is $\mathcal{O}\left( {N_{\mathrm{S}}}+{N_{\mathrm{I}}^2}{N_{\mathrm{B}}}+{N_{\mathrm{I}}}{N_{\mathrm{B}}}K+C^2{N_{\mathrm{I}}^2}+{N_{\mathrm{I}}^3}+K{N_{\mathrm{I}}^2} \right)$. Thus, the overall computational complexity of \textbf{Algorithm} \ref{algo:2} is $\mathcal{O}({N_{\mathrm{S}}^3}+{N_{\mathrm{I}}^3}+{N_{\mathrm{B}}^3}K^3+(C^2+K^2)({N_{\mathrm{S}}^2}+{N_{\mathrm{I}}^2})+C^2{N_{\mathrm{B}}^2} +{N_{\mathrm{I}}^2}({N_{\mathrm{B}}}+{N_{\mathrm{S}}})+{N_{\mathrm{B}}}{N_{\mathrm{I}}}({N_{\mathrm{S}}}+K)+{N_{\mathrm{B}}}{N_{\mathrm{S}}}K+ \frac{I_{\max}MA^2}{\eta_1\eta_2}(N_{\mathrm{Bsen}}+N_{\mathrm{I}}^2+N_{\mathrm{I}}N_{\mathrm{B}}+N_{\mathrm{B}}K))$.
}
\begin{algorithm}[htb!]
		\caption{The proposed beamforming and position design scheme for multi-user with multi-path}\label{algo:2}
\scalebox{1}{ 
        \parbox{0.45\textwidth}{ 
\begin{algorithmic}[1]
		\State Input $t=i=1$, $\mathbf{r}^{\mathrm{I}} \in \mathcal{C}$, $\mathbf{W}^0$, $\mathbf{r}_{\mathrm{I}}^0$, $\bf{\Theta}^0$ and $\mathbf{r}_{\mathrm{co}}^0$.
				\Repeat
					\State Update $\mathbf{x}_{\mathrm{fp}}^t$ and $y_k^t$ by Eq. (\ref{eq:eqx}) and Eq. (\ref{eq:eqy}).
					\State Update $\lambda_{{r_\mathrm{co}}}$ by bisection search method  and sensing beamforming vector $\mathbf{r}_{\mathrm{co}}^t$ by Eq. (\ref{eq:eqrco}), respectively.		
					\Repeat 
					\State	Update the position of all the movable elements by Eq. (\ref{eq:eqri}), and set ${\lambda}_{r^{\mathrm{I}}}=\eta_1{\lambda}_{r^{\mathrm{I}}}$.
					\If {the solution of $\left(\mathbf{r}^{\mathrm{I}}\right)^{i+1}$ is feasible and the value of Eq. (\ref{eq:eq55a}) increase}
					\State Set $\mathbf{r}_{\mathrm{I}}^{\mathrm{fes}}=\left(\mathbf{r}^{\mathrm{I}}\right)^{i+1}$ and ${\lambda}_{r^{\mathrm{I}}}^{\mathrm{fes}}={\lambda}_{r^{\mathrm{I}}}$.
					\ElsIf {$i>I$}
					\State Set $i=I-S$, $\left(\mathbf{r}^{\mathrm{I}}\right)^{i}=\mathbf{r}_{\mathrm{I}}^{\mathrm{fes}}$, and ${\lambda}_{r^{\mathrm{I}}}=\eta_2{\lambda}_{r^{\mathrm{I}}}^{\mathrm{fes}}$.
					\EndIf
					\State  Set $i=i+1$.
				   \Until  {the increase of Eq. (\ref{eq:eqri}) is below $\epsilon_1$ or $i= I_{\max}$}.			
					\State Update transmit beamforming matrix $\mathbf{W}^t$ by solving (P4).
					\Repeat 
					\State	Update $\lambda_{\mathrm{\theta},k}$ by bisection search method.  Update ${\boldsymbol{\xi }}$, $\{\mathbf{z}_k\}_{k=1}^{K+1}$, and ${{\boldsymbol{\mu}}_i}$ by Eq. (\ref{eq:eqxi}) $\sim$ Eq. (\ref{eq:eqmu}). 
				   \Until  {the increase of Eq. (\ref{eq:eq74a}) is below $\epsilon_2$}.	
				\Until {the increase of the objective value in Eq. (\ref{eq:eq35a}) is below $\epsilon$}.	
	\end{algorithmic}}}
	\end{algorithm} \vspace{-0mm} 
\vspace{-0mm}\section{Numerical results}\label{se:se5}
\begin{figure*}[t]\vspace{-0mm}
\begin{minipage}[t]{0.66\linewidth}
\centering
\subfigure[The beampattern gain at IRS with fixed point elements.]
{\includegraphics[width=2.3in]{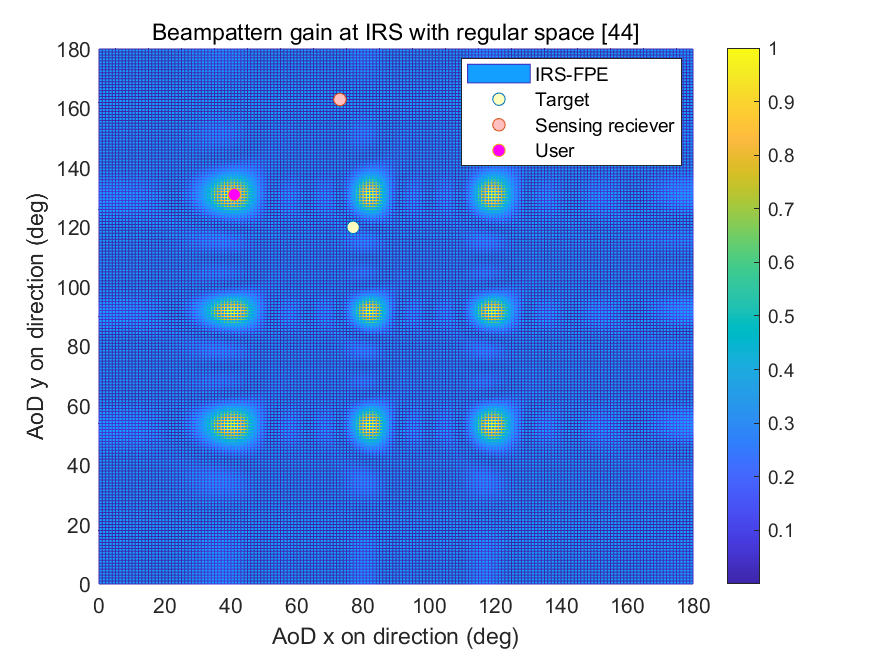}}
\subfigure[The beampattern gain at IRS with movable elements.]
{\includegraphics[width=2.3in]{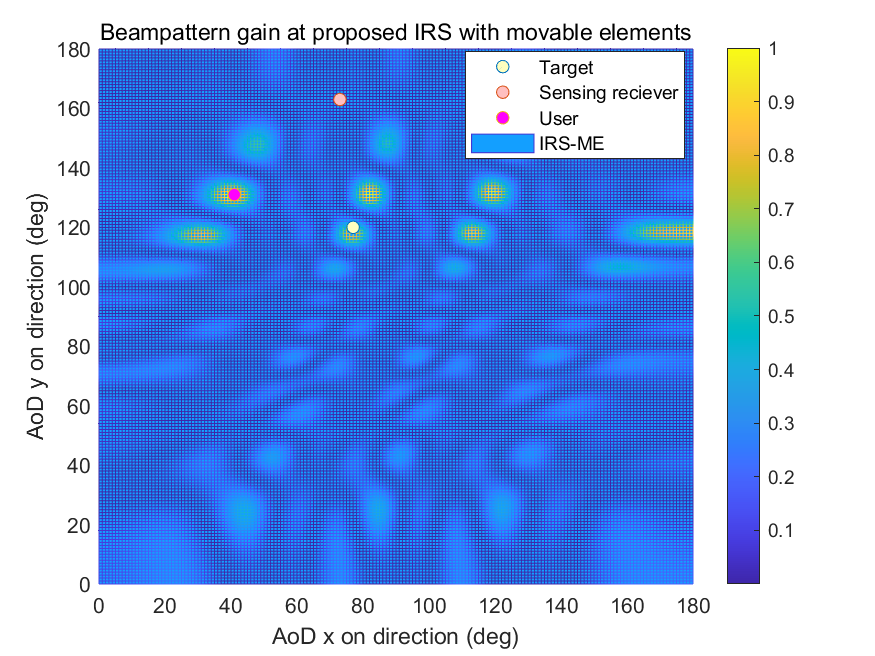}}
\caption{Beampattern gain of the IRS with fixed point elements and with movable elements.}\label{fig:fig3}
\end{minipage}
\begin{minipage}[t]{0.33\linewidth}
\centering
{\includegraphics[width=2.3in]{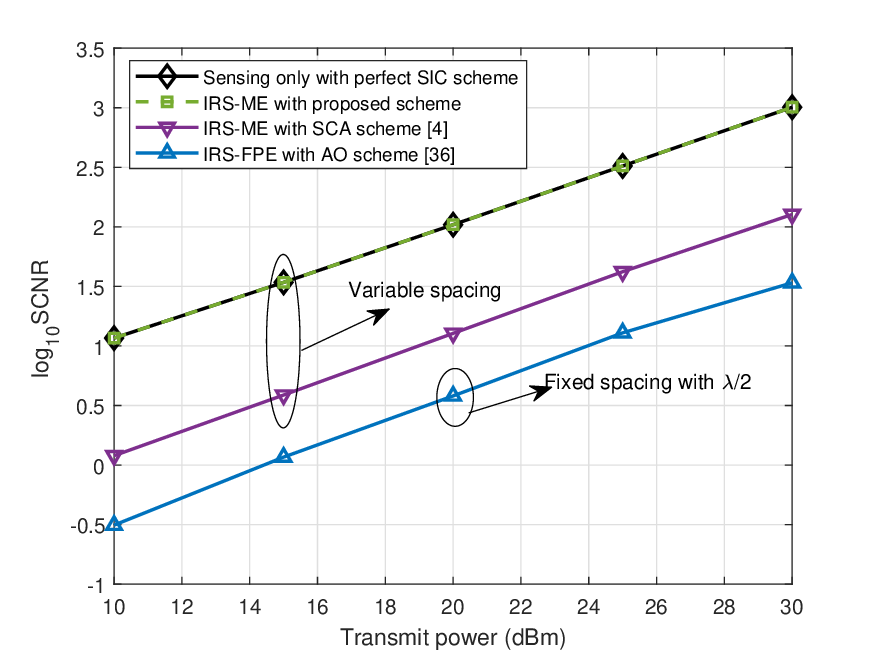}}
\caption{Sensing performance of the proposed scheme.}\label{fig:fig4}
\end{minipage}
\end{figure*}
In this section, simulation results are presented to demonstrate the effectiveness of the proposed beamforming and
position design schemes.  The ISAC transmitter and sensing receiver are located at coordinates (0m,0m,0m) and (40m,0m,0m), respectively, and the IRS is deployed near the user side at (30m,30m,0m), where the distance between the IRS and the center of user cluster is 10 meters, while the 
distance between the IRS and the target is 15 meters \cite{WQQ}.
 The channel power gain is modeled as $\sigma_{\mathrm{i}}^2=c_0d_i^{-\alpha_i}$, where $c_0=-30$ dB denotes the expected value of the average channel power gain at the reference distance of 1 m, and $\alpha_{i}=2.2$ represents the path-loss exponent \cite{c0}. The carrier frequency is 28 GHz. The parameters for Algorithm 1 are set as $T_1=100$ and $T_2=10$. For Algorithm 2, the parameters are defined as $\epsilon = \epsilon_{1} =\epsilon_{2}= 10^{-6}$, $I_{\max}=100$, $S=10$, ${\lambda}_{r^{\mathrm{I}}}=0.05\lambda$, $\rho_p=100$, $\rho_{\mathrm{c},k}=10$, $\eta_1=0.9$, $\eta_2=0.1$, and $\rho=1.5$. 
{Unless otherwise specified, we set $N_{\mathrm{I}}=4\times4$, $A=8\lambda$, $N_{\mathrm{B}}=16$, $N_{\mathrm{S}}=16$, $K=3$, $L_{\mathrm{BI},k}=L_{\mathrm{IU},k}=C=4$,  $\sigma^2=-90$ $\mathrm{dBm}$.
\vspace{-0mm}\subsection{Single-user Scenario}
Fig. \ref{fig:fig3} illustrates the beampatterns of IRS with fixed-point elements and movable elements.\footnote{These highlight circle regions exhibit a certain periodicity in the angular domain, which depends on the number and distribution characteristics of the IRS elements. For a detailed proof, please refer to \cite{R6}, as it is omitted here due to page limitations.} It is observed that 
the IRS with fixed-point elements sacrifices sensing performance by reducing the beampattern gain at the target to guarantee communication quality. In contrast, the IRS with movable elements can improve the beampattern gain for both the user and the target by adjusting the position of elements and its phase shifts, thus achieving superior performance in both communication and sensing.

Fig. \ref{fig:fig4} compares the performance of the proposed algorithm with different schemes, including the ``IRS-FPE with AO scheme'', which optimizes phase shifts for the IRS with fixed-point elements using perfect CSI based on the algorithm in \cite{t-irs}; the ``IRS-ME with SCA scheme'', which applies the position optimization algorithm in \cite{MA-CHANNEL}.
Simulation results indicate that the IRS with movable elements outperforms the traditional fixed-spacing IRS. As expected in Corollary 1, adjusting the positions of the movable elements allows for the simultaneous achievement of upper bounds in both communication and sensing.
This is expected since the movable IRS enables each element to move across the IRS surface, while retaining the ability to control both amplitude and phase of the reflecting signals. Therefore, by leveraging this extra spatial flexibility, an IRS with movable elements can reposition its elements to locations with more favorable channel conditions, thereby enhancing both sensing and communication performance.\footnote{The movement resolution of elements is not considered in the ISAC system. However, this can be accounted for by discretizing the movable region and selecting the nearest grid point to the optimized position. Due to page limitations, we omit the related simulations.}
\vspace{-0mm}
\subsection{Multi-user Scenario}
\begin{figure*}[htbp]\vspace{-0mm}
\centering
\begin{minipage}[t]{0.32\linewidth}
{\includegraphics[width=2.3in]{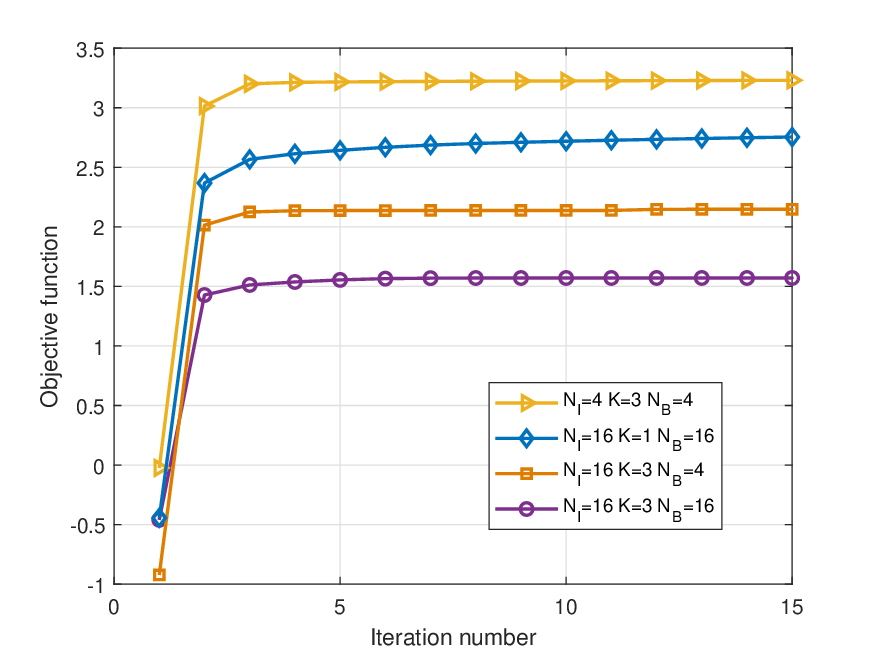}}
\caption{The convergence of the proposed subspace-based algorithms.}\label{fig:fig6}
\end{minipage}
\begin{minipage}[t]{0.32\linewidth}
{\includegraphics[width=2.3in]{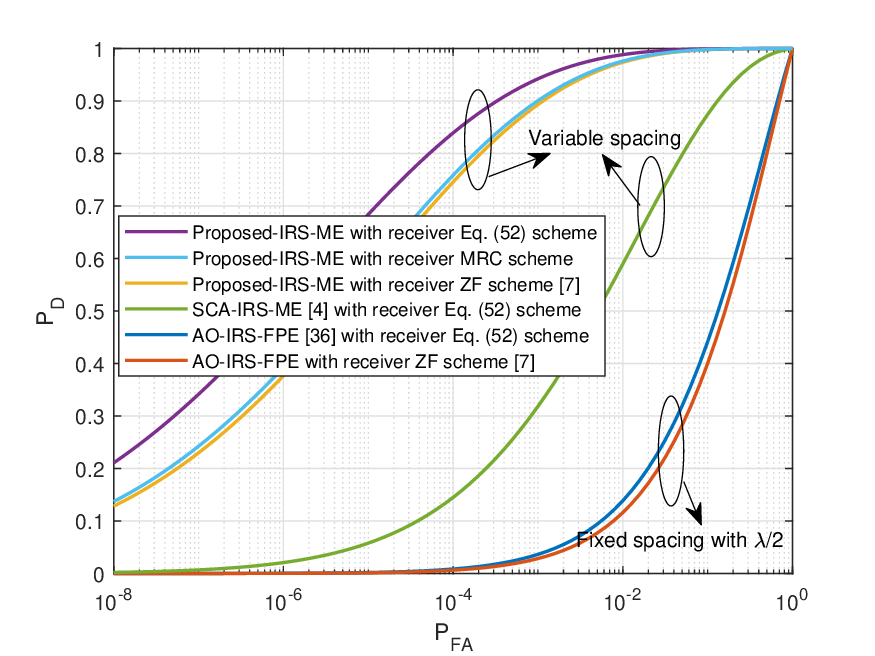}}
\caption{ROC comparison of different beamforming and element position schemes.}\label{fig:fig7}
\end{minipage}
\begin{minipage}[t]{0.32\linewidth}
{\includegraphics[width=2.3in]{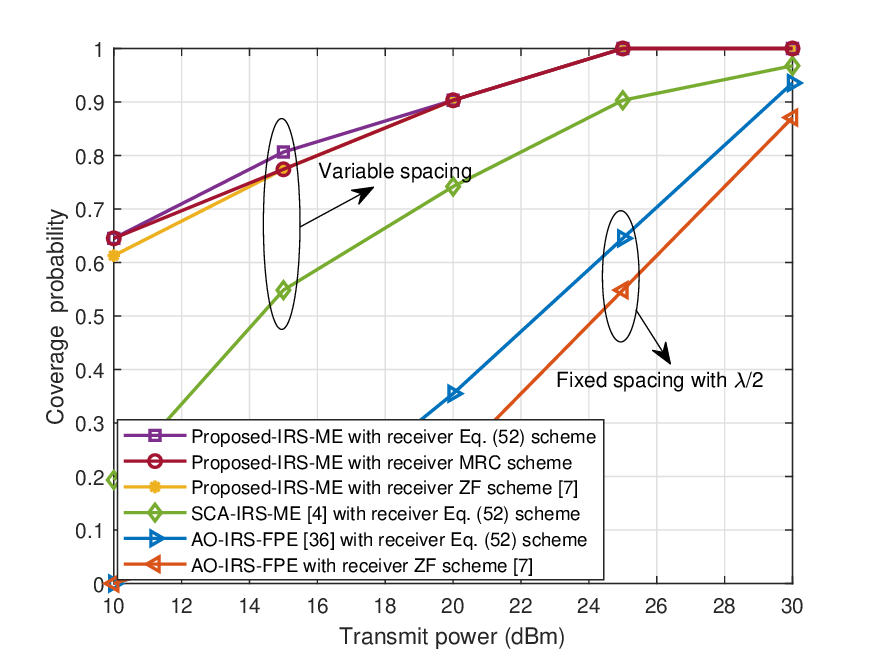}}
\caption{Sensing coverage probability of the proposed beamforming and element position scheme.}\label{fig:fig9}
\end{minipage}
\end{figure*}

\begin{figure*}[htbp]
\centering
\begin{minipage}[t]{0.32\linewidth}
{\includegraphics[width=2.3in]{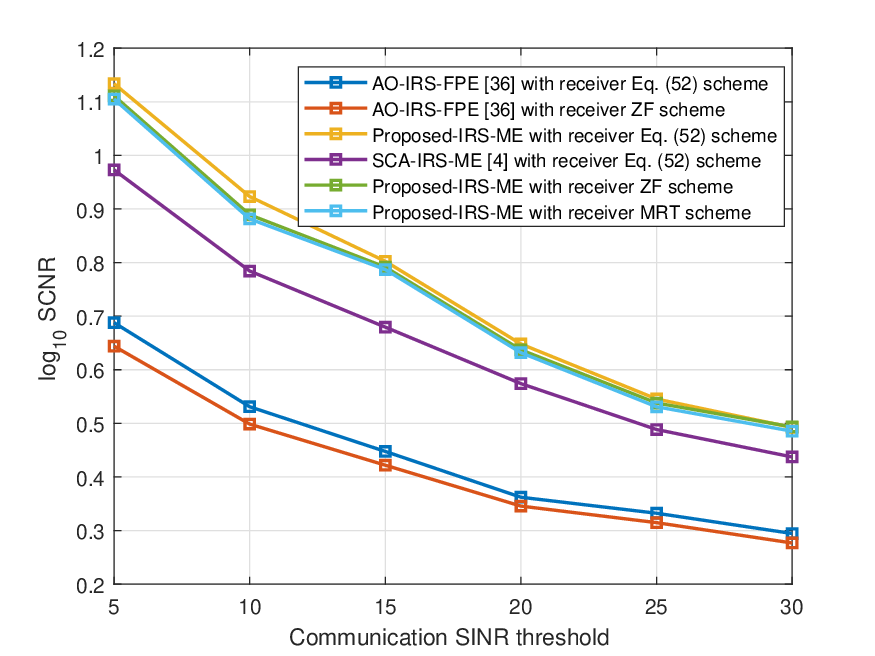}}
\caption{ISAC trade-offs under different beamforming and element position schemes.}\label{fig:tradeoff}
\end{minipage}
\begin{minipage}[t]{0.32\linewidth}
{\includegraphics[width=2.3in]{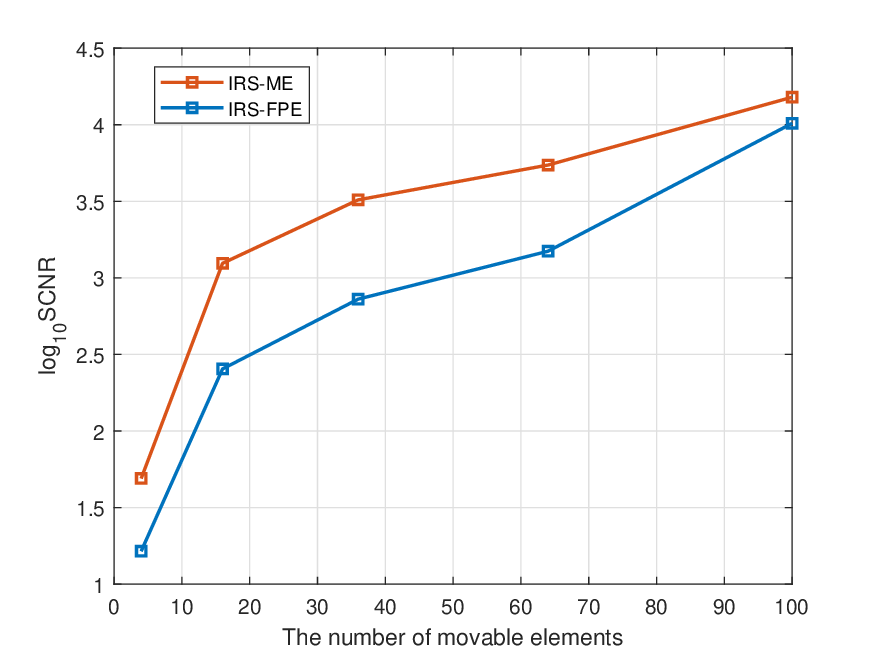}}
\caption{Sensing performance versus the number of IRS elements.}\label{fig:ni_ed1}
\end{minipage}
\begin{minipage}[t]{0.32\linewidth}
{\includegraphics[width=2.3in]{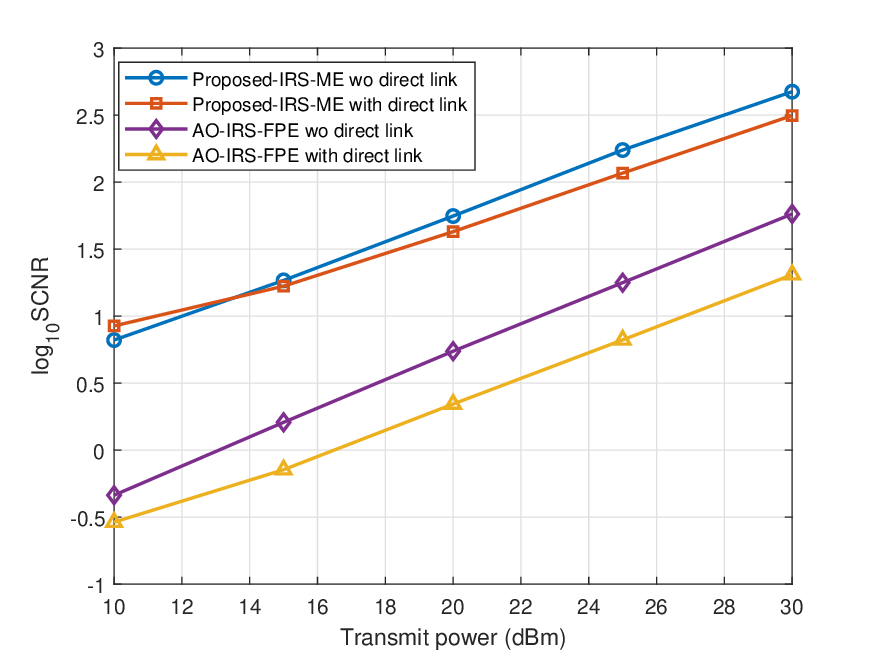}}
\caption{Sensing performance comparison of with/without direct link.}\label{fig:lospath_ed1}
\end{minipage}
\end{figure*}

For multi-user scenarios, we first validate the convergence behavior of the proposed algoritm.
It is observed in Fig. \ref{fig:fig6} that Algorithm 2 converges within a small number of iterations, regardless of the number of users, movable elements, transmit and receive antennas.   This is attributed to the closed-form solutions, which significantly enhance overall performance.

Fig. \ref{fig:fig7} compares the sensing performance using the receiver operating characteristic (ROC) \cite{ROC} as the metric across various beamforming and position schemes with $P=15$ dBm. The ``Proposed-IRS-ME with ZF scheme'' and ``Proposed-IRS-ME with MRC scheme'' employ the sensing beamforming design in \cite{MA_Full} and MRC scheme, with $\bf{W}$, $\boldsymbol{\Theta}$, and $\mathbf{r}^{\mathrm{I}}$ optimized by the proposed algorithm. The ``AO-IRS-FPE with ZF scheme'' employs the sensing beamforming scheme proposed in \cite{MA_Full} with $\bf{W}$ and $\boldsymbol{\Theta}$, optimized by the algorithm in \cite{t-irs}.
Simulation results indicate that compared with the traditional fixed-spacing IRS, the IRS with movable elements demonstrates a huge performance improvement. Furthermore, it surpasses the SCA-based scheme, which optimizes the positions of movable elements one by one \cite{MIMO-CAPACITY} and is prone to becoming trapped in locally optimal solutions.


Fig. \ref{fig:fig9} compares the sensing coverage probability of the IRS with movable elements, where the sensing coverage is defined as the probability that SCNR is not less than the threshold value $\Gamma$ 
, i.e.,
\begin{equation}\begin{array}{l}
\mathcal{P}=\mathbb{P}\left(\gamma_{\mathrm{s}}\left( {{{\bf{W}}},{\bf{r}^{\mathrm{I}}},{\bf{\Theta }}},{\bf{r}_{\mathrm{co}}} \right)\ge\Gamma\right).\end{array}
\end{equation}
It is demonstrated that by adjusting the positions of the reflecting elements, the coverage probability can be further enhanced. This is expected since the IRS enables its element to move across the movable region where the beampattern gain can be reconfigured as shown in Fig. \ref{fig:fig3}b.

Fig. \ref{fig:tradeoff} compares the sensing performance of the proposed algorithm with various schemes under different communication constraints. The results reveal the trade-off between communication and sensing performance, with sensing performance deteriorating as the communication SINR threshold increases. Notably, the proposed algorithm outperforms the baseline schemes, demonstrating that IRS with movable elements effectively balances the communication-sensing trade-off through position adjustments.

Fig. \ref{fig:ni_ed1} presents simulation results for the sensing performance across different numbers of IRS elements. It is demonstrated that the sensing performance of the IRS with movable elements is better than that of the IRS with fixed-spacing elements. However, due to the limited size of the movable region, the available range for each element decreases as the number of elements increases, leading to a reduction in the performance gap between the traditional IRS and that with movable elements.
This is expected, as the distance constraint between adjacent elements, i.e., $\left \| \mathbf{r}^{\mathrm{I}}_m - \mathbf{r}^{\mathrm{I}}_n \right\| \ge D, \forall n \neq m\in[1, N_{\mathrm{I}}]$, limits the movable range of the elements. Consequently, as the number of elements increases, the IRS with movable elements eventually transforms into the traditional fixed-spacing IRS.

Fig. \ref{fig:lospath_ed1} illustrates the sensing performance in the presence of a direct link. The results demonstrate that the proposed algorithm remains effective under such conditions. Notably, the IRS with movable elements outperforms the conventional fixed-spacing IRS, confirming that the algorithm can be applied to scenarios where a direct link exists with small modifications.

\begin{figure*}[htbp]
\centering
\begin{minipage}[t]{0.32\linewidth}
{\includegraphics[width=2.3in]{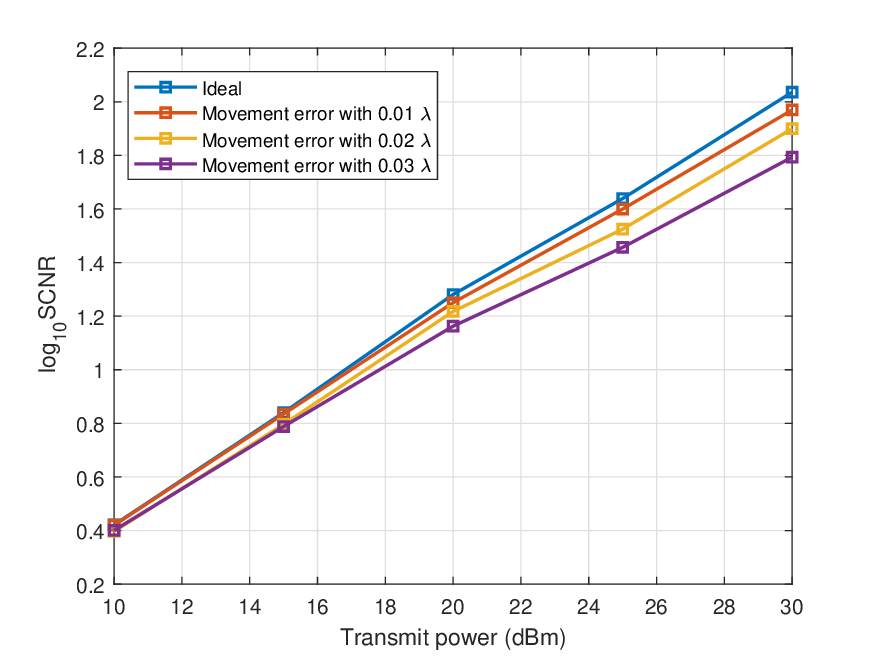}}
\caption{The impact of element movement error on sensing performance.}\label{fig:mrerror_ed1}
\end{minipage}
\begin{minipage}[t]{0.32\linewidth}
{\includegraphics[width=2.3in]{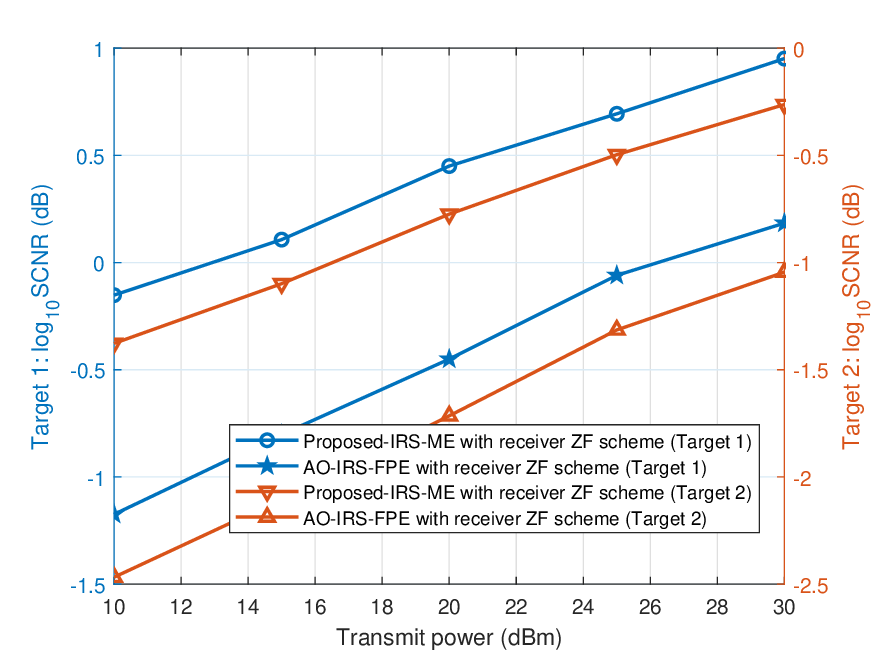}}
\caption{Sensing performance in multiple targets scenario.}\label{fig:multitarget}
\end{minipage}
\begin{minipage}[t]{0.32\linewidth}
{\includegraphics[width=2.3in]{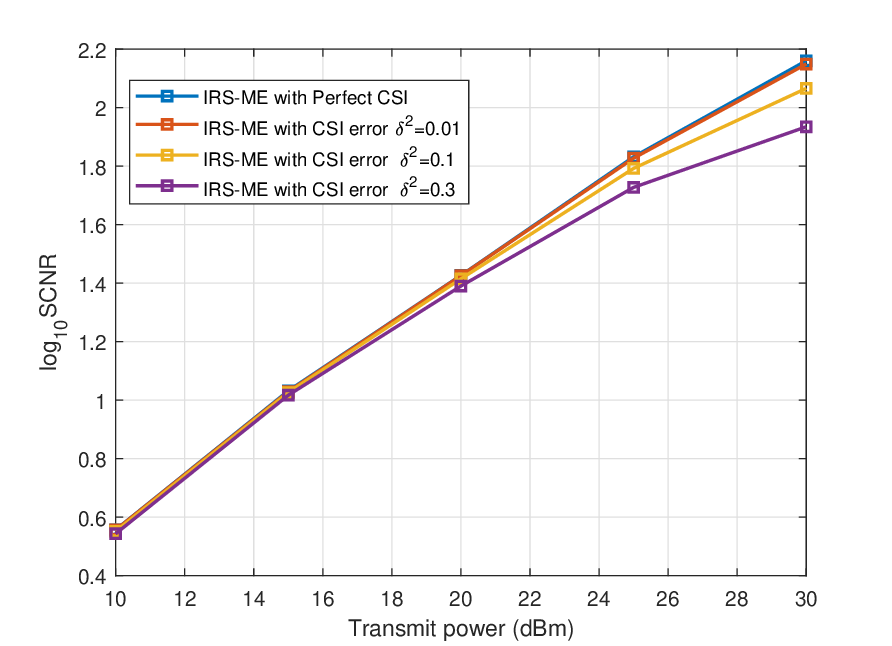}}
\caption{The impact of CSI error on sensing performance.}\label{fig:CSIno15_ed1}
\end{minipage}
\end{figure*}

\subsection{Extension to Real-World Scenarios}

In practical wireless systems, achieving accurate position of the movable element is challenging due to the hardware limitations. To illustrate the impact of element movement errors on system performance, simulation results are provided in Fig. \ref{fig:mrerror_ed1}. The figure compares sensing performance for different element movement errors, set at $0.01\lambda$, $0.02\lambda$, and $0.03\lambda$ based on the practical movement resolution of MA \cite{me}. As shown in Fig. \ref{fig:mrerror_ed1}, an increase in movement error leads to performance degradation, which is expected, as slight adjustments in MA positions within the wavelength range can significantly affect system behavior. However, when movement errors are small, the proposed algorithm can still yield a good performance.

Rather than the time-division narrow-beam method, the proposed algorithm for BS beamforming, IRS phase shift, and element position design  remains applicable for simultaneous multi-target detection. To mitigate interference among multiple target echoes, a zero-forcing (ZF) beamformer strategy should be applied at the sensing receiver based on the received signals, thereby enhancing performance. Fig. \ref{fig:multitarget} illustrates the performance of the ZF-based algorithm with two targets. As shown in the figure, the system assisted by an IRS with movable elements consistently outperforms the one with fixed elements, highlighting the significant advantages of movable-element IRS.

The simulation results presented above section assume perfect CSI. In practical wireless systems, achieving perfect CSI is challenging due to noise and limited training resources. Consequently, evaluating the effects of imperfect CSI on ISAC performance is essential.
Fig. \ref{fig:CSIno15_ed1} shows the impact of CSI error on sensing performance, where the CSI error can be modeled as \cite{CSIerror}, i.e.,
\begin{equation}
{{\bf{H}}_{{\rm{BI}}}} = {{\bf{\tilde H}}_{{\rm{BI}}}} + \Delta{{\bf{H}}_{{\rm{BI}}}},{{\bf{h}}_{{\rm{IU},k}}} = {{\bf{\tilde h}}_{{\rm{IU,k}}}} + \Delta{{\bf{h}}_{{\rm{IU,k}}}}, {{\bf{H}}_{{\rm{t}}}} = {{\bf{\tilde H}}_{{\rm{t}}}} + \Delta{{\bf{H}}_{{\rm{t}}}},
\end{equation}
where 
\begin{equation}
\mathbf{H}_{\mathrm{t}}={\alpha _{{\rm{IS,T}}}}{\bf{a}}_{{\rm{S}}}^*(\varphi _{{\rm{T}}}^{\rm{S}}){\bf{a}}_{\rm{I}}^T(\varphi _{{\rm{T}}}^{\rm{IS}},\omega _{{\rm{T}}}^{\rm{IS}})+\sum\limits_{c=0}^{C}{\alpha _{{\rm{IS,c}}}}{\bf{a}}_{{\rm{S}}}^*(\varphi _{{\rm{c}}}^{\rm{S}}){\bf{a}}_{\rm{I}}^T(\varphi _{{\rm{c}}}^{\rm{IS}},\omega _{{\rm{c}}}^{\rm{IS}}),
\end{equation}
${{\bf{\tilde H}}_{{\rm{BI}}}}$, ${{\bf{\tilde h}}_{{\rm{IU,k}}}}$, and ${{\bf{\tilde H}}_{{\rm{t}}}}$ denote the estimated channels.
 $\Delta {{\bf{H}}_{{\rm{BI}}}}$, $\Delta {{\bf{h}}_{{\rm{IU},k}}}$, and  $\Delta {{\bf{H}}_{{\rm{t}}}}$ denote the corresponding CSI errors, which follow the CSCG distribution, i.e., $\mathrm{vec}\left(\Delta\mathbf{H}_{\mathrm{BI}}\right)\sim \mathcal{CN}\left(0,\delta^{2}\mathbf{I}_{N_{I}N_{B}}\right)$, $\mathrm{vec}\left(\Delta\mathbf{h}_{\mathrm{IU,k}}\right)\sim \mathcal{CN}\left(0,\delta^{2}\mathbf{I}_{N_{I}}\right)$, and $\mathrm{vec}\left(\Delta\mathbf{H}_{\mathrm{t}}\right)\sim \mathcal{CN}\left(0,\delta^{2}\mathbf{I}_{N_{I}N_{S}}\right)$, where $\delta^{2}$ quantifies the channel estimation inaccuracy.
As shown in Fig. \ref{fig:CSIno15_ed1}, sensing performance degrades with increasing CSI error. Nevertheless, when errors are small, the IRS with movable element maintains robust performance. Thus, the proposed algorithm can still yield a good performance for the channels with errors.

\vspace{-0mm}\section{Conclusion}\label{se:se6}
This paper exploited the spatial degree of freedom in the movable elements of IRS to enhance the performance of ISAC. Performance was analyzed for two typical scenarios and the corresponding optimization schemes were proposed for beamforming and element position design. For the single-user scenario, we derived the upper bounds of sensing and communication performance, and optimized the elements position based on angular information. Different from the traditional fixed-spacing IRS, both the communication and sensing upper bounds can be achieved. For the multi-user with multi-path scenario, we derived the lower bounds on ergodic rates for sensing and communication and developed an MPPGD-based element position algorithm. 
Simulation results demonstrated that the coverage probability can be further enhanced, highlighting the significant potential of IRS with movable elements in ISAC systems.
Future research could focus on the development of low-complexity, low pilot overhead elements position design algorithms, aiming to reduce system costs and avoid suboptimal solutions within the movable region.
Moreover, investigating the performance of IRS with movable elements in more practical scenarios, such as systems with imperfect CSI, movement error, and phase noise, and designing robust beamforming and element position algorithm represent another promising research direction for further research.
\vspace{0mm}\appendices
\section{Proof of Theorem \ref{theorem:0}}\label{app:theorem:0}
A tight upper bound on the SCNR is the full array gain over the desired direction, $\varphi _{{\rm{T}}}^{\rm{S}}$, and under the null-steering constraint over the interference direction, $\varphi _{{{0}}}^{\rm{S}}$, which can be derived by the ZF beamformer \cite{MA_Full}\vspace{-0mm}
\begin{equation}\vspace{-0mm}\scalebox{1}{$
\begin{array}{l}
{{\bf{r}_{\mathrm{co}}^{\mathrm{ZF}}}=\frac{{\bf{r}_{\mathrm{co}}}}{\left\|{\bf{r}_{\mathrm{co}}}\right\|},}\\
{{\bf{r}_{\mathrm{co}}}=\left[\mathbf{I}_{N_{\mathrm{S}}}-{\bf{a}}_{{\rm{S}}}^*(\varphi _{{{0}}}^{\rm{S}})\left({\bf{a}}_{{\rm{S}}}^T(\varphi _{{{0}}}^{\rm{S}}){\bf{a}}_{{\rm{S}}}^*(\varphi _{{{0}}}^{\rm{S}})\right)^{-1}{\bf{a}}_{{\rm{S}}}^T(\varphi _{{{0}}}^{\rm{S}})\right]{\bf{a}}_{{\rm{S}}}^*(\varphi _{{\rm{T}}}^{\rm{S}}).}
\end{array}$}
\end{equation}
Therefore, the SCNR is bounded by\vspace{-0mm}
\begin{equation}\vspace{-0mm}\scalebox{1}{$
\begin{array}{l}
 \gamma_{\mathrm{s}}\left( {{{\bf{W}}},{\bf{r}_m^{\mathrm{I}}},{\bf{\Theta }}},{\bf{r}_{\mathrm{co}}^{\mathrm{ZF}}} \right)=\\
\left|{\bf{a}}_{{\rm{S}}}^T(\varphi _{{\rm{T}}}^{\rm{S}}){\mathbf{r}_{\mathrm{co}}^{\mathrm{ZF}}}\right|^2
\frac{{{{\left\| {\alpha _{{\rm{IS,T}}}}{\alpha _{{\rm{BI}}}}{\bf{a}}_{\rm{I}}^T(\varphi _{{\rm{T}}}^{\rm{IS}}\hspace{-0.5mm},\hspace{-0.5mm}\omega _{{\rm{T}}}^{\rm{IS}}){\bf{\Theta }}
{\bf{a}}_{\rm{I}}^*(\varphi ^{\rm{BI}}\hspace{-0.5mm},\hspace{-0.5mm}\omega ^{\rm{BI}}){\bf{a}}_{{\rm{B}}}^T(\varphi^{\rm{B}}){\mathbf{W}} \right\|}^2}}}{{ {\sigma ^2}}}\\
\leq  \frac{N_{\mathrm{S}}{{{\left|{\alpha _{{\rm{IS,T}}}}{\alpha _{{\rm{BI}}}}{\bf{a}}_{\rm{I}}^T(\varphi _{{\rm{T}}}^{\rm{IS}}\hspace{-0.5mm},\hspace{-0.5mm}\omega _{{\rm{T}}}^{\rm{IS}}){\bf{\Theta }}
{\bf{a}}_{\rm{I}}^*(\varphi ^{\rm{BI}}\hspace{-0.5mm},\hspace{-0.5mm}\omega ^{\rm{BI}})\right|^2
\left\| {\bf{a}}_{{\rm{B}}}^T(\varphi^{\rm{B}}){\mathbf{W}} \right\|}^2}}}{{ {\sigma ^2}}}
\\
\triangleq\gamma_{\mathrm{s}}^0\left( {{{\bf{W}}},{\bf{r}^{\mathrm{I}}},{\bf{\Theta }}} \right).
\end{array}$}
\end{equation}

Define $\mathbf{R}_\mathrm{c}=\mathbf{w}_\mathrm{c}\mathbf{w}_\mathrm{c}^H$ and $\mathbf{R}_\mathrm{s}=\mathbf{w}_\mathrm{s}\mathbf{w}_\mathrm{s}^H$, the SINR and SCNR can be rewritten into\vspace{-0mm}
\begin{equation}\vspace{-0mm}\scalebox{1}{$
\begin{array}{l}
\gamma_{\mathrm{s}}^0\left( {\mathbf{R}_\mathrm{c},\mathbf{R}_\mathrm{s},{\bf{r}_m^{\mathrm{I}}},{\bf{\Theta }}} \right) =
\frac{N_{\mathrm{S}}{{{\left|{\alpha _{{\rm{IS,T}}}}{\alpha _{{\rm{BI}}}}{\bf{a}}_{\rm{I}}^T(\varphi _{{\rm{T}}}^{\rm{IS}}\hspace{-0.5mm},\hspace{-0.5mm}\omega _{{\rm{T}}}^{\rm{IS}}){\bf{\Theta }}
{\bf{a}}_{\rm{I}}^*(\varphi ^{\rm{BI}}\hspace{-0.5mm},\hspace{-0.5mm}\omega ^{\rm{BI}})\right|^2}}}}{{ {\sigma ^2}}}\\
\qquad\qquad\times {\bf{a}}_{{\rm{B}}}^T(\varphi^{\rm{B}})\left({{\bf{R}}_{\mathrm{s}}}+{{\bf{R}}_{\mathrm{c}}}\right)
{\bf{a}}_{{\rm{B}}}^*(\varphi^{\rm{B}}),
\end{array}$}
\end{equation}
and\vspace{-0mm}
\begin{equation}\vspace{-0mm}\scalebox{1}{$
\begin{aligned}
&{\gamma_{\mathrm{c}}}\left( {{{\bf{R}}_{\mathrm{s}}},{{\bf{R}}_{\mathrm{c}}},{\bf{r}_m^{\mathrm{I}}},{\bf{\Theta }}} \right)=  \\
&\frac{{\left | {\alpha _{{\rm{IU}}}}{\alpha _{{\rm{BI}}}}{\bf{a}}_{\rm{I}}^T(\varphi^ {\rm{IU}},\omega^{\rm{IU}})
{\bf{\Theta }}{\bf{a}}_{\rm{I}}^*(\varphi ^{\rm{BI}},\omega ^{\rm{BI}})
 \right | ^2{\bf{a}}_{{\rm{B}}}^T(\varphi^{\rm{B}}){{\bf{R}}_{\mathrm{c}}}{\bf{a}}_{{\rm{B}}}^*(\varphi^{\rm{B}})}}
{{  \left | {{\alpha _{{\rm{IU}}}}{\alpha _{{\rm{BI}}}}{\bf{a}}_{\rm{I}}^T(\varphi^ {\rm{IU}},\omega^{\rm{IU}}){\bf{\Theta }}{\bf{a}}_{\rm{I}}^*(\varphi ^{\rm{BI}},\omega ^{\rm{BI}})} \right | ^2 {\bf{a}}_{{\rm{B}}}^T(\varphi^{\rm{B}}){{\bf{R}}_{\mathrm{s}}}{\bf{a}}_{{\rm{B}}}^*(\varphi^{\rm{B}}) + {\sigma ^2}}}.
\end{aligned}$}
\end{equation}
Let $\{\mathbf{R}_{\mathrm{c}}^*, \mathbf{R}_{\mathrm{s}}^*\}$ denote one of the schemes for the beamforming at the ISAC transmitter. It can be verified that $\{\mathbf{R}_{\mathrm{c}}^*+ \mathbf{R}_{\mathrm{s}}^*, \bf{0}\}$ satisfies the upper bound for communication and achieves the same value for SCNR. Moreover, according to the power constraints, i.e., $\mathrm{Tr}\left(\mathbf{R}_{\mathrm{c}}+ \mathbf{R}_{\mathrm{s}}\right)\leq P$, the communication and sensing performance can be further bounded when \vspace{-0mm}
\begin{equation}\label{eq:eqw}\vspace{-0mm}\scalebox{1}{$
{{{\bf{w}}_{\mathrm{c}}} = \sqrt P \frac{{\bf{a}}_{{\rm{B}}}^*(\varphi^{\rm{B}})}{{\left\| {\bf{a}}_{{\rm{B}}}^*(\varphi^{\rm{B}}) \right\|}},{{\bf{w}}_{\mathrm{s}}}=\bf{0},}$}\end{equation}
Therefore, the system performance can be upper bounded by\vspace{-0mm}
\begin{equation}\vspace{-0mm}\scalebox{1}{$
\begin{array}{l}
\gamma_{\mathrm{c}}\left( {\mathbf{R}_\mathrm{c},\mathbf{R}_\mathrm{s},{\bf{r}_m^{\mathrm{I}}},{\bf{\Theta }}} \right)=  \\
\frac{{\left | {\alpha _{{\rm{IU}}}}{\alpha _{{\rm{BI}}}}{\bf{a}}_{\rm{I}}^T(\varphi^ {\rm{IU}},\omega^{\rm{IU}})
{\bf{\Theta }}{\bf{a}}_{\rm{I}}^*(\varphi ^{\rm{BI}},\omega ^{\rm{BI}})
 \right | ^2{\bf{a}}_{{\rm{B}}}^T(\varphi^{\rm{B}})\left({{\bf{R}}_{\mathrm{c}}}+{{\bf{R}}_{\mathrm{s}}}\right){\bf{a}}_{{\rm{B}}}^*(\varphi^{\rm{B}})}}
{{ {\sigma ^2}}}\\
\overset{\text{(a)}}{\leq} \frac{PN_{\mathrm{B}}{\left | {\alpha _{{\rm{IU}}}}{\alpha _{{\rm{BI}}}}{\bf{a}}_{\rm{I}}^T(\varphi^ {\rm{IU}},\omega^{\rm{IU}})
{\bf{\Theta }}{\bf{a}}_{\rm{I}}^*(\varphi ^{\rm{BI}},\omega ^{\rm{BI}})
 \right | ^2}}
{{ {\sigma ^2}}}
\triangleq\gamma_{\mathrm{c}}^{\mathrm{upp}},
\end{array}$}
\end{equation}
and\vspace{-0mm}
$$
\gamma_{\mathrm{s}}^0\left( {\mathbf{R}_\mathrm{c},\mathbf{R}_\mathrm{s},{\bf{r}_m^{\mathrm{I}}},{\bf{\Theta }}} \right) =
\frac{N_{\mathrm{S}}{{{\left|{\alpha _{{\rm{IS,T}}}}{\alpha _{{\rm{BI}}}}{\bf{a}}_{\rm{I}}^T(\varphi _{{\rm{T}}}^{\rm{IS}}\hspace{-0.5mm},\hspace{-0.5mm}\omega _{{\rm{T}}}^{\rm{IS}}){\bf{\Theta }}
{\bf{a}}_{\rm{I}}^*(\varphi ^{\rm{BI}}\hspace{-0.5mm},\hspace{-0.5mm}\omega ^{\rm{BI}})\right|^2}}}}{{ {\sigma ^2}}}
$$
\begin{equation}\vspace{-0mm}\scalebox{1}{$
\begin{array}{l}
\times {\bf{a}}_{{\rm{B}}}^T(\varphi^{\rm{B}})\left({{\bf{R}}_{\mathrm{s}}}+{{\bf{R}}_{\mathrm{c}}}\right)
{\bf{a}}_{{\rm{B}}}^*(\varphi^{\rm{B}})\\
\overset{\text{(b)}}{\leq} \frac{PN_{\mathrm{B}}N_{\mathrm{S}}{{{\left|{\alpha _{{\rm{IS,T}}}}{\alpha _{{\rm{BI}}}}{\bf{a}}_{\rm{I}}^T(\varphi _{{\rm{T}}}^{\rm{IS}}\hspace{-0.5mm},\hspace{-0.5mm}\omega _{{\rm{T}}}^{\rm{IS}}){\bf{\Theta }}
{\bf{a}}_{\rm{I}}^*(\varphi ^{\rm{BI}}\hspace{-0.5mm},\hspace{-0.5mm}\omega ^{\rm{BI}})\right|^2}}}}{{ {\sigma ^2}}}
\triangleq\gamma_{\mathrm{s}}^{\mathrm{upp}},
\end{array}$}
\end{equation}
where (a) and (b) can be respectively obtained by the phase alignment method \cite{WQQ}, i.e., \vspace{-0mm}
\begin{equation}\label{eq:co:only}\vspace{-0mm}\scalebox{1}{$
\begin{array}{l}
\boldsymbol{\xi}=\mathrm{diag}\left({\bf{a}}_{\rm{I}}(\varphi^ {\rm{IU}},\omega^{\rm{IU}})\right)
{\bf{a}}_{\rm{I}}^H(\varphi ^{\rm{BI}},\omega ^{\rm{BI}}),
\end{array}$}
\end{equation} 
and \vspace{-0mm}
\begin{equation}\label{eq:se:only}\vspace{-0mm}\scalebox{1}{$
\begin{array}{l}
\boldsymbol{\xi}=\mathrm{diag}\left({\bf{a}}_{\rm{I}}(\varphi^ {\rm{IS}},\omega^{\rm{IS}})\right)
{\bf{a}}_{\rm{I}}^H(\varphi ^{\rm{BI}},\omega ^{\rm{BI}}).
\end{array}$}
\end{equation}  

This completes the proof.
\vspace{-0mm}\section{Proof of Corollary \ref{coro:1}}\label{app:coro:1}
Since the diagonal element of PRM follows a CSCG distribution, i.e., $\scalebox{1}{$\mathcal{CN} \left(0,\frac{\sigma_{\mathrm{i}}^2}{L_{\mathrm{i}}}\right)$}$, therefore, $\scalebox{1}{${\left| {{\alpha _{{\rm{I2U}},l,k}}} \right|}$}$ and $\scalebox{1}{${\left| {{\alpha _{{\rm{BI}},l}}} \right|}$}$ follows the Rayleigh distribution, with 
$\scalebox{1}{$\begin{array}{l}{\mathbb{E}\left\{ {{{\left| {{\alpha _{{\rm{I2U}},l,k}}} \right|}^2}} \right\}}=\frac{{\sigma _{{\rm{IU}},k}^{\rm{2}}}}{{{L_{{\rm{IU}},k}}}}\end{array}$}$ and $\scalebox{1}{$\mathbb{E}\left\{ {{{\left| {{\alpha _{{\rm{BI}},l}}} \right|}^2}} \right\}=\frac{{\sigma _{{\rm{BI}}}^{\rm{2}}}}{{{L_{{\rm{BI}}}}}}$}$.  Let $\scalebox{1}{${g_{m,k = }}\sum\limits_{l = 1}^{{L_{{\rm{IU,}}k}}} {\alpha _{{\rm{IU}},l,k}^*} {e^{ - j\frac{{2\pi }}{\lambda }\rho_{l,k}^{\mathrm{IU}}(\mathbf{r}_{m}^{\mathrm{I}})}} \sim CN\left( {0,\sigma _{{\rm{IU}},k}^{\rm{2}}} \right)$}$, and $\scalebox{1}{${h_{m,n}} = \sum\limits_{l = 1}^{{L_{{\rm{BI}}}}} {{\alpha _{{\rm{BI}},l}}} {e^{ - j\frac{{2\pi }}{\lambda }\left( {
\rho_{l}^{\mathrm{BI}}(\mathbf{r}_{m}^{\mathrm{I}}) - \rho_{l}^{\mathrm{B}}(\mathbf{r}_{n}^{\mathrm{B}})
} \right)}} \sim CN\left( {0,\sigma _{{\rm{BI}}}^{\rm{2}}} \right)$}$, the effective channel between the ISAC transmitter and $k$-th user can be rewritten into\vspace{-0mm}
\begin{equation}\label{eq:eqh}\vspace{-0mm}\scalebox{1}{$
{\bf{h}}_{{\rm{IU,}}k}^{{H}}{\rm{diag}}\left( {\bf{\theta }} \right){{\bf{H}}_{{\rm{BI}}}}\hspace{-1mm} =\hspace{-1mm} \left[ \hspace{-1mm}{\sum\limits_{m = 1}^{{N_{\mathrm{I}}}} {g_{m,k}^*{h_{m,1}}} {e^{j{\theta _m}}}\hspace{-1mm}, \hspace{-1mm}\cdots\hspace{-1mm} ,\hspace{-1mm}\sum\limits_{m = 1}^{{N_{\mathrm{I}}}} {g_{m,k}^*{h_{m,{N_{\mathrm{B}}}}}} {e^{j{\theta _m}}}} \hspace{-1mm}\right].$}
\end{equation}
By applying the MRT method, $A^{\mathrm{c}}_k$ can be rewritten into\vspace{-0mm}
\begin{equation}\vspace{-0mm}\scalebox{1}{$
\begin{array}{l}
A^{\mathrm{c}}_k= {{\left| {{\eta _k}} \right|}^2}\mathbb{E}\left\{ {{{\left\| {{\bf{h}}_{{\rm{IU,}}k}^{{H}}{\bf{\Theta }}{{\bf{H}}_{{\rm{BI}}}}} \right\|}^4}} \right\} \\
+ \sum\limits_{i \ne k}^K {{\left| {{\eta _i}} \right|}^2}{\mathbb{E}\left\{ {{{\left| {{\bf{h}}_{{\rm{IU,}}k}^{{H}}{\bf{\Theta }}{{\bf{H}}_{{\rm{BI}}}}{{\left( {{\bf{h}}_{{\rm{IU,}}i}^{{H}}{\bf{\Theta }}{{\bf{H}}_{{\rm{BI}}}}} \right)}^H}} \right|}^2}} \right\}}.
\end{array}$}
\end{equation}
According to Eq. (\ref{eq:eqh}), the first term of $A^{\mathrm{c}}_k$ can be rewritten into\vspace{-0mm}
\begin{equation}\vspace{-0mm}\scalebox{1}{$
\begin{array}{l}
\mathbb{E}\left\{ {{{\left\| {{\bf{h}}_{{\rm{IU,}}k}^{{H}}{\bf{\Theta }}{{\bf{H}}_{{\rm{BI}}}}} \right\|}^4}} \right\}=
{\left| {{\eta _k}} \right|^2}\sum\limits_{n = 1}^{{N_{\mathrm{B}}}} {\mathbb{E}\left\{ {{{\left| {\sum\limits_{m = 1}^{{N_{\mathrm{I}}}} {g_{m,k}^*{h_{m,n}}} {e^{j{\theta _m}}}} \right|}^4}} \right\}} \\
+ {\left| {{\eta _k}} \right|^2}\hspace{-1mm}\sum\limits_{n = 1}^{{N_{\mathrm{B}}}}\hspace{-1mm} {\sum\limits_{n_{1} \ne n}^{{N_{\mathrm{B}}}}\hspace{-2mm} {\mathbb{E}\hspace{-1mm}\left\{\hspace{-1mm} {{{\left| {\sum\limits_{m = 1}^{{N_{\mathrm{I}}}} {g_{m,k}^*{h_{m,n}}} {e^{j{\theta _m}}}} \right|}^2}{{\left| {\sum\limits_{m = 1}^{{N_{\mathrm{I}}}} {g_{m,k}^*{h_{m,n_{1}}}} {e^{j{\theta _m}}}} \right|}^2}} \hspace{-1mm}\right\}} },
\end{array}$}
\end{equation}
where the first part can be derived as\vspace{-0mm}
\begin{equation}\vspace{-0mm}\vspace{-0mm}\scalebox{1}{$
\begin{array}{l}
\mathbb{E}\left\{ {{{\left| {\sum\limits_{m = 1}^{{N_{\mathrm{I}}}} {g_{m,k}^*{h_{m,n}}} {e^{j{\theta _m}}}} \right|}^4}} \right\}=
 \sum\limits_{m = 1}^{{N_{\mathrm{I}}}} \sum\limits_{m_{1} = 1}^{{N_{\mathrm{I}}}} \sum\limits_{m_{2} = 1}^{{N_{\mathrm{I}}}} \sum\limits_{m_{3} = 1}^{{N_{\mathrm{I}}}} \\
\qquad\times \mathbb{E}\left\{ {g_{m,k}^*{g_{m_{1},k}}{g_{m_{2},k}}g_{m_{3},k}^*} \right\}\mathbb{E}\left\{ {{h_{m,n}}h_{m_{1},n}^*{h_{m_{3},n}}h_{m_{2},n}^*} \right\} \\
\qquad\times{e^{j\left( {{\theta _m} - {\theta _{m_{1}}} - {\theta _{m_{2}}} + {\theta _{m_{3}}}} \right)}},
\end{array}$}
\end{equation}
where\begin{equation}\label{eq:eqsybeg}\vspace{-0mm}\scalebox{1}{$
\begin{array}{l}
\mathbb{E}\left\{ {g_{m,k}^*{g_{m_{1},k}}{g_{m_{2},k}}g_{m_{3},k}^*} \right\} 
={{{\bf{ \bar{G}}}}_{k,k}}\left[ {{{\bf{r}}_m^{\mathrm{I}}},{{\bf{r}}_{m_{1}}^{\mathrm{I}}},{{\bf{r}}_{m_{2}}^{\mathrm{I}}},{{\bf{r}}_{m_{3}}^{\mathrm{I}}}} \right]\\
=\sum\limits_{l = 1}^{{L_{{\rm{IU,}}k}}} {\sum\limits_{l_{1} = 1}^{{L_{{\rm{IU,}}k}}} {\frac{{\sigma _{{\rm{IU}}}^{\rm{4}}}}{{L_{{\rm{IU}}}^2}}} } {e^{j\frac{{2\pi }}{\lambda }\left( {
\rho_{l,k}^{\mathrm{IU}}(\mathbf{r}_{m}^{\mathrm{I}}) - \rho_{l_{1},k}^{\mathrm{IU}}(\mathbf{r}_{m_{1}}^{\mathrm{I}}) - \rho_{l,k}^{\mathrm{IU}}(\mathbf{r}_{m_{2}}^{\mathrm{I}}) + \rho_{l_{1},k}^{\mathrm{IU}}(\mathbf{r}_{m_{3}}^{\mathrm{I}})
} \right)}}\\
 + \sum\limits_{l = 1}^{{L_{{\rm{IU,}}k}}} {\sum\limits_{l_{2} = 1}^{{L_{{\rm{IU,}}k}}} {\frac{{\sigma _{{\rm{IU}}}^{\rm{4}}}}{{L_{{\rm{IU}}}^2}}} } {e^{j\frac{{2\pi }}{\lambda }\left( {
\rho_{l,k}^{\mathrm{IU}}(\mathbf{r}_{m}^{\mathrm{I}}) - \rho_{l,k}^{\mathrm{IU}}(\mathbf{r}_{m_{1}}^{\mathrm{I}}) - \rho_{l_{2},k}^{\mathrm{IU}}(\mathbf{r}_{m_{2}}^{\mathrm{I}}) + \rho_{l_{2},k}^{\mathrm{IU}}(\mathbf{r}_{m_{3}}^{\mathrm{I}})
} \right)}},
\end{array}$}
\end{equation}
Similarly,  $\mathbb{E}\left\{ {{h_{m,n}}h_{m_{1},n}^*h_{m_{2},n}^*{h_{m_{3},n}}} \right\}$ can be derived as\vspace{-0mm}
$$
\mathbb{E}\left\{ {{h_{m,n}}h_{m_{1},n}^*h_{m_{2},n}^*{h_{m_{3},n}}} \right\}={{{\bf{ \bar {F}}}}_{n,n}}\left[ {{{\bf{r}}_m^{\mathrm{I}}},{{\bf{r}}_{m_{1}}^{\mathrm{I}}},{{\bf{r}}_{m_{2}}^{\mathrm{I}}},{{\bf{r}}_{m_{3}}^{\mathrm{I}}}} \right]
$$
$$
 = \sum\limits_{p = 1}^{{L_{{\rm{BI}}}}} {\sum\limits_{p_{2} = 1}^{{L_{{\rm{BI}}}}} {\frac{{\sigma _{{\rm{BI}}}^{\rm{4}}}}{{L_{{\rm{BI}}}^2}}} } {e^{ - j\frac{{2\pi }}{\lambda }\left( {
\rho_{p}^{\mathrm{BI}}(\mathbf{r}_m^{\mathrm{I}}) - \rho_{p}^{\mathrm{BI}}(\mathbf{r}_{m_{1}}^{\mathrm{I}}) - \rho_{p_{2}}^{\mathrm{BI}}(\mathbf{r}_{m_{2}}^{\mathrm{I}}) + \rho_{p_2}^{\mathrm{BI}}(\mathbf{r}_{m_3}^{\mathrm{I}})
} \right)}} 
$$
\begin{equation}\vspace{-0mm}\scalebox{1}{$
\begin{array}{l}
+ \sum\limits_{p = 1}^{{L_{{\rm{BI}}}}} {\sum\limits_{p_{1} = 1}^{{L_{{\rm{BI}}}}} {\frac{{\sigma _{{\rm{BI}}}^{\rm{4}}}}{{L_{{\rm{BI}}}^2}}} } {e^{ - j\frac{{2\pi }}{\lambda }\left( {
\rho_{p}^{\mathrm{BI}}(\mathbf{r}_{m}^{\mathrm{I}}) - \rho_{p_1}^{\mathrm{BI}}(\mathbf{r}_{m_1}^{\mathrm{I}}) - \rho_{p}^{\mathrm{BI}}(\mathbf{r}_{m_2}^{\mathrm{I}}) + \rho_{p_1}^{\mathrm{BI}}(\mathbf{r}_{m_3}^{\mathrm{I}})
} \right)}},
\end{array}$}
\end{equation}
Then, 
\begin{equation}\vspace{-0mm}\scalebox{1}{$
\begin{array}{l}
\mathbb{E}\left\{ {{h_{m,n}}h_{m_{1},n_{1}}^*h_{m_{2},n}^*{h_{m_{3},n_{1}}}} \right\} ={{{\bf{ \bar {F}}}}_{n,n_{1}}}\left[ {{{\bf{r}}_m^{\mathrm{I}}},{{\bf{r}}_{m_{1}}^{\mathrm{I}}},{{\bf{r}}_{m_{2}}^{\mathrm{I}}},{{\bf{r}}_{m_{3}}^{\mathrm{I}}}} \right]\\
 = \sum\limits_{p = 1}^{{L_{{\rm{BI}}}}} {\sum\limits_{p_{2} = 1}^{{L_{{\rm{BI}}}}} {\frac{{\sigma _{{\rm{BI}}}^{\rm{4}}}}{{L_{{\rm{BI}}}^2}}} } {e^{ - j\frac{{2\pi }}{\lambda }\left( {
\rho_{p}^{\mathrm{BI}}(\mathbf{r}_{m}^{\mathrm{I}}) - \rho_{p}^{\mathrm{BI}}(\mathbf{r}_{m_1}^{\mathrm{I}}) - \rho_{p_2}^{\mathrm{BI}}(\mathbf{r}_{m_2}^{\mathrm{I}}) + \rho_{p_2}^{\mathrm{BI}}(\mathbf{r}_{m_3}^{\mathrm{I}})
} \right)}}\\
 + \sum\limits_{p = 1}^{{L_{{\rm{BI}}}}} {\sum\limits_{p_{1} = 1}^{{L_{{\rm{BI}}}}} {\frac{{\sigma _{{\rm{BI}}}^{\rm{4}}}}{{L_{{\rm{BI}}}^2}}} } {e^{ - j\frac{{2\pi }}{\lambda }\left( {
\rho_{p}^{\mathrm{BI}}(\mathbf{r}_{m}^{\mathrm{I}}) - \rho_{p_1}^{\mathrm{BI}}(\mathbf{r}_{m_1}^{\mathrm{I}}) - \rho_{p}^{\mathrm{BI}}(\mathbf{r}_{m_2}^{\mathrm{I}}) + \rho_{p_1}^{\mathrm{BI}}(\mathbf{r}_{m_3}^{\mathrm{I}}) } \right)}}  \\
 \qquad \times e^{ - j\frac{{2\pi }}{\lambda }\left(- \rho_{p}^{\mathrm{B}}(\mathbf{r}_n^{\mathrm{B}}) + \rho_{p_1}^{\mathrm{B}}(\mathbf{r}_n^{\mathrm{B}}) + \rho_{p}^{\mathrm{B}}(\mathbf{r}_{n_1}^{\mathrm{B}}) - \rho_{p_1}^{\mathrm{B}}(\mathbf{r}_{n_1}^{\mathrm{B}})\right)}.
\end{array}$}
\end{equation}
Similarly, the second term of $A^{\mathrm{c}}_k$ can be derived as\vspace{-0mm}
\begin{equation}\vspace{-0mm}\scalebox{1}{$
\begin{array}{l}
\mathbb{E}\left\{ {{{\left| {{\bf{h}}_{{\rm{IU,}}k}^{{H}}{\bf{\Theta }}{{\bf{H}}_{{\rm{BI}}}}{{\left( {{\bf{h}}_{{\rm{IU,}}i}^{{H}}{\bf{\Theta }}{{\bf{H}}_{{\rm{BI}}}}} \right)}^H}} \right|}^2}} \right\}\\
 = \sum\limits_{n = 1}^{{N_{\mathrm{B}}}} \sum\limits_{n_{1} = 1}^{{N_{\mathrm{B}}}}\sum\limits_{m = 1}^{{N_{\mathrm{I}}}} \sum\limits_{m_{1} = 1}^{{N_{\mathrm{I}}}}  \sum\limits_{m_{2} = 1}^{{N_{\mathrm{I}}}} \sum\limits_{m_{3} = 1}^{{N_{\mathrm{I}}}} \mathbb{E}\left\{ {g_{m,k}^*{g_{m_{1},i}}{g_{m_{2},k}}g_{m_{3},i}^*} \right\}\\
\qquad \times \mathbb{E}\left\{ {{h_{m,n}}h_{m_{1},n}^*h_{m_{2},n_{1}}^*{h_{m_{3},n_{1}}}} \right\}{e^{j\left( {{\theta _m} - {\theta _{m_{1}}} - {\theta _{m_{2}}} + {\theta _{m_{3}}}} \right)}},
\end{array}$}
\end{equation}
where \vspace{-0mm}
\begin{equation}\vspace{-0mm}\scalebox{1}{$
\begin{array}{l}
\mathbb{E}\left\{ {g_{m,k}^*{g_{m_{1},i}}{g_{m_{2},k}}g_{m_{3},i}^*} \right\}={{{\bf{ \bar{G}}}}_{k,i}}\left[ {{{\bf{r}}_m^{\mathrm{I}}},{{\bf{r}}_{m_{1}}^{\mathrm{I}}},{{\bf{r}}_{m_{2}}^{\mathrm{I}}},{{\bf{r}}_{m_{3}}^{\mathrm{I}}}} \right]\\
= {\frac{{\sigma _{{\rm{IU,k}}}^{\rm{2}}}}{{{L_{{\rm{IU,k}}}}}}\frac{{\sigma _{{\rm{IU,}}i}^{\rm{2}}}}{{{L_{{\rm{IU}},i}}}}}  \\
 \quad\times\sum\limits_{l = 1}^{{L_{{\rm{IU,}}k}}} {\sum\limits_{l_{1} = 1}^{{L_{{\rm{IU,}}k}}} } {e^{j\frac{{2\pi }}{\lambda }\left( {
\rho_{l,k}^{\mathrm{IU}}(\mathbf{r}_{m}^{\mathrm{I}}) - \rho_{l_{1},i}^{\mathrm{IU}}(\mathbf{r}_{m_{1}}^{\mathrm{I}}) - \rho_{l,k}^{\mathrm{IU}}(\mathbf{r}_{m_{2}}^{\mathrm{I}}) + \rho_{l_{1},i}^{\mathrm{IU}}(\mathbf{r}_{m_{3}}^{\mathrm{I}})
} \right)}},
\end{array}$}
\end{equation}
Similarly, $B^{\mathrm{c}}_k$ can be rewritten as\vspace{-0mm}
\begin{equation}\vspace{-0mm}\scalebox{1}{$
\begin{array}{l}
B^{\mathrm{c}}_k
= \sum\limits_{m = 1}^{{N_{\mathrm{I}}}} {\sum\limits_{m_{1} = 1}^{{N_{\mathrm{I}}}} {\mathbb{E}\left\{ {g_{m,k}^*{g_{m_{1},k}}} \right\}\mathbb{E}\left\{ {{h_{m,n}}h_{m_{1},n}^*} \right\}} } {e^{j\left( {{\theta _m} - {\theta _{m_{1}}}} \right)}},
\end{array}$}
\end{equation}
where \vspace{-0mm}
\begin{equation}\vspace{-0mm}\scalebox{1}{$
\begin{array}{l}
\hspace{-4mm}\mathbb{E}\left\{ {g_{m,k}^*{g_{m_{1},k}}} \right\}={{\bf{G}}_k}\left[ {{{\bf{r}}_m^{\mathrm{I}}},{{\bf{r}}_{m_{1}}^{\mathrm{I}}}} \right]\\
 = \sum\limits_{l = 1}^{{L_{{\rm{IU,}}k}}} {\frac{{\sigma _{{\rm{IU}}}^{\rm{2}}}}{{{L_{{\rm{IU}}}}}}{e^{  j\frac{{2\pi }}{\lambda }\left(\rho_{l,k}^{\mathrm{IU}}(\mathbf{r}_m^{\mathrm{I}}) -\rho_{l,k}^{\mathrm{IU}}(\mathbf{r}_{m_{1}}^{\mathrm{I}}) \right)}}},
\end{array}$}
\end{equation}
and\vspace{-0mm}
\begin{equation}\vspace{-0mm}\scalebox{1}{$
\begin{array}{l}
\mathbb{E}\left\{ {{h_{m,n}}h_{m_{1},n}^*} \right\}={\bf{F}}\left[ {{{\bf{r}}_m^{\mathrm{I}}},{{\bf{r}}_{m_{1}}^{\mathrm{I}}}} \right]\\
 = \sum\limits_{l = 1}^{{L_{{\rm{BI}}}}} {\frac{{\sigma _{{\rm{BI}}}^{\rm{2}}}}{{{L_{{\rm{BI}}}}}}{e^{j\frac{{2\pi }}{\lambda }\left( \rho_{l,k}^{\mathrm{IU}}(\mathbf{r}_{m_{1}}^{\mathrm{I}}) -\rho_{l,k}^{\mathrm{IU}}(\mathbf{r}_{m}^{\mathrm{I}}) \right)}}{e^{j\left( {{\theta _m} - {\theta _{m_{1}}} } \right)}}},
\end{array}$}
\end{equation}
which completes the proof of calculating the first part of $B^{\mathrm{c}}_k$.
Then, $C^{\mathrm{c}}_k$ can be calculated by\vspace{-0mm}
\begin{equation}\vspace{-0mm}\scalebox{1}{$
\begin{array}{l}
C^{\mathrm{c}}_k=
  {\left| {{\eta _{\mathrm{s}}}} \right|^2}\mathbb{E}\left\{\left|  {{\alpha _{{\rm{IS,T}}}}} \right|^2\right\}\sum\limits_{n = 1}^{{N_{\mathrm{B}}}} \sum\limits_{m = 1}^{{N_{\mathrm{I}}}} \sum\limits_{m_{1} = 1}^{{N_{\mathrm{I}}}} \sum\limits_{n_{1} = 1}^{{N_{\mathrm{B}}}} \sum\limits_{m_{2} = 1}^{{N_{\mathrm{I}}}} \sum\limits_{m_{3} = 1}^{{N_{\mathrm{I}}}}\\
\qquad \times {\mathbb{E}\left\{ {g_{m,k}^*{g_{m_{2},k}}} \right\}\mathbb{E}\left\{ {{h_{m,n}}h_{m_{1},n}^*h_{m_{2},n_{1}}^*{h_{m_{3},n_{1}}}} \right\}} \\
\qquad \times {e^{j\left( {{\theta _m} - {\theta _{m_{1}}} - {\theta _{m_{2}}} + {\theta _{m_{3}}} - \frac{{2\pi }}{\lambda }\rho_{\mathrm{T}}^{\mathrm{IS}}(\mathbf{r}_{m_1}^{\mathrm{I}}) + \frac{{2\pi }}{\lambda }\rho_{\mathrm{T}}^{\mathrm{IS}}(\mathbf{r}_{m_3}^{\mathrm{I}})
} \right)}}.
\end{array}$}
\end{equation}
This completes the proof of the lower bound of SINR.
Similarly for sensing, $A^{\mathrm{s}}$, can be calculated as
\begin{equation}\vspace{-0mm}\scalebox{1}{$
\begin{array}{l}
A^{\mathrm{s}}
\hspace{-1mm} =\hspace{-1mm} \sum\limits_{k = 1}^K \hspace{-1mm}{\mathbb{E}\hspace{-1mm}\left\{ {{{\hspace{-1mm}\left| {\sum\limits_{c = 0}^C {{\alpha _{{\rm{IS,}}i}}{\bf{a}}_{{\rm{S}}}^T(\varphi_{\mathrm{T}}^{\rm{IS}}){\bf{a}}_{{\rm{S}}}^*(\varphi_{c}^{\rm{IS}}){\bf{a}}_{\rm{I}}^T(\varphi_{c}^{\rm{IS}},\omega _{c}^{\rm{IS}}){\bf{\Theta }}{{\bf{H}}_{{\rm{BI}}}}{{\bf{w}}_{\mathrm{c},k}}} } \right|}^2}} \hspace{-1mm}\right\}} \\
 + \mathbb{E}\left\{ {{{\left| {\sum\limits_{c = 0}^C {{\alpha _{{\rm{IS,}}i}}{\bf{a}}_{{\rm{S}}}^T(\varphi_{\mathrm{T}}^{\rm{IS}}){\bf{a}}_{{\rm{S}}}^*(\varphi_{c}^{\rm{IS}}){\bf{a}}_{\rm{I}}^T(\varphi_{c}^{\rm{IS}},\omega _{c}^{\rm{IS}}){\bf{\Theta }}{{\bf{H}}_{{\rm{BI}}}}{{\bf{w}}_{\mathrm{s}}}} } \right|}^2}} \right\},
\end{array}$}
\end{equation}
define \vspace{-0mm}
\begin{equation}\vspace{-0mm}\scalebox{1}{$
{{\bf{J}}_c}\left[ {{{\bf{r}}_{m_{2}}^{\mathrm{I}}},{{\bf{r}}_{m_{}}^{\mathrm{I}}}} \right] = {e^{j\frac{{2\pi }}{\lambda }\left( {
\rho_{c}^{\mathrm{IS}}(\mathbf{r}_{m_{}}^{\mathrm{I}}) - \rho_{c}^{\mathrm{IS}}(\mathbf{r}_{m_2}^{\mathrm{I}}),
} \right)}}$}
\end{equation}
and the first term of $A^{\mathrm{s}}$ can be calculated by\vspace{-0mm}
\begin{equation}\vspace{-0mm}\scalebox{1}{$
\begin{array}{l}
\sum\limits_{k = 1}^K {\mathbb{E}\left\{ {{{\left| {\sum\limits_{c = 0}^C {{\alpha _{{\rm{IS,}}c}}{\bf{a}}_{{\rm{S}}}^T(\varphi_{\mathrm{T}}^{\rm{IS}}){\bf{a}}_{{\rm{S}}}^*(\varphi_{c}^{\rm{IS}}){\bf{a}}_{\rm{I}}^T(\varphi_{c}^{\rm{IS}},\omega _{c}^{\rm{IS}}){\bf{\Theta }}{{\bf{H}}_{{\rm{BI}}}}{{\bf{w}}_{\mathrm{c},k}}} } \right|}^2}} \right\}} \\
=\left|\eta _k\right|^2\sum\limits_{c = 0}^C {{\left| {\bf{a}}_{{\rm{S}}}^T(\varphi_{\mathrm{T}}^{\rm{IS}}){\bf{a}}_{{\rm{S}}}^*(\varphi_{c}^{\rm{IS}}) \right|}^2}\sum\limits_{n = 1}^{{N_{\mathrm{B}}}} \sum\limits_{m = 1}^{{N_{\mathrm{I}}}} \sum\limits_{m_{1} = 1}^{{N_{\mathrm{I}}}} \sum\limits_{n_{1} = 1}^{{N_{\mathrm{B}}}} \sum\limits_{m_{2} = 1}^{{N_{\mathrm{I}}}} \sum\limits_{m_{3} = 1}^{{N_{\mathrm{I}}}} \\
 \times {\mathbb{E}\left\{ {{{\left| {{\alpha _{{\rm{IS,}}c}}} \right|}^2}} \right\}} \mathbb{E}\left\{ {{h_{m,n}}h_{m_{1},n}^*h_{m_{2},n_{1}}^*{h_{m_{3},n_{1}}}} \right\} \mathbb{E}\left\{ {{g_{m_{1},k}}g_{m_{3},k}^*} \right\}\\
\times{{\bf{J}}_c}\left[ {{{\bf{r}}_{m_{2}}^{\mathrm{I}}},{{\bf{r}}_{m_{}}^{\mathrm{I}}}} \right] {e^{j\left( {
  {\theta _m} - {\theta _{m_{1}}} - {\theta _{m_{2}}} + {\theta _{m_{3}}}
} \right)}}, 
\end{array}$}
\end{equation}
define 
\begin{equation}\label{eq:eqsyend}\vspace{-0mm}\scalebox{1}{$
\begin{array}{l}
{{{\bf{\bar {J}}}}_{\mathrm{T},c}}\left[ {{{\bf{r}}_m^{\mathrm{I}}},{{\bf{r}}_{m_{1}}^{\mathrm{I}}},{{\bf{r}}_{m_{2}}^{\mathrm{I}}},{{\bf{r}}_{m_{3}}^{\mathrm{I}}}} \right] \\
\qquad= {e^{j\frac{{2\pi }}{\lambda }\left( {\rho_{c}^{\mathrm{IS}}(\mathbf{r}_{m}^{\mathrm{I}}) - \rho_{\mathrm{T}}^{\mathrm{IS}}(\mathbf{r}_{m_1}^{\mathrm{I}}) - \rho_{c}^{\mathrm{IS}}(\mathbf{r}_{m_2}^{\mathrm{I}}) + \rho_{\mathrm{T}}^{\mathrm{IS}}(\mathbf{r}_{m_3}^{\mathrm{I}})
} \right)}}.
\end{array}$}
\end{equation}
and the second term is\vspace{-0mm}
\begin{equation}\vspace{-0mm}\scalebox{1}{$
\begin{array}{l}
\mathbb{E}\left\{ {{{\left| {\sum\limits_{c = 0}^C {{\alpha _{{\rm{IS,}}c}}{\bf{a}}_{{\rm{S}}}^T(\varphi_{\mathrm{T}}^{\rm{IS}}){\bf{a}}_{{\rm{S}}}^*(\varphi_{c}^{\rm{IS}}){\bf{a}}_{\rm{I}}^T(\varphi_{c}^{\rm{IS}},\omega _{c}^{\rm{IS}}){\bf{\Theta }}{{\bf{H}}_{{\rm{BI}}}}{{\bf{w}}_{\mathrm{s}}}} } \right|}^2}} \right\}\\
 = \left|\eta _{\mathrm{s}}\right|^2\hspace{-1mm}\sum\limits_{c = 0}^{C} \left|{\bf{a}}_{{\rm{S}}}^T(\theta _{{\rm{IS,T}}}^{\rm{r}}){\bf{a}}_{{\rm{S}}}^*(\theta _{{\rm{IS,}}c}^{\rm{r}})\right|^2\hspace{-1mm}\sum\limits_{n = 1}^{{N_{\mathrm{B}}}} \hspace{-0.5mm}\sum\limits_{m = 1}^{{N_{\mathrm{I}}}} \hspace{-0.5mm}\sum\limits_{m_{1} = 1}^{{N_{\mathrm{I}}}} \hspace{-0.5mm}\sum\limits_{n_{1} = 1}^{{N_{\mathrm{B}}}} \hspace{-0.5mm}\sum\limits_{m_{2} = 1}^{{N_{\mathrm{I}}}} \hspace{-0.5mm}\sum\limits_{m_{3} = 1}^{{N_{\mathrm{I}}}} \\
 \times
{\mathbb{E}\left\{ {{{\left| {{\alpha _{{\rm{IS,}}c}}} \right|}^2}} \right\}\mathbb{E}\left\{ {{{\left| {{\alpha _{{\rm{IS,T}}}}} \right|}^2}} \right\}\mathbb{E}\left\{ {{h_{m,n}}h_{m_{1},n}^*h_{m_{2},n_{1}}^*{h_{m_{3},n_{1}}}} \right\}}\\
 \times {{{\bf{\bar {J}}}}_{\mathrm{T},c}}\left[ {{{\bf{r}}_m^{\mathrm{I}}},{{\bf{r}}_{m_{1}}^{\mathrm{I}}},{{\bf{r}}_{m_{2}}^{\mathrm{I}}},{{\bf{r}}_{m_{3}}^{\mathrm{I}}}} \right]{e^{j\left( { {\theta _m} - {\theta _{m_{1}}} - {\theta _{m_{2}}} + {\theta _{m_{3}}}} \right)}}\\
 + \left|\eta _{\mathrm{s}}\right|^2\left\|{\bf{a}}_{{\rm{S}}}^T(\theta _{{\rm{IS,T}}}^{\rm{r}})\right\|^4\sum\limits_{n = 1}^{{N_{\mathrm{B}}}} \sum\limits_{m = 1}^{{N_{\mathrm{I}}}} \sum\limits_{m_{1} = 1}^{{N_{\mathrm{I}}}} \sum\limits_{n_{1} = 1}^{{N_{\mathrm{B}}}} \sum\limits_{m_{2} = 1}^{{N_{\mathrm{I}}}} \sum\limits_{m_{3} = 1}^{{N_{\mathrm{I}}}} \\
  \times
{\mathbb{E}\left\{ {{{\left| {{\alpha _{{\rm{IS,T}}}}} \right|}^4}} \right\}\mathbb{E}\left\{ {{h_{m,n}}h_{m_{1},n}^*h_{m_{2},n_{1}}^*{h_{m_{3},n_{1}}}} \right\}}  \\
  \times    {{{\bf{\bar {J}}}}_{\mathrm{T},c}}\left[ {{{\bf{r}}_m^{\mathrm{I}}},{{\bf{r}}_{m_{1}}^{\mathrm{I}}},{{\bf{r}}_{m_{2}}^{\mathrm{I}}},{{\bf{r}}_{m_{3}}^{\mathrm{I}}}} \right]{e^{j\left( {  {\theta _m} - {\theta _{m_{1}}} - {\theta _{m_{2}}} + {\theta _{m_{3}}}} \right)}},
\end{array}$}
\end{equation}
Similarly, the desired signal of sensing, i.e., $B^{\mathrm{s}}$, can be rewritten into\vspace{-0mm}
\begin{equation}\vspace{-0mm}\scalebox{1}{$
\begin{array}{l}
B^{\mathrm{s}}={{{\left\| {\mathbb{E}\left\{ {\alpha _{{\rm{IS,T}}}}{{{\bf{r}}^H_{\mathrm{co}}}{\bf{a}}_{{\rm{S}}}^*(\varphi _{{\rm{T}}}^{\rm{S}}){\bf{a}}_{\rm{I}}^T(\varphi _{{\rm{T}}}^{\rm{IS}},\omega _{{\rm{T}}}^{\rm{IS}}){\bf{\Theta }}{{\bf{H}}_{{\rm{BI}}}}{\bf{W}}} \right\} } \right\|}^2}}\\
={\eta _{\mathrm{s}}}N_{\mathrm{S}}\mathbb{E}\left\{ {{{\left| {{\alpha _{{\rm{IS,T}}}}} \right|}^2}} \right\}\sum\limits_{n = 1}^{{N_{\mathrm{B}}}} \sum\limits_{m = 1}^{{N_{\mathrm{I}}}} \sum\limits_{m_{1} = 1}^{{N_{\mathrm{I}}}} \mathbb{E}\left\{ {{h_{m,n}}h_{m_{1},n}^*} \right\}\\
\qquad \times{{\bf{J}}_c}\left[ {{{\bf{r}}_{m_{1}}^{\mathrm{I}}},{{\bf{r}}_{m_{}}^{\mathrm{I}}}} \right]{e^{j\left( { {\theta _m} - {\theta _{m_{1}}}} \right)}}.
\end{array}$}
\end{equation}

This completes the proof.
\vspace{-0mm}
\section{Proof of Corollary \ref{coro:2}}\label{app:coro:2}
For the considered traditional fixed-spacing IRS, the expectations for communication and sensing can be rewritten into\vspace{-0mm}
\begin{equation}\vspace{-0mm}\scalebox{1}{$
\begin{array}{l}
A^{\mathrm{c}}_k = 8{p_{c,k}}\sigma _{{\rm{IU,k}}}^{\rm{2}}\sigma _{{\rm{BI}}}^2N_B^{}{{\left( {\frac{{\sin \left( {\frac{\pi }{\lambda }{\rm{d}}{N_I}\left( {\varphi _{\rm{k}}^{{\rm{IU}}} - {\varphi ^{{\rm{BI}}}}} \right)} \right)}}{{\sin \left( {\frac{\pi }{\lambda }{\rm{d}}\left( {\varphi _{\rm{k}}^{{\rm{IU}}} - {\varphi ^{{\rm{BI}}}}} \right)} \right)}}} \right)}^2} \\
+ 4\sigma _{{\rm{BI}}}^2\sigma _{{\rm{IU,k}}}^{\rm{2}}N_B^{}{{\left( {\frac{{\sin \left( {\frac{\pi }{\lambda }{\rm{d}}{N_I}\left( {\varphi _{\rm{k}}^{{\rm{IU}}} - {\varphi ^{{\rm{BI}}}}} \right)} \right)}}{{\sin \left( {\frac{\pi }{\lambda }{\rm{d}}\left( {\varphi _{\rm{k}}^{{\rm{IU}}} - {\varphi ^{{\rm{BI}}}}} \right)} \right)}}} \right)}^2}\sum\limits_{i \ne k}^K {{p_{c,i}}},
\end{array}$}
\end{equation}
\begin{equation}\vspace{-0mm}\scalebox{1}{$
\begin{array}{l}
B^{\mathrm{c}}_k={{p_{c,k}}N_B^{}\sigma _{{\rm{IU,k}}}^2\sigma _{{\rm{BI}}}^2{{\left( {\frac{{\sin \left( {\frac{\pi }{\lambda }{\rm{d}}{N_I}\left( {\varphi _{\rm{k}}^{{\rm{IU}}} - {\varphi ^{{\rm{BI}}}}} \right)} \right)}}{{\sin \left( {\frac{\pi }{\lambda }{\rm{d}}\left( {\varphi _{\rm{k}}^{{\rm{IU}}} - {\varphi ^{{\rm{BI}}}}} \right)} \right)}}} \right)}^2}},
\end{array}$}
\end{equation}
and\vspace{-0mm}
\begin{equation}\vspace{-0mm}\scalebox{1}{$
\begin{array}{l}
C^{\mathrm{c}}_k={2{p_s}\sigma _{{\rm{BI}}}^{\rm{2}}\sigma _{{\rm{IU}}}^{\rm{2}}\sigma _{{\rm{IS}}}^{\rm{2}}N_B^{}{{\left( {\frac{{\sin \left( {\frac{\pi }{\lambda }{\rm{d}}{N_I}\left( {\varphi _{\rm{k}}^{{\rm{IU}}} - {\varphi ^{{\rm{BI}}}}} \right)} \right)}}{{\sin \left( {\frac{\pi }{\lambda }{\rm{d}}\left( {\varphi _{\rm{k}}^{{\rm{IU}}} - {\varphi ^{{\rm{BI}}}}} \right)} \right)}}} \right)}^2}}.
\end{array}$}
\end{equation}

Similarly, the expectations for sensing can be expressed by\vspace{-0mm}
\begin{equation}\vspace{-0mm}\scalebox{1}{$
\begin{array}{l}
A^{\mathrm{s}}=2\sum\limits_{l = 1}^{1,T} {{{\left| {{\bf{a}}_{\rm{S}}^T(\theta _{\rm{T}}^S){\bf{a}}_{\rm{S}}^*(\theta _l^S)} \right|}^2}{{\left( {\frac{{\sin \left( {\frac{\pi }{\lambda }{\rm{d}}{N_I}\left( {\varphi _l^{{\rm{IS}}} - {\varphi ^{{\rm{BI}}}}} \right)} \right)}}{{\sin \left( {\frac{\pi }{\lambda }{\rm{d}}\left( {\varphi _l^{{\rm{IS}}} - {\varphi ^{{\rm{BI}}}}} \right)} \right)}}} \right)}^2}} \sum\limits_{k = 1}^K {{p_{c,k}}} \\
 + {p_s}{{\left| {{\bf{a}}_{\rm{S}}^T(\theta _{{\rm{IS,T}}}^{\rm{r}}){\bf{a}}_{\rm{S}}^*(\theta _{{\rm{IS,1}}}^{\rm{r}})} \right|}^2}\sigma _{{\rm{IS,0}}}^{\rm{2}}{{\left( {\frac{{\sin \left( {\frac{\pi }{\lambda }{\rm{d}}{N_I}\left( {\varphi _{\rm{0}}^{{\rm{IS}}} - {\varphi ^{{\rm{BI}}}}} \right)} \right)}}{{\sin \left( {\frac{\pi }{\lambda }{\rm{d}}\left( {\varphi _{\rm{0}}^{{\rm{IS}}} - {\varphi ^{{\rm{BI}}}}} \right)} \right)}}} \right)}^2}\\
 + 2\sigma _{{\rm{IS,T}}}^{\rm{2}}{p_s}{{\left\| {{\bf{a}}_{\rm{S}}^T(\theta _{{\rm{IS,T}}}^{\rm{r}})} \right\|}^4}{{\left( {\frac{{\sin \left( {\frac{\pi }{\lambda }{\rm{d}}{N_I}\left( {\varphi _{\rm{T}}^{{\rm{IS}}} - {\varphi ^{{\rm{BI}}}}} \right)} \right)}}{{\sin \left( {\frac{\pi }{\lambda }{\rm{d}}\left( {\varphi _{\rm{T}}^{{\rm{IS}}} - {\varphi ^{{\rm{BI}}}}} \right)} \right)}}} \right)}^2},
\end{array}$}
\end{equation}
and\vspace{-0mm}
\begin{equation}\vspace{-0mm}\scalebox{1}{$
\begin{array}{l}
B^{\mathrm{s}}={N_{\rm{S}}^2{p_s}\frac{{\sigma _{{\rm{IS,T}}}^2}}{4}{{\left( {\frac{{\sin \left( {\frac{\pi }{\lambda }{\rm{d}}{N_I}\left( {\varphi _{\rm{T}}^{{\rm{IS}}} - {\varphi ^{{\rm{BI}}}}} \right)} \right)}}{{\sin \left( {\frac{\pi }{\lambda }{\rm{d}}\left( {\varphi _{\rm{T}}^{{\rm{IS}}} - {\varphi ^{{\rm{BI}}}}} \right)} \right)}}} \right)}^2}}.
\end{array}$}
\end{equation}

It can be observed that $\gamma_{\mathrm{c},k}^{\mathrm{lb}}\left( {{{{\bf{W}}},{\bf{r}^{\mathrm{I}}},{\bf{\Theta }}}} \right)$ is an increasing function of ${{\left( {\frac{{\sin \left( {\frac{\pi }{\lambda }{\rm{d}}{N_I}\left( {\varphi _{\rm{k}}^{{\rm{IU}}} - {\varphi ^{{\rm{BI}}}}} \right)} \right)}}{{\sin \left( {\frac{\pi }{\lambda }{\rm{d}}\left( {\varphi _{\rm{k}}^{{\rm{IU}}} - {\varphi ^{{\rm{BI}}}}} \right)} \right)}}} \right)}^2}$, of which the $L^{c}$-th zero point is derived as $d = \frac{{L^{c}}}{{{N_I}\left( {\varphi _{\rm{k}}^{{\rm{IU}}} - {\varphi ^{{\rm{BI}}}}} \right)}}\lambda$.
Similarly, $\gamma_{\mathrm{s}}^{\mathrm{lb}}\left( {{{\bf{W}}},{\bf{r}^{\mathrm{I}}},{\bf{\Theta }}},{\bf{r}_{\mathrm{co}}} \right)$ is an decreasing function of $\left( {\frac{{\sin \left( {\frac{\pi }{\lambda }{\rm{d}}\left( {\varphi _{\rm{T}}^{{\rm{IS}}} - {\varphi ^{{\rm{BI}}}}} \right)} \right)}\sin \left( {\frac{\pi }{\lambda }{\rm{d}}{N_I}\left( {\varphi _{\rm{0}}^{{\rm{IS}}} - {\varphi ^{{\rm{BI}}}}} \right)} \right)}{{\sin \left( {\frac{\pi }{\lambda }{\rm{d}}{N_I}\left( {\varphi _{\rm{T}}^{{\rm{IS}}} - {\varphi ^{{\rm{BI}}}}} \right)} \right)}{\sin \left( {\frac{\pi }{\lambda }{\rm{d}}\left( {\varphi _{\rm{0}}^{{\rm{IS}}} - {\varphi ^{{\rm{BI}}}}} \right)} \right)}}} \right)^2$, of which the maximal point is derived at $d = \frac{{L^{s}_1}}{{{N_I}\left( {\varphi _{\rm{T}}^{{\rm{IS}}} - {\varphi ^{{\rm{BI}}}}} \right)}}\lambda = \frac{{L^{s}_2}}{{\left( {\varphi _{\rm{0}}^{{\rm{IS}}} - {\varphi ^{{\rm{BI}}}}} \right)}}\lambda$. This thus completes the proof.

\end{document}